\documentclass{JHEP3}
\usepackage[all]{xy}
\usepackage{latexsym,graphics,epsfig}
\usepackage{rotate}
\usepackage{amsfonts,amssymb}

\title{Higher Gauge Theory: 2-Connections on 2-Bundles}

\author{John Baez
\\ Department of Mathematics
\\ University of California
\\ Riverside, CA 92521, USA 
\\ E-mail: \email{baez@math.ucr.edu}}

\author{Urs Schreiber 
\\ Fachbereich Physik
\\ Universit{\"a}t Duisburg-Essen 
\\ Essen, 45117, Germany
\\ E-mail: \email{Urs.Schreiber@uni-essen.de}}

\def\bigdisunion {\bigsqcup}

\newcommand{\maps}{\colon}
\def\ker {{\rm ker}}
\def\image {{\rm im}}

\def\stackto #1 { {\stackrel{#1}{\longrightarrow}} }
\def\stackTo #1 { {\stackrel{#1}{\Longrightarrow}} }

\newcommand{\U}{{\rm U}}

\newcommand{\Aut}{{\rm Aut}}

\def\g {\mathfrak{g}}
\def\h {\mathfrak{h}}

\def\Set {{\mathrm{Set}}}
\def\Diff {{\mathrm{Diff}}}
\def\DiffInfty {{\mathrm{Diff}_\infty}}
\def\Top {{\mathrm{Top}}}
\def\Cat {{\mathrm{Cat}}}
\def\Grp {{\mathrm{Grp}}}
\def\twoDiff {{\mathfrak{C}}}
\def\Lie {{\mathrm{Lie}}}

\def\Ob {{\rm Ob}}
\def\Mor {{\rm Mor}}

\def\paths {\mathcal{P}}
\def\loops {\mathrm{L}}

\renewcommand{\P}{\paths}

\def\hol {{\rm hol}}

\def\trans {{\rm Trans}}

\def\annotation #1 {\marginpar{\footnotesize {#1}}}

   \def\twogroup {\mathcal{G}}

   \def\crossedmodule {\mathcal{C}}

   \def\H {H}

   \def\standardpathspaceconnection {\mathcal{A}}

   \def\manifold {{M}}

   \def\twoB {{B}}

   \def\twoF {{F}}

   \def\twoU {{U}}

   \def\twoP {{P}}

\newtheorem{theorem}{Theorem}[section]
\newtheorem{definition}{Definition}[section]{}
{}
\newtheorem{proposition}{Proposition}[section]

\newtheorem{corollary}[theorem]{Corollary}

\def\comment #1{}
\def\cf {{\it cf. }}

\def\refer #1{{(\ref{#1})}}
\def\fullref #1{\ref{#1} (p.\pageref{#1})}
\def\refdef #1{{(def. \ref{#1})}}

\def\of #1{\!\left({#1}\right)}

\def\set #1{\left\lbrace{#1}\right\rbrace}
\def\id {\mathrm{id}}

\def\brackets #1{\left[{#1}\right]}
\def\braces #1{\left\lbrace{#1}\right\rbrace}

\def\order #1{\mathcal{O}\of{#1}}

\def\commutator #1#2{\brackets{{#1},{#2}}}
\def\antiCommutator #1#2{\braces{{#1},{#2}}}

\def\defas {\equiv}

\def\equalby #1{\stackrel{\refer{#1}}{=}}

\def\extd {\mathbf{{d}}}

\def\endofproof {\hfill$\Box$\newline\newline}

\def\eigenspace #1#2 {\mathrm{eig}\of{#1,#2}}

\newlength{\skiplength}
\def\skiph #1{\settowidth{\skiplength}{#1}\hspace{\skiplength}}

\def\inner {\!\cdot\!}

\abstract{
  Connections and curvings on gerbes are beginning to play a vital role in 
  differential geometry and mathematical physics --- first 
  abelian gerbes, and more recently nonabelian gerbes.  These concepts
  can be elegantly understood using the concept of `2-bundle' recently 
  introduced by Bartels.  A 2-bundle is a generalization of a bundle in 
  which the fibers are categories rather than sets.  Here we 
  introduce the concept of a `2-connection' on a principal 2-bundle.
  We describe principal 2-bundles with connection in terms of local data,
  and show that under certain conditions this reduces to the cocycle data 
  for nonabelian gerbes with connection and curving subject to a certain
  constraint --- namely, the vanishing of the `fake curvature', as defined
  by Breen and Messing.   This constraint also turns out to guarantee the 
  existence of `2-holonomies': that is, parallel transport over both curves 
  and surfaces, fitting together to define a 2-functor from the `path
  2-groupoid' of the base space to the structure 2-group.  We give a general 
  theory of 2-holonomies and show how they are related to ordinary parallel 
  transport on the path space of the base manifold.}

\begin{document}

\newpage

\section{Introduction}

The gauge principle and the concept of connection is at the very heart
of modern physics, and is also central to much of modern mathematics. 
It is all about parallel transport along \emph{curves}.  Due to the 
influence of string theory on
the one hand (see \S\fullref{section: Nonabelian 2-Holonomies and Physics}) 
and higher category theory on the other (see \S\fullref{section: Higher Gauge Theory}), 
there are compelling reasons to generalize this concept to higher dimensions and 
find a notion of parallel transport along \emph{surfaces}.

\subsection{Outline of the Results}

In this paper we categorify
the notion of a \emph{principal bundle with connection},
defining \emph{principal 2-bundles with 2-connections}.
We work out their description in terms of local data and show that 
under certain conditions this is equivalent to the cocycle description of 
(possibly twisted) nonabelian gerbes satisfying a certain constraint ---
the vanishing of the `fake curvature'.  We show that this constraint is 
also sufficient to guarantee the existence of \emph{2-holonomies}, i.e.,
parallel transport over surfaces.  We examine these 2-holonomies
in detail using 2-functors into 2-groups on the one hand,
and connections on path space on the other hand.

Several aspects of this have been studied before.   
Categorification is described in \cite{BaezDolan:1998}
and its application to groups and Lie algebras, which yields 
2-groups and Lie 2-algebras, is discussed in 
\cite{BaezLauda:2003,BaezCrans:2003,Baez:2002}.  The concept of
2-group is incorporated in the definition of principal 2-bundles 
(without connection) in \cite{Bartels:2004}.  Local 2-connections as 
2-functors from path 2-groupoids to 2-groups were considered in 
\cite{Pfeiffer:2003,GirelliPfeiffer:2004}.  Connections on path space
were discussed in \cite{AlvarezFerreiraSanchezGuillen:1998,Schreiber:2004e},
and reparametrization invariance for a special case was investigated 
by \cite{Alvarez:2004}.  Cocycle data for nonabelian gerbes with
connection and curving were obtained originally in 
\cite{BreenMessing:2001} using algebraic geometry
and later by Aschieri and Jur{\v c}o 
\cite{AschieriCantiniJurco:2003,AschieriJurco:2004}
using a nonabelian generalization of the bundle gerbes originally
introduced in \cite{Murray:1994}.


Here we extend this work by:
\begin{itemize}
  \item 
    defining the concept of a principal 2-bundle with 2-connection,
  \item showing that a 2-connection on a trivial principal 2-bundle
    has 2-holonomies defining a 2-functor into the structure 2-group
    when the 2-connection has vanishing `fake curvature' (a concept 
    already defined for nonabelian gerbes by Breen and Messing 
    \cite{BreenMessing:2001}),
  \item
    clarifying the relation between connections on a trivial
    principal bundle over the path space of
    a manifold and 2-connections on a trivial principal 2-bundle over the 
    manifold itself, showing that a connection on the path space whose
    holonomies are invariant under arbitrary surface reparameterizations 
    defines a 2-connection on the original manifold,
   \item
    deriving the local `gluing data' that describe how a nontrivial 
    2-bundle with 2-connection can be built from trivial 2-bundles with 
    2-connection on open sets that cover the base manifold,
   \item
    demonstrating that these gluing data for 2-bundles with
    2-connection coincide with the cocycle
    description of (possibly twisted) nonabelian gerbes, subject to the
    constraint of vanishing fake curvature.
\end{itemize}

The starting point for all these considerations is the ordinary concept
of a principal fiber bundle.
Such a bundle can be specified using the following
`gluing data': 
\begin{itemize}
  \item
    a base manifold $B$,
  \item
    a cover of $B$ by open sets $\set{U_i}_{i\in I}$,
  \item
    a Lie group $G$ (the `gauge group' or `structure group'), 
  \item
    on each double overlap $U_{ij} = U_i \cap U_j$ a 
    $G$-valued function $g_{ij}$, 
  \item
    such that on triple overlaps
    the following transition law holds:
    \begin{eqnarray}
       g_{ij} g_{jk} = g_{ik}.
       \nonumber
    \end{eqnarray}
\end{itemize}
Such a bundle is augmented with a connection by specifying:
\begin{itemize}
  \item
    in each open set $U_i$ a smooth functor $\mathrm{hol}_i \maps
    \P_1(U_i) \to G$ from the path groupoid of $U_i$ to the gauge group, 
  \item
   such that for all paths $\gamma$ in double overlaps $U_{ij}$
   the following transition law holds:
   \begin{eqnarray}
     \mathrm{hol}_i\of{\gamma} &=& g_{ij} \, \mathrm{hol}_j\of{\gamma} \,  g_{ij}^{-1}.
     \nonumber
   \end{eqnarray}
\end{itemize}
Here the `path groupoid' $\P_1(X)$ of a manifold $X$ has points of $X$
as objects and certain equivalence classes of smooth paths in 
$X$ as morphisms.  There are various ways to work out the technical 
details; see \cite{MackaayPicken:2000} for the approach we adopt, which 
uses `thin homotopy classes' of smooth paths.  Technical details aside, the 
basic idea is that a connection on a trivial $G$-bundle over $X$ gives a 
well-behaved map assigning to each path $\gamma$ the holonomy 
$\hol(\gamma) \in G$ of the connection along that path.   Saying this
map is a `smooth functor' means that these holonomies compose when 
we compose paths, and that the holonomy $\hol(\gamma)$ depends smoothly on 
the path $\gamma$.

Our task shall be to categorify all of this and to work out the 
consequences.  The basic tool will be \emph{internalization}:
given a mathematical concept $X$ defined solely in terms of 
sets, functions and commutative diagrams involving these, and given 
any category $C$, one obtains the concept of an `$X$ in $C$' by 
replacing all these sets, functions and commutative diagrams by
corresponding objects, morphisms, and commutative diagrams in $C$.

For example, take $X$ to be the concept of `group'.  A group in $\Diff$ 
(the category with finite-dimensional smooth spaces as objects and smooth
maps as morphisms) is nothing but a \emph{Lie group}.  In other words, a
Lie group is a group that is a manifold, for which all the group operations
are smooth maps.  Similarly, a group in $\Top$ (the category with 
topological spaces as objects and continuous maps as morphisms)
is a \emph{topological group}.  

These examples are standard, but we will need some slightly less familiar 
ones.  In particular, we will need the concept of `strict 2-group', which 
is a group in  $\Cat$ (the category with categories as objects and functors as 
morphisms).  By a charming principle called `commutativity of 
internalization', strict 2-groups can also be thought of as categories
in $\Grp$ (the category with groups as objects and homomorphisms as
morphisms).   We will also need the concept of a `2-space', which is a 
category either in $\Diff$, or perhaps in 
some more general category of smooth spaces,
allowing for infinite-dimensional examples.  Finally, combining the concepts
of strict 2-group and 2-space, we obtain that of `strict Lie 2-group'.  
This is a strict 2-group in $\Diff$, or equivalently, a group in the 
category of 2-spaces.   

To arrive at the definition of a 2-bundle $P \to M$, the first steps are to 
replace the total space $P$ and base space $M$ by 2-spaces, and to replace 
the structure group by a strict Lie 2-group.  Often it is interesting to 
keep the base space an ordinary space, which can be regarded as a 2-space 
with only identity morphisms.  However, for applications to string theory 
we will also be interested in more general 2-bundles, where the base 2-space 
has some manifold $X$ as its space of objects and the free loop space $LX$ 
as its space of morphisms \cite{Schreiber:2004}.  
This sort of application also requires that we 
consider smooth spaces that are more general than finite-dimensional smooth 
manifolds.

Just as a connection on a trivial principal bundle over $M$ gives a 
functor $\hol$ from the path groupoid of $M$ to the structure group, one 
might hope that a 2-connection on a trivial principal 2-bundle would define
a 2-functor from some sort of `path 2-groupoid' to the structure 2-group. 
This has already been studied in \cite{Pfeiffer:2003,GirelliPfeiffer:2004} 
in the context of lattice 2-gauge theory.  
Thus, the main issues not yet addressed are those 
involving differentiability. 

To address these issues, we define for any smooth space $M$ a smooth 
2-groupoid $\P_2(M)$ such that:
\begin{itemize}
\item
the objects of $\P_2(M)$ are points of $M$: 
$ \quad   \bullet \, {\textstyle{\small{\textit{x}}}}   $
\item 
the morphisms of $\P_2(M)$ are smooth paths $\gamma \maps [0,1] \to M$ such
that $\gamma(t)$ is constant in a neighborhood of $t = 0$ and $t = 1$:
$
\xymatrix{
   \bullet \ar@/^1pc/[rr]^{\gamma}
&& \bullet
}
$
\item 
the 2-morphisms of $\P_2(M)$ are thin homotopy classes of smooth maps 
$f \maps [0,1]^2 \to M$ such that $f(s,t)$ is independent of $s$ in a 
neighborhood of $s = 0$ and $s = 1$, and constant in a neighborhood of 
$t = 0$ and $t = 1$:
$
\xymatrix{
   \bullet \ar@/^1pc/[rr]^{\gamma_1}="0"
           \ar@/_1pc/[rr]_{\gamma_2}="1"
&& \bullet \ar@{=>}"0";"1"^{f}
}
$
\end{itemize}
We call the 2-morphisms in $\P_2(M)$ `bigons'.
The `thin homotopy' equivalence relation guarantees that two maps differing 
only by a reparametrization define the same bigon.  This is important because 
we seek a {\it reparametrization-invariant notion of surface holonomy}.

We show that any 2-connection on a principal 2-bundle over $M$ with vanishing
fake curvature yields a smooth 2-functor $\hol \maps \P_2(M) \to \twogroup$, 
where $\twogroup$ 
is the structure 2-group.   We call this 2-functor the \emph{2-holonomy} of 
the 2-connection.
In simple terms, its existence means the 2-connection has
well-defined holonomies both for paths and surfaces, compatible with the
standard operations of composing paths and surfaces, and depending smoothly
on the path or surface in question.  

To expand on this slightly, one must recall \cite{BaezLauda:2003}
that any strict Lie 2-group $\twogroup$ determines a `crossed module' of Lie groups
$(G,H,t,\alpha)$, where:
\begin{itemize}
\item $G$ is the Lie group of objects of $\twogroup$,
\item $H$ is the Lie group of morphisms of $\twogroup$ with source equal to $1 \in G$,
\item $t \maps H \to G$ is the homomorphism sending each morphism
in $H$ to its target,
\item $\alpha$ is the action of $G$ as automorphisms of $H$ defined
using conjugation in $H$ as follows: $\alpha(g) h = 1_g h {1_g}^{-1}$. 
\end{itemize}
In these terms, a 2-connection on a trivial principal 2-bundle over $M$
with structure 2-group $\twogroup$ is nothing but a $\mathrm{Lie}\of{G}$-valued
1-form $A$ together with a  $\mathrm{Lie}\of{H}$-valued 2-form $B$ on $M$. 
Translated into this framework, Breen and Messing's `fake curvature' is the 
$\mathrm{Lie}\of{G}$-valued 2-form 
\[
dt\of{B} + F_A, 
\]
where $F_A = dA + A \wedge A$ is the usual curvature of $A$.
We show that {\it when the fake curvature vanishes}, 
but not in general otherwise,
there is a well-defined 2-holonomy $\hol \maps \P_2(M) \to \twogroup$.
The importance of vanishing fake curvature in the framework of lattice gauge
theory was already emphasized in \cite{GirelliPfeiffer:2004}. 
The special case where also $F_A = 0$ was studied in 
\cite{AlvarezFerreiraSanchezGuillen:1998}, 
while a discussion of this constraint in terms of loop space in the case 
$G = H$ was given in \cite{Schreiber:2004e}.  Our result subsumes these 
cases in a common framework.

This framework for 2-connections on trivial 2-bundles is sufficient
for local considerations.  Thus, all that remains 
is to turn it into a global notion
by categorifying the \emph{transition laws} for a principal bundle with 
connection, which in terms of local data read:
\begin{eqnarray}
    g_{ij} g_{jk} &=& g_{ik}
  \nonumber\\
  \mathrm{hol}_i\of{\gamma} &=& g_{ij} \mathrm{hol}_j\of{\gamma} g_{ij}^{-1}
  \nonumber
  \,.
\end{eqnarray}
The basic idea is to replace these equations by specified isomorphisms, 
using the fact that a 2-group $\twogroup$ has not only objects (forming the group $G$)
but also morphisms (which can be describing using elements of the group $H$).
These isomorphisms should in turn satisfy certain coherence laws of their
own.  These coherence laws have already been worked out for 2-bundles without
connection \cite{Bartels:2004} and for twisted nonabelian gerbes with 
connection and curving 
\cite{BreenMessing:2001,AschieriCantiniJurco:2003,AschieriJurco:2004};
here we put these ideas together.  We do this not for the most general 
2-bundle with 2-connection, but for the special case where:
\begin{itemize}
  \item 
     the structure 2-group is a strict Lie 2-group
    (indeed, we have not even mentioned the more general
     `coherent' Lie 2-groups),
  \item
     the base 2-space is either one with only identity morphisms
     (that is, an ordinary space) or one where the morphisms are loops,
  \item
    the arrow-part of the categorified $g_{ij}$ is either trivial
    or nontrivial on `infinitesimal' loops (where the latter are really
    antisymmetric rank $(2,0)$ tensors at a given base point).
\end{itemize}
The reason for these restriction is that, as we show,
the local data describing such 2-bundles with 2-connections
is equivalent to the cocycle data describing possibly twisted nonabelian 
bundle gerbes with connection and curving, subject to the constraint of 
vanishing fake curvature.

More precisely,
in terms of the notation in \cite{AschieriCantiniJurco:2003,AschieriJurco:2004}
we find that:
\begin{itemize}
  \item
    the point part of $g_{ij}$ is the gerbe transition function $\varphi_{ij}$,
  \item
    the arrow part of $g_{ij}$ defines a 2-form which transforms as the gerbe's $d_{ij}$,
  \item
    the natural transformation in the transition law for $g_{ij}$ encodes the
    gerbe 0-forms $f_{ijk}$,
  \item
    a natural transformation between the restriction maps on multiple overlaps
    define the twist $\lambda_{ijkl}$ of the gerbe,
  \item
    the point part of $\mathrm{hol}_i$ defines the gerbe 1-form $A_i$
  \item
    the arrow part of $\mathrm{hol}_i$ defines the gerbe 2-form $B_i$,
  \item
    the natural transformation in the transition law for $\mathrm{hol}_i$
    encodes the 1-form $a_{ij}$ of the gerbe,
  \item
    the curvature of $\hol_i$ defines the gerbe curvature 3-form $H_i$
\end{itemize}
and:
\begin{itemize}
  \item
    the point part of the transition law for $g_{ij}$ gives the gerbe transition law of the
    $\varphi_{ij}$,
  \item
    the arrow part of the transition law for the $g_{ij}$ gives the gerbe transition law
    of $d_{ij}$,
  \item
    the coherence law for the natural transformation defining the transition law
    for the $g_{ij}$ gives the gerbe transition law for the $f_{ijk}$,
  \item
    the point part of the transition law for $\mathrm{hol}_i$ gives the gerbe transition
    law for $A_i$,
  \item
    the arrow part of the transition law for $\mathrm{hol}_i$ gives the 
    gerbe transition law for $B_i$,
  \item
    the coherence law for the transition law of $\mathrm{hol}_i$ gives the
    gerbe transition law for $a_{ij}$,
  \item
    the transition law for $\hol_i$ also gives the gerbe transition law for $H_i$.
\end{itemize}

This is our main result. In summary, we find that categorifying the 
notion of a principal bundle with connection and with strict structure
2-group gives a structure that 
includes as a special case that of a twisted nonabelian bundle gerbe 
with connection and curving, together with a notion 
of nonabelian surface holonomy when the fake curvature vanishes.

\subsection{Nonabelian 2-Holonomies and Physics}
\label{section: Nonabelian 2-Holonomies and Physics}

The motivation for the work in this paper comes to a good part
from certain open questions in theoretical physics, which we briefly
indicate here.

In the context of `M-theory'
(the, only partially understood, 
expected 11-dimensional completion of string theory, which again
is a widely studied candidate theory of quantum gravity and unification of forces)
spatially 2-dimensional objects called  membranes, or M2-branes, 
are fundamental, as are their magnetic duals, the M5-branes.

The general configuration of M2-branes and M5-branes involves
membranes whose boundaries are attached to the 5-branes, 
generalizing the way how open strings
may end on various types of branes. 
(In fact, in type II string theory, networks of $(p,q)$-strings 
may end on webs of $(p,q)$-5-branes and these configurations 
lift in M-theory to configurations of M2s on M5s.)

While the bulk of the membrane couples to the supergravity 3-form
potential, its boundary, when ending on a 5-brane, couples to the
2-form that is part of the field content on that brane.

By comparison with the analogous situation for strings, a stack of
coinciding 5-branes should carry a \emph{nonabelian} 2-form. Therefore
the boundary part of the action for a membrane ending on that stack
should involve some notion of \emph{nonabelian surface holonomy}
with respect to the worldsheet of the membrane's boundary.

But a general definition and theory of nonabelian surface holonomy 
is essentially missing.
It is well understood how \emph{abelian} gerbes give
rise to a notion of \emph{abelian} surface holonomy. Recently
abelian gerbes have been generalized to \emph{nonabelian}
gerbes, but for the latter so far no concept of surface holonomy
is known. On the other hand, using 2-groups it is straightforward to
come up with a local notion of surface holonomy. However, in order to
construct a well defined action for a membrane ending on a stack
of 5-branes, it is crucial to take care of global issues.

For these reasons the construction of a globally defined notion
of nonabelian surface holonomy, which is the subject of this paper,
is a necessary prerequisite for fully understanding the
fundamental objects of M-theory.

This is interesting for ordinary field theory,
quite independently of whether strings really exist or not: Configurations
of membranes ending on 5-branes can alternatively be described in terms of
certain (superconformal) field theories of 
(possibly nonabelian)
2-form fields defined on the six-dimensional  worldvolume of 5-branes.
When these field theories are compactified on a torus they give rise
to (super)Yang-Mills theory in four dimensions. Interestingly, the famous
Montonen-Olive $\mathrm{SL}\of{2,\mathbb{Z}}$ duality of such 4-dimensional
gauge theory should arise this way simply as the modular transformations
on the internal torus which act as symmetries on
the conformally invariant six-dimensional theory. 

From this point of view nonabelian 2-form gauge theory in six dimensions
appears as a tool
for understanding (strongly coupled) ordinary gauge theory in four dimensions.

It is interesting to note that these six-dimensional
theories require the curvature 3-form of the 2-form field to be
Hodge-\emph{self-dual}. In the nonabelian case this is subtle because this 3-form should
obey the local transition laws of nonabelian gerbes, which are not
(at least not in any obvious way) compatible with Hodge-self-duality in general,
since they involve corrections to a covariant transformation of the 3-form.
The only \emph{obvious} solution 
to this compatibility problem is to require
the so-called `\emph{fake curvature}' of the gerbe to vanish
(but it is not known if this is the most general solution). If this
is the case the 3-form field strength transforms covariantly and can hence
consistenly be chosen to be self-dual.

A major theme of the present paper is to show that 
what we call 2-bundles with 
2-connections and with strict structure 2-group gives rise to
general nonabelian gerbes, but necessarily subject to the constraint
of vanishing `fake curvature'. This vanishing condition is 
a direct consequence of the very nature of strict 2-groups and tightly related to 
our ability to construct a global nonabelian surface-holonomy. 

For these reasons 2-bundles with 2-connections can be expected to 
play a role for the description of 
nonabelian gauge theories in four and in six dimensions.

An overview of the relation between four- and six-dimensional conformal
field theories is given in \cite{Witten:2002}.
A list of references on the physics of 5-branes is given
in \cite{Schreiber:2004e}. 
For $(p,q)$ webs of 5-branes see
\cite{AharonyHananyKol:1997,KolRahmfeld:1998,BerensteinLeigh:1998}.

\vfill

\paragraph{Structure of this paper.}

We approach the subject
in \S\fullref{section: Higher Gauge Theory} 
by informally considering some general aspects
of gauge theory, holonomy, and its higher dimensional generalization
using categorification.

This leads us to recall some aspects of 2-bundles in 
\S\fullref{section: 2-bundles (without Connection)} 
and to continue their study by working out
the local meaning of the existence of 2-transitions.

After 2-bundles without 2-connection have thus been understood,
the problem of globally constructing 2-holonomy in a 2-bundle is tackled in 
\S\fullref{section: Path space} 
by a thorough investigation of local connections
and holonomies on path spaces, which allows us to construct the
local 2-functor $\hol \maps \paths_2\of{\manifold} \to \twogroup$ 
in terms of a $\g$-valued 1-form $A$ and an $\h$-valued
2-form $B$ on $\manifold$.

With this functor in hand it is straightforward to categorify the
notion of connection in a bundle and obtain the definition of a
(global!) 2-connection in 
\S\fullref{section 2-bundles with 2-connection}, 
where we again
translate the abstract diagrams into local data and find that
the 2-transitions for the 2-connection define, under certain conditions,
the cocycle data of a nonabelian gerbe with connection and `curving'.

A summary and discussion of these constructions and results is 
given in 
\S\fullref{Summary and Discussion}.

\newpage

\section{2-Connections}

\subsection{Higher Gauge Theory}
\label{section: Higher Gauge Theory}

Before entering the details of the theory of 2-bundles with 2-connections,
it is helpful to recall some general facts about the nature of 
gauge theory and its generalization to the parallel transport of
higher dimensional objects. (See also \cite{Baez:2002}.)

\subsubsection{Algebra from Topology}

Ordinary gauge theory describes how 0-dimensional particles 
transform as we move them along 1-dimensional
paths. It is natural to assign Lie group elements to 
each path:
\[
\xymatrix{
   \bullet \ar@/^1pc/[rr]^{g}
&& \bullet
}
\]
The reason is that composition of paths then corresponds to
multiplication in the group:
\[
\xymatrix{
   \bullet \ar@/^1pc/[rr]^{g}
&& \bullet \ar@/^1pc/[rr]^{g'}
&& \bullet
}
\]
while reversing the direction of a path corresponds to taking
inverses:
\[
\xymatrix{
   \bullet
&& \bullet \ar@/_1pc/[ll]_{g^{-1}}
}
\]
and the associative law makes the holonomy along a triple
composite unambiguous:
\[
\xymatrix{
   \bullet \ar@/^1pc/[rr]^{g}
&& \bullet \ar@/^1pc/[rr]^{g'}
&& \bullet \ar@/^1pc/[rr]^{g''}
&& \bullet
}
\]
In short, the topology dictates the algebra.

To \emph{really} let the topology dictate the algebra, we should
replace the Lie group by a `smooth groupoid': a groupoid
in some convenient category of smooth spaces. Mackaay and
Picken \cite{MackaayPicken:2000} have noted that for any manifold
$\manifold$ there is a smooth groupoid $\paths_1\of{\manifold}$,
the {\bf path groupoid}, for which:
\begin{itemize}
  \item
    objects are points $x \in \manifold$
  \item
    morphisms are thin homotopy classes of smooth paths
   $\gamma : [0,1] \to \manifold$ such that $\gamma\of{t}$
   is constant near $t = 0,1$.
\end{itemize}

For any Lie group $G$, a principal $G$-bundle $P \to \manifold$
gives a smooth groupoid $\trans\of{P}$, the {\bf transport groupoid},
for which:
\begin{itemize}
  \item
    objects are torsors $P_x$ for $x \in \manifold$
  \item
    morphisms are torsor morphisms.
\end{itemize}

Via parallel transport, any connection on $P$ gives a smooth
functor called its {\bf holonomy}
\begin{eqnarray}
  \label{ordinary holonomy functor}
  \hol \maps \paths_1\of{\manifold} \to \trans\of{P}
  \,.
\end{eqnarray}

A trivialization of $P$ makes $\trans\of{P}$ equivalent to $G$, so it
gives:
\begin{eqnarray}
  \hol \maps \paths_1\of{\manifold} \to G
  \,.
  \nonumber
\end{eqnarray}

Now suppose we wish to do something similar
for `strings' that move along surfaces.
Naively
we might wish our holonomy to assign a group element to each
surface like this:
\[
\xymatrix{
   \bullet \ar@/^1pc/[rr]_{}="0"
           \ar@/_1pc/[rr]_{}="1"
           \ar@{=>}"0";"1"^{g}
&& \bullet
}
\]
There are two obvious ways to compose surfaces of this sort, vertically:
\[
\xymatrix{
   \bullet \ar@/^2pc/[rr]_{}="0"
           \ar[rr]_{}="1"
           \ar@{=>}"0";"1"^{g}
           \ar@/_2pc/[rr]_{}="2"
           \ar@{=>}"1";"2"^{g'}
&& \bullet
}
\]
and horizontally:
\[
\xymatrix{
   \bullet \ar@/^1pc/[rr]_{}="0"
           \ar@/_1pc/[rr]_{}="1"
           \ar@{=>}"0";"1"^{g}
&& \bullet \ar@/^1pc/[rr]_{}="2"
           \ar@/_1pc/[rr]_{}="3"
           \ar@{=>}"2";"3"^{g'}
&& \bullet
}
\]
Suppose that both of these correspond to multiplication
in the group $G$.  Then to obtain well-defined holonomy for this
surface regardless of whether we do vertical or horizontal composition
first:
\[
\xymatrix{
   \bullet \ar@/^2pc/[rr]_{}="0"
           \ar[rr]_{}="1"
           \ar@{=>}"0";"1"^{g_1}
           \ar@/_2pc/[rr]_{}="2"
           \ar@{=>}"1";"2"^{g_1^\prime}
&& \bullet \ar@/^2pc/[rr]_{}="3"
           \ar[rr]_{}="4"
           \ar@{=>}"3";"4"^{g_2}
           \ar@/_2pc/[rr]_{}="5"
           \ar@{=>}"4";"5"^{g'_2}
&& \bullet
}
\]
we must have
\[      (g_1 g_2)(g_1' g_2') = (g_1 g_1')(g_2 g_2') . \]
This forces $G$ to be abelian!

In fact, this argument goes back to a classic paper by
Eckmann and Hilton \cite{EckmannHilton}.  Moreover, they showed
that even if we allow $G$ to be equipped with two products,
say $g \circ g'$ for vertical composition and $g \cdot g'$ for
horizontal, so long as both products share the
same unit and satisfy this `{\bf interchange law}':
\[  
  (g_1 \circ g_1') \cdot (g_2 \circ g_2') =
      (g_1 \cdot g_2) \circ (g_1' \cdot  g_2')  
\]
then in fact they must agree --- so by the previous argument,
both are abelian.  The proof is very easy:
\[  
   g\cdot g' 
   = (g \circ 1)\cdot (1 \circ g') 
   = (g \cdot 1) \circ (1 \cdot g') 
   =  g \circ g'. 
\]

Pursuing this approach, we ultimately get the theory of
connections on `abelian gerbes'. If $G= \U\of{1}$, such a
connection looks locally like a 2-form - and it shows up
naturally in string theory, satisfying equations very much like
those of electromagnetism.

To go beyond this and get \emph{nonabelian} higher gauge fields,
we should let the topology dictate the algebra, and consider a connection
that gives holonomies \emph{both for paths and for surfaces}.

We can replace the path groupoid by the {\bf path 2-groupoid}, where
\begin{itemize}

  \item
    objects are points in $\manifold$: 
\[
\xymatrix{
   \bullet
}
\]

\item
 morphisms are smooth paths $\gamma \maps [0,1] \to \manifold$ 
 with $\gamma\of{\sigma}$ constant near $\sigma = 0,1$
\[
\xymatrix{
   \bullet \ar@/^1pc/[rr]^{\gamma}
&& \bullet
}
\]

\item
  2-morphisms are 
  thin homotopy classes of smooth maps $f \maps [0,1]^2 \to \manifold$
  with $f\of{\sigma,\tau}$ independent of $\tau$ near $\tau=0,1$ and constant near
  $\sigma = 0,1$
\[
\xymatrix{
   \bullet \ar@/^1pc/[rr]_{}="0"
           \ar@/_1pc/[rr]_{}="1"
&& \bullet \ar@{=>}"0";"1"^{f}
}
\]
\end{itemize}

Assume that for each path we have a holonomy taking values in some
group $G$:
where composition of paths corresponds to multiplication in $G$.
Assume also that for each 1-parameter family of paths with
fixed endpoints we have a holonomy taking values in some other
group $H$:
where vertical composition corresponds to multiplication in
$H$:
\[
\xymatrix{
   \bullet \ar@/^2pc/[rr]_{}="0"
           \ar[rr]_{}="1"
           \ar@{=>}"0";"1"^{h}
           \ar@/_2pc/[rr]_{}="2"
           \ar@{=>}"1";"2"^{h'}
&& \bullet
}
\]

Next, assume that we can parallel transport an element $g \in G$
along a 1-parameter family of paths to get a new element $g' \in G$:
\[
\xymatrix{
   \bullet \ar@/^1pc/[rr]^{g}_{}="0"
           \ar@/_1pc/[rr]_{g'}_{}="1"
           \ar@{=>}"0";"1"^{h}
&& \bullet
}
\]

Now, the picture above suggests that we should think of $h$ as a kind of
`arrow' or `morphism' going from $g$ to $g'$.  We can use category theory to
formalize this.  However, in category theory, when a morphism
goes from an object $x$ to an object $y$, we think of the morphism as
determining both its source $x$ and its target $y$.  The group
element $h$ does not determine $g$ or $g'$.  However, the pair $(h,g)$
does.  Thus it is useful to create a category $\twogroup$ where the set 
$\twogroup^1$ of
objects is just $G$, while the set $\twogroup^2$ of morphisms consists
of pairs $f = (h,g) \in H \times G$.
Switching our notation to reflect this, we rewrite
the above picture as
\[
\xymatrix{
   \bullet \ar@/^1pc/[rr]^{g}_{}="0"
           \ar@/_1pc/[rr]_{g'}_{}="1"
           \ar@{=>}"0";"1"^{f}
&& \bullet
}
\]
and write $f \maps g \to g'$ for short.  We have source and target
maps
\[  s,t \maps \twogroup^2 \to \twogroup^1\]
with $s(f) = g$ and $t(f) = g'$.

In this new notation we can vertically compose $f \maps g \to g'$ and
$f' \maps g' \to g''$ to get $f \circ f' \maps g \to g''$, as follows:
\[
\xymatrix{
   \bullet \ar@/^2pc/[rr]^{g}_{}="0"
           \ar[rr]^<<<<<<{g'}_{}="1"
           \ar@{=>}"0";"1"^{f}
           \ar@/_2pc/[rr]_{g''}_{}="2"
           \ar@{=>}"1";"2"^{f'}
&& \bullet
}
\]
This is just composition of morphisms in the category $\twogroup$.  However, we
can also horizontally compose $f_1 \maps g_1 \to g_1'$ and $f_2 \maps
g_2 \to g_2'$ to get $f_1 \cdot f_2 \maps g_1g_2 \to g_1'g_2'$, as follows:
\[
\xymatrix{
   \bullet \ar@/^1pc/[rr]^{g_1}_{}="0"
           \ar@/_1pc/[rr]_{g_1'}_{}="1"
           \ar@{=>}"0";"1"_{f_1}
&& \bullet \ar@/^1pc/[rr]^{g_2}_{}="2"
           \ar@/_1pc/[rr]_{g_2'}_{}="3"
           \ar@{=>}"2";"3"_{f_2}
&& \bullet
}
\]
We assume this operation makes $\twogroup^2$ into a group with the pair
$(1,1) \in H \times G$ as its multiplicative unit.

The good news is that now we can assume an interchange law saying this
holonomy is well-defined:
\[
\xymatrix{
   \bullet \ar@/^2pc/[rr]^{g_1}_{}="0"
           \ar[rr]^<<<<<<{g_2}_{}="1"
           \ar@{=>}"0";"1"^{f_1}
           \ar@/_2pc/[rr]_{g_3}_{}="2"
           \ar@{=>}"1";"2"^{f_1'}
&& \bullet \ar@/^2pc/[rr]^{g_1'}_{}="3"
           \ar[rr]^<<<<<<{g_2'}_{}="4"
           \ar@{=>}"3";"4"^{f_2}
           \ar@/_2pc/[rr]_{g_3'}_{}="5"
           \ar@{=>}"4";"5"^{f_2'}
&& \bullet
}
\]
namely:
\begin{eqnarray}
  \label{the exchange law} 
  (f_1 \circ f_1') \cdot (f_2 \circ f_2') =  (f_1 \cdot f_2) \circ (f_1' \cdot f_2')
\end{eqnarray}
without forcing either $\twogroup^2$ or $\twogroup^1$ to be abelian!   Instead,
$\twogroup^2$ is forced to be a semidirect product of the groups $G$ and
$H$.

The structure we are rather roughly describing here is in fact already
known to mathematicians under the name of a `categorical group'
\cite{Forrester-Barker:2002,MacLane}.
The reason is that $\twogroup$ turns out to be a category
whose set of objects $\twogroup^1$ is a group, 
whose set of morphisms $\twogroup^2$ is a
group, and where all the usual category operations are group
homomorphisms.  To keep the terminology succinct and to hint at
generalizations to still higher-dimensional holonomies, we prefer to
call this sort of structure a `2-group'.  Moreover, we shall focus most
of our attention on `Lie 2-groups', where $\twogroup^1$ and 
$\twogroup^2$ are Lie groups
and all the operations are smooth.  More details on Lie 2-groups can
be found in a paper by Baez and Lauda \cite{BaezLauda:2003}.

A Lie 2-group $\twogroup$ amounts to the same thing as a `Lie
crossed module': a pair of Lie groups $G$ and $H$ together with a
homomorphism $t \maps H \to G$ and an action $\alpha$ of $G$ on $H$
satisfying a couple of compatibility conditions.  The idea is to let $G
= \twogroup^1$, let $H$ be the subgroup of $\twogroup^2$ consisting of morphisms
with source equal to $1$, and let $t$ be the map sending each such
morphism to its target.  The action $\alpha$ is defined by letting
$\alpha(g)f$ be this horizontal composite:
\[
\xymatrix{
   \bullet \ar@/^1pc/[rr]^{g}_{}="0"
           \ar@/_1pc/[rr]_{g}_{}="1"
           \ar@{=>}"0";"1"^{1_g}
&& \bullet \ar@/^1pc/[rr]^{1}_{}="2"
           \ar@/_1pc/[rr]_{t(f)}_{}="3"
           \ar@{=>}"2";"3"^{f}
&& \bullet \ar@/^1pc/[rr]^{g^{-1}}_{}="4"
           \ar@/_1pc/[rr]_{g^{-1}}_{}="5"
           \ar@{=>}"4";"5"^{1_{g^{-1}}}
&& \bullet
}
\]

It appears that one can develop a full-fledged theory of bundles,
connections, curvature, and so on with a Lie
2-group taking the place of a Lie group.  So far most work has focused
on the special case when $G$ is trivial and $H = \U(1)$, using
the language of $\U(1)$ gerbes
\cite{Brylinski,CareyJohnsonMurray,CMR,Hitchin,Keurentjes,MackaayPicken:2000}.
Here, however, we really want
$H$ to be nonabelian.  Some important progress in this direction can be
found in Breen and Messing's paper on the differential geometry of
nonabelian gerbes \cite{BreenMessing:2001}.
While they use different
terminology, their work basically develops the theory of connections and
curvature for Lie 2-groups where $H$ is an arbitrary Lie group, $G =
\Aut(H)$ is its group of automorphisms, $t$ sends each element of $H$ to the corresponding inner
automorphism, and the action of $G$ on $H$ is the obvious one.  We call
this sort of Lie 2-group the `automorphism 2-group' of $H$.  Luckily, it
is easy to extrapolate the whole theory from this case.  

In particular, for any Lie 2-group $\twogroup$ one can define the notion of a
`principal 2-bundle' having $\twogroup$ as its gauge 2-group; this has
recently been done by Bartels \cite{Bartels:2004}.  One can also 
define connections and curvature for these principal 2-bundles.

One of our goals in this paper is to show
that given a connection of this sort (satisfying
certain conditions), we may define
holonomies for paths and surfaces which behave just as one would
like.   The other is to relate this notion of 2-connection 
to that of a connection on the space 
$\paths_1\of{M}$ whose points are paths in $M$.    A connection on $\paths_1\of{M}$ 
assigns a holonomy to any path in $\paths_1\of{M}$, but such a path traces
out a surface in $M$.  Such a connection 
thus gives a concept of (nonabelian) surface holonomy, which however
will depend on the \emph{parameterization} of the surface
unless we impose extra conditions.

Intuitively it is clear that these two points of view should be closely 
related, but little is known about the details of this relation.
Motivated by the recent discovery 
\cite{Schreiber:2004e}
that a certain consistency condition for surface holonomy
appearing in the loop space approach is discussed also in the literature on
2-groups \cite{GirelliPfeiffer:2004}, 
while other such consistency conditions have exclusively been discussed in the 
loop space context \cite{AlvarezFerreiraSanchezGuillen:1998}, 
the aim here is to clarify this issue.

Hence in analogy to the situation 
\refer{ordinary holonomy functor}
for ordinary bundles with ordinary connections,
in the following we want to make precise the following statement:

A principal $\twogroup$-2-bundle $\twoP \to \twoB$ gives a smooth
2-groupoid $\trans\of{P}$ where
\begin{itemize}
  \item
    objects are 2-torsors $P_x$
  \item
    morphisms are 2-torsor morphisms $f \maps P_x \to P_y$
  \item
    2-morphisms are 2-torsor 2-morphisms $\theta \maps f \Rightarrow g$.
\end{itemize}
Via parallel transport, a 2-connection on $P$ gives a
smooth 2-functor called its {\bf holonomy}:
\[
  \hol \maps \paths_2\of{\manifold} \to \trans\of{P}
  \,.
\]
A trivialization of $\twoP$ makes $\trans\of{P}$ equivalent to 
$\twogroup$, so that it gives
\[
  \hol \maps \paths_2\of{\manifold} \to \twogroup
  \,.
\]

In order to understand how such a structure is included in the
framework of categorified bundles, it is very helpful to use a technique called
`internalization'.

\subsubsection{Internalization}
\label{internalization}

The crucial concept for categorification is `internalization'.
  Ehresmann and Lawvere 
showed 
\cite{Ehresmann:1966,Borceux:1994}
how to `internalize' concepts
by defining them in terms of commutative diagrams
(see section 2 of \cite{BaezCrans:2003} for more details): 

A {\bf small category}, say $C$,
has a \underline{set} of objects $\Ob(C) \defas C^1$, a
\underline{set} of morphisms $\Mor(C) \defas C^2$, source and target 
\underline{functions}
\[               s,t \maps \Ob(C) \to \Mor(C)  , \]
a composition \underline{function}
\[     \circ \maps \Mor(C) {{}_{s}\times_{t}} \Mor(C) \to \Mor(C) \]
and an identity--assigning \underline{function}
\[      \id \maps \Ob(C) \to \Mor(C) \]
making diagrams commute which describe associativity of
composition and behaviour of source and target maps under composition.

Internalization means 
letting these diagrams live within some category $K$:

A {\bf category in $K$}, say $C$,
has an \underline{object} $\Ob(C) \in K$, an
\underline{object} $\Mor(C) \in K$, source and target 
\underline{morphisms}
\[               s,t \maps \Ob(C) \to \Mor(C)  , \]
a composition \underline{morphism}
\[     \circ \maps \Mor(C) {{}_{s}\times_{t}} \Mor(C) \to \Mor(C) \]
and an identity-assigning \underline{morphism}
\[      \id \maps \Ob(C) \to \Mor(C) \]
making the above diagrams commute in $K$.

Similarly we can define
{\bf functors in $K$} and {\bf natural transformations in $K$}, 
obtaining a 2-category \textbf{\textit{K}Cat}.  
We can also define {\bf groups in 
$K$} and {\bf homomorphisms in $K$}, obtaining a category 
\textbf{\textit{K}Grp}.

\subsubsection{Smooth Categories, 2-Groups, and Lie 2-Groups}

Using the above, we can categorify concepts from differential 
geometry with the help of internalization:

\begin{itemize}
\item
A {\bf smooth category} (called {\bf 2-space} in the following)
is a category in $\Diff$, the category of smooth spaces
with smooth maps between them.
\item
A {\bf strict 2-group} (or {\bf categorical group}) is a \hfill \break
category in $\Grp$, the category of groups with homomorphisms between them.
\item
A {\bf strict Lie 2-group} is a category in $\Lie\Grp$, the category
of Lie groups.
\end{itemize}

Important examples of strict 2-groups are the following:

\vskip 2em

1)  Any abelian group $A$ gives a strict 2-group with one object
and $A$ as the automorphisms of this object.  Lie 2-groups of this
kind will be structure 2-groups of 2-bundles giving rise to \emph{abelian gerbes}.

\vskip 2em
\noindent
2)  Any category $C$ gives a 2-group $\Aut(C)$ whose objects
are equivalences $f \maps C \to C$ and whose morphisms are natural
isomorphisms between these.  

\vskip 2em
\noindent
3)  A group $H$ is a category with one object and all
morphisms invertible.  In this case, 2)
gives a strict 2-group $\Aut(H)$ whose objects are automorphisms
of $H$ and whose morphisms from $f$ to $f'$ are 
elements $k \in H$ with $f'(h) = kf(h)k^{-1}$.

\vskip 2em
\noindent
4) Any Lie group $H$ gives a strict Lie 2-group $\Aut(H)$ defined
as in 3) but with everything smooth.  Lie 2-groups
of this sort will be structure 2-groups of 2-bundles that give rise
to \emph{nonabelian gerbes}.

\paragraph{Properties of Strict 2-Groups.}

When it comes to explicit descriptions the 
most important fact about strict 2-groups is that they are characterized
by \emph{crossed modules}, in a way to be made precise below in 
prop. \ref{strict 2-group coming from crossed module}.

\begin{definition}
 \label{Lie crossed module}
  A {\bf Lie crossed module} is a quadruple $(G,H,t,\alpha)$ consisting
of Lie groups $G$ and $H$, a homomorphism $t : H \to G$, and an action
of $G$ on $H$ (that is, a homomorphism $\alpha: G \to \mathrm{Aut}\of{H}$)
satisfying
\begin{eqnarray}
  t\of{\alpha\of{g}\of{h}} &=& g \, t\of{h}\, g^{-1}
  \nonumber
\end{eqnarray}
and
\begin{eqnarray}
  \alpha\of{t\of{h}}\of{h^\prime} &=& h\, h^\prime\, h^{-1}
  \nonumber
\end{eqnarray}
for all $g \in G$ and $h,h^\prime \in H$.
\end{definition}
(\cf \cite{Baez:2002}, definition 3.)

\begin{definition}
  \label{differential crossed module}
  A {\bf differential crossed module} is a quadruple 
 $(\g,\h,dt,d\alpha)$ 
consisting
of Lie algebras $\g$ and $\h$, 
a homomorphism $dt : \h \to \g$, and an action
of $\g$ on $\h$ 
(that is, a homomorphism $d\alpha: \g \to \mathrm{Der}\of{\h}$)
satisfying
\[
  dt\of{d\alpha\of{x}\of{y}} = \commutator{x}{dt\of{y}}
\]
and
\begin{eqnarray}
  \label{notation for action of h on h'}
  d\alpha\of{dt\of{y}}\of{y^\prime} &=& \commutator{y}{y^\prime}
\end{eqnarray}
for all $x \in \g$ and $y,y^\prime \in \h$.

For convenience we will also write
\[
  d\alpha\of{x}\of{x^\prime} \defas \commutator{x}{x^\prime}    
\]
for $x,x^\prime in \g$.
\end{definition}
(\cf \cite{Baez:2002}, definition 15.)

Now the relation between crossed modules and strict 2-groups is the following:

\begin{proposition}
  \label{strict 2-group coming from crossed module}
  Every strict 2-group
  comes from a crossed module \refdef{Lie crossed module}
  $(G,H,\alpha,t)$
  such that 2-group elements are labeled by pairs 
  \[
    (g,h) \hspace{1cm}\mbox{with $g \in G$ and $h\in H$}\,,
  \]
  such that the source $d_0$ and target $d_1$ are given by
  \begin{eqnarray}
    \label{source and target of strict 2-group element}
    d_0\of{(g,h)} &=& g
    \nonumber\\
    d_1\of{(g,h)} &=& t\of{h}g
  \end{eqnarray}
  and such that horizontal and vertical composition is given by
  \begin{eqnarray}
    \label{horizontal and vertical product in strict 2-group}
    (g,h) \cdot (g^\prime, h^\prime) &=& (gg^\prime, h \,\alpha\of{g}\of{h^\prime})
    \nonumber\\
    (g,h) \circ (t\of{h}g,h^1) &=& (g,h^\prime h)
    \,,
  \end{eqnarray}
  respectively.
\end{proposition}
\Proof 
The proof is given for instance in \cite{Forrester-Barker:2002}.\endofproof

It is the property \refer{source and target of strict 2-group element}
of strict 2-groups that leads to the constraint of vanishing fake
curvature in 2-bundles with 2-connections, as has first been explicitly
disussed in \cite{GirelliPfeiffer:2004} using lattice 2-gauge theory,
and as we will derive using path space connections in 
\S\fullref{section: Local 2-Holonomy from local Path Space Holonomy}.

\vskip 2em

We end this section by mentioning a couple of simple facts about
2-groups that are needed in calculations throughout this paper:

\begin{definition}
  \label{notions of 2-group inverse}
  There are two kinds of {\bf inverses} for 2-groups, those with
respect to the horizontal and those with respect to the vertical
product. The inverse with respect to the horizontal product will be
denoted by $(\cdot)^{-1}$, while that with respect to the
vertical product by $(\cdot)^r$ (for ``\emph{r}eversion'').
\end{definition}

\begin{corollary}
  The following lists a couple of useful facts about strict 2-groups:
  \begin{enumerate}

  \item
  With respect to the horizontal product 
  \refer{horizontal and vertical product in strict 2-group} of a
  strict 2-group 
  the {\bf identity} element is
  \[
    1 = (1,1)
  \]
  and the {\bf horizontal inverse} \refdef{notions of 2-group inverse} of $(g,h)$ is
  \[
    (g,h)^{-1} = (g^{-1}, \alpha\of{g^{-1}}\of{h^{-1}})
    \,.
  \]
  The {\bf vertical identity} with base $g$ is
  \begin{eqnarray}
    \label{vertical identity 2-group element}
    1_{g} &\defas& (g,1)
  \end{eqnarray}
  and the {\bf vertical inverse} \refdef{notions of 2-group inverse} of $(g,h)$ is
  \[
    (g,h)^r = \left(t\of{h} g, h^{-1}\right)
     \,.
  \]

 \item
    
  {\bf Horizontal conjugation} with a vertical identity 
   \refer{vertical identity 2-group element}
   yields
  \[
    1_g \cdot (g^\prime, h^\prime) \cdot (1_g)^{-1}
    =
    \left(
      g\,g^\prime\, g^{-1},
      \alpha\of{g}\of{h^\prime}
    \right)
  \]

  \item
    \label{ker t is abelian}
    The 2-group elements with source and target the identity in $1 \in G$ form an
  abelian subgroup. 
  By prop. \ref{strict 2-group coming from crossed module}
  this are elements of the form $(1,a)$ with $t\of{a} = 1$.
  Equivalently $\mathrm{ker}\of{t} \in H$ is a normal abelian
  subgroup of $H$. 

  \end{enumerate}
\end{corollary}
\Proof

\ref{ker t is abelian}.: 
  This is a special case of a famous argument by Eckmann-Hilton 
 (\cf \cite{BaezLauda:2003}, p. 55).

\endofproof

We conclude our review of categorified gauge theory
by briefly discussion nonabelian gerbes in the next subsection.

\subsubsection{Nonabelian Gerbes}
\label{gerbes}

Categorification, while being a systematic
procedure, can usually be performed in several ways, in particular
when there are several equivalent formulations of the original structure 
to be categorified. 

The 2-bundles to be discussed in the present paper arise as the
categorification of ordinary bundles, being maps $E \stackto{p} B$
from some total space to a base space. On the other hand, an ordinary
bundle is alternatively characterized by its sheaf of sections.
Gerbes, which go back to work by Giraud,
arise roughly as the categorification of the notion of a sheaf
\cite{Brylinski,Moerdijk:2002,BreenMessing:2001}.

The theory of abelian gerbes has been motivated by 
the desire to find a higher dimensional
version of the fact that line bundles over $\manifold$ 
provide a geometric realization of elements
of $H^2\of{\manifold,\mathbb{Z}}$. Abelian gerbes
provide a geometric realization of the elements of
$H^3\of{\manifold,\mathbb{Z}}$.

As a more concrete description of gerbes than the original one 
in the language of algebraic geometry, Murray introduced 
(abelian) `bundle gerbes' \cite{Murray:1994}, Chatterjee
introduced (abelian) `gerbs' \cite{Chatterjee:1998} and recently
Aschieri, Cantini and Jur{\v c}o studied the nonabelian
generalization of bundle gerbes \cite{AschieriCantiniJurco:2003}.

`Bundle gerbes' involve bundles over bundles of paths over
a base space and hence can be handled in terms of differential
geometry, which is more conveniently applied to physics
(e.g. \cite{CareyJohnsonMurray, AschieriJurco:2004,CareyJohnsonMurrayStevensonWang:2004})
than algebraic geometry. Their relation to path spaces already
indicates that (bundle) gerbes should be closely related to 2-bundles over 2-spaces,
since the archetypical 2-space is a path space. In the following
we shall study how close this connection really is.

The term `gerb' introduced by Chatterjee is supposed to refer to the
description of gerbes in terms of local transition functions and
{\v C}hech cocycles.
This is the most elementary description and also the only perspective
on gerbes considered in the present paper, a good part which 
is devoted to working out the description of
2-bundles with 2-connections in terms of local data and showing that the
2-transition laws reproduce the cocycle descrition of nonabelian gerbes.
In order to be self-contained, this cocycle data is listed here.

For nonabelian gerbes this was derived in terms of bundle gerbes 
in \cite{AschieriCantiniJurco:2003}. Our
notation follows that in \cite{AschieriJurco:2004}. For vanishing `twist' the
same in different notation was first given in \cite{BreenMessing:2001}.

The cocycle data of a nonabelian gerbe consists of

\begin{itemize}
  \item
    a base space $\manifold$
  \item
    a good cover $U$ of $\manifold$ (with $U^{[n]}$ denoting the space of $n$-fold 
        intersections of patches in $U$)
  \item
    a crossed module $(G,H,\alpha,t)$ with differential crossed module
    $(\g,\h,d\alpha,dt)$ (usually considered to be an automorphism crossed module) 
  \item
    transition functions
    \begin{eqnarray}
      U^{[2]} &\to&  \Omega^0\of{\manifold,G}
      \nonumber\\
      (x,i,j) &\mapsto& g_{ij}\of{x}
    \end{eqnarray}
   \item
    connection 1-forms
    \begin{eqnarray}
      U^{[1]} &\to&  \Omega^1\of{\manifold,\g}
      \nonumber\\
        (x,i) &\mapsto& A_i\of{x}
    \end{eqnarray}
  \item
    curving 2-forms
    \begin{eqnarray}
      U^{[1]} &\to&  \Omega^2\of{\manifold,\h}
      \nonumber\\
        (x,i) &\mapsto& B_i\of{x}
    \end{eqnarray}
   \item
    transition transformation 0-forms
    \begin{eqnarray}
      U^{[3]} &\to& \Omega^0\of{\manifold,H}
      \nonumber\\
      (x,i,j,k) &\mapsto& f_{ijk}\of{x}
    \end{eqnarray}
   \item
    connection transformation 1-forms
    \begin{eqnarray}
      U^{[2]} &\to&  \Omega^1\of{\manifold,\h}
      \nonumber\\
        (x,i,j) &\mapsto& a_{ij}\of{x}
    \end{eqnarray}
  \item
   curving transformation 2-forms
    \begin{eqnarray}
      \label{curving transformation 2-forms of nonabelian gerbe}
      U^{[2]} &\to& \Omega^2\of{\manifold,\h}
      \nonumber\\
        (x,i,j) &\mapsto& d_{ij}\of{x}
    \end{eqnarray}
  \item
   twist $p$-forms
    \begin{eqnarray}
      \label{gerbe twist p-forms}
      U^{[4]} &\to&  \Omega^0\of{\manifold,\ker\of{t} \subset H}
      \nonumber\\
      (x,i,j,k,l) &\mapsto& \lambda_{ijkl}\of{x}
      \nonumber\\
      \nonumber\\
      U^{[3]} &\to&  \Omega^1\of{\manifold,\ker\of{dt} \subset \h}
      \nonumber\\
      (x,i,j,k) &\mapsto& \alpha_{ijk}\of{x}
      \nonumber\\
      \nonumber\\
      U^{[2]} &\to& \Omega^2\of{\manifold,\ker\of{dt} \subset \h}
      \nonumber\\
      (x,i,j) &\mapsto& \beta_{ij}\of{x}
      \nonumber\\
      \nonumber\\
      U^{[1]} &\to&  \Omega^3\of{\manifold,\ker\of{dt} \subset \h}
      \nonumber\\
      (x,i) &\mapsto& \gamma_{i}\of{x}
    \end{eqnarray}
\end{itemize}
such that the following transition laws are satisfied:
\begin{itemize}
  \item transition law for the transition functions
\begin{eqnarray}
  \label{gerbe transition law for the transition functions}
  \phi_{ij}\of{x}\phi_{jk}\of{x} &=& t\of{f_{ijk}\of{x}}\phi_{ik}\of{x}\,,\;\;\forall (x,i,j,k)\in U^{[3]}
\end{eqnarray}
  \item transition law for the connection 1-forms
\begin{eqnarray}
  \label{gerbe trans law for connection 1-form}
  A_i\of{x} + dt\of{a_{ij\of{x}}} 
  &=& 
  \phi_{ij}\of{x}A_j\of{x}\phi_{ij}^{-1}\of{x} + 
  \phi_{ij}\of{x}(\extd \phi_{ij}^{-1})\of{x}
  \,,\;\;\forall (x,i,j)\in U^{[2]}
  \nonumber\\
\end{eqnarray}

  \item transition law for the curving 2-forms
\begin{eqnarray}
  \label{gerbe trans law for curving 2-forms}
  B_i\of{x}
  &=& 
  \alpha\of{\phi_{ij}\of{x}}\of{B_j\of{x}}
  +
  k_{ij}\of{x}
  -
  d_{ij}\of{x}
  -
  \beta_{ij}\of{x}
  \,,\;\;\forall (x,i,j)\in U^{[2]}
  \,.
  \nonumber\\
\end{eqnarray}

 \item transition law for the curving transformation 2-forms
  \begin{eqnarray}
  \label{gerbe trans law for curving transformation 2-forms}
    d_{ij} + \phi_{ij}\of{d_{jk}}
    &=&
    f_{ijk} \,d_{ik}\, f_{ijk}^{-1}
    +
    f_{ijk}d\alpha\of{dt\of{B_i} + F_{A_i}}{f_{ijk}^{-1}}
  \,,\;\;\forall (x,i,j)\in U^{[2]}
  \,.
  \end{eqnarray}

\end{itemize}

In addition to these there are what for reasons explained in 
\S\fullref{section: 2-bundles (without Connection)} 
and \S\fullref{section 2-bundles with 2-connection}
we here shall call \emph{coherence laws}, since they
ensure that compositions of the above transformations are well defined:
\begin{itemize}
  \item
   coherence law for the transformators of the transition functions
\begin{eqnarray}
  \label{gerbe coherence law for transformators of transition functions}
  f_{ikl}^{-1}\of{x}f_{ijk}^{-1}\of{x}\,\alpha\of{\phi_{ij}\of{x}}\of{f_{jkl}\of{x}}f_{ijl}\of{x}
  &=&
  \lambda_{ijkl}\of{x}
  \,,\;\;\forall (x,i,j,k,l)\in U^{[4]}
\end{eqnarray}

  \item
  coherence law for the transformators of the connection 1-form
  \begin{eqnarray}
    \label{gerbe coherence law for transformators of connection forms}    
    \alpha_{ijk}
    &=&
    a_{ij} + \phi_{ij}\of{a_{jk}}
    -
    f_{ijk} \,a_{ik}\, f_{ijk}^{-1}
    -
    f_{ijk}df^{-1}_{ijk}
    -
    f_{ijk}d\alpha\of{A_i}\of{f_{ijk}^{-1}}
    \,.
  \end{eqnarray}
\end{itemize}

Finally the \emph{curvature 3-form} of the nonabelian gerbe is defined as
\begin{eqnarray}
  \label{gerbe curvature 3-form}
  H_i 
   &\defas&
 d_{A_i}B_i + \gamma_i
\end{eqnarray}
and its transformation law is
\begin{eqnarray}
  \label{gerbe curvature transition law}
  H_i
  &=&
  \phi_{ij}\of{H_j}
  -
  \extd d_{ij}
  -
  \commutator{a_{ij}}{d_{ij}}
  -
  d\alpha\of{dt\of{B_i} + F_{A_i}}\of{a_{ij}}
  -
  d\alpha\of{A_i}\of{d_{ij}}
  \,.
\end{eqnarray}

\newpage

\subsection{2-Bundles (without Connection)}
\label{section: 2-bundles (without Connection)}

2-bundles have been defined by Toby Bartels \cite{Bartels:2004}
by categorification of the concept of an ordinary bundle. 

An
orinary bundle consists of two sets, the {\bf total space}
$E$ and the {\bf base space} $B$, together with a {\bf projection map} 
\[
  E \stackto{p} B
  \,.
\]
We can think of $p$ as a morphism in the category $\Set$, whose objects are
sets and whose morphisms are functions between sets.

Usually one wants $E$ and $B$ not to be general sets, but to be sets that are
smooth spaces. There is a subcategory $\Diff$ of $\Set$ whose objects are
smooth spaces and whose morphisms are smooth maps between these.
A \emph{bundle in} $\Diff$ is hence a smooth space $E$ and a smooth space
$B$ together with a smooth map $E \stackto{p} B$.

The category $\Set$ as well as its sub-category $\Diff$ are 1-categories. 
This means that they have objects and (1-)morphisms going 
between objects, but no nontrivial 2-morphisms going between 1-morphisms. 
We get a categorified
notion of bundle by considering a \emph{bundle in} $C$, for $C$ a proper
2-category which has objects, 1-morphisms between objects and 
2-morphisms between 1-morphisms (but no nontrivial 3-morphisms between
2-morphisms).

The natural choice for a 2-category replacement for the 1-category $\Set$ is
the 2-category $\Cat$, the 2-category whose objects are categories, whose
morphisms are functors and whose 2-morphisms are natural transformations.

Hence a \emph{bundle in} $\Cat$ is a category $E$, a category $B$ and a functor
$E \stackto{p} B$. 

As with ordinary bundles, usually one wants categorified bundles to have some
smoothness property. In analogy to the subcategory $\Diff$ of $\Set$ there
should be a sub-2-category $\twoDiff$ of $\Cat$ whose objects are `smooth categories'
in an appropriate sense. The notion of smooth category, called \emph{2-space},
is itself obtained by internalization:

\begin{definition}
  A {\bf 2-space} is a category internalized in $\DiffInfty$, 
  where $\DiffInfty$
  is the category of (possibly infinite dimensional) smooth spaces
  with smooth maps between them as morphisms.
  A {\bf 2-map} is a smooth functor between two 2-spaces.
\end{definition}
We write a 2-space $S$ as $S = (S^1,S^2)$ with $S^1 = \Ob\of{S}$ the
space of objects, also called the {\bf point space} and $S^2 = \Mor\of{S}$ the
space of morphisms, also called the {\bf arrow space}. Being a category
implies that there are smooth maps $d_0 : S^2 \to S^1$ and $d_1 : S^2 \to S^1$
which map each arrow in $S^2$ to its {\bf source} and {\bf target} point
in $S^1$, respectively. Moreover, there is a smooth map which sends two
composable arrows in $S^2$ to their composition in $S^2$.

Two important subsets of 2-spaces that play a role in this paper are 
`trivial' and `simple' 2-spaces:
\begin{definition}
  \label{simple 2-space}
  $\,$
  \begin{itemize}
  \item
  A 2-space for which all morphisms are identity morphisms shall be called a
  {\bf trivial 2-space}.
  \item
  A 2-space for which the source and the target map coincide shall be called a
  {\bf simple 2-space}.
  \end{itemize}
\end{definition}

Smooth maps $f = (f^1,f^2)$ between 2-spaces,
consist of an ordinary smooth map $f^1$ taking the point space smoothly to another 
point space,
together with a smooth map $f^2$ taking arrows to arrows in such a way
that it respects the action of the point map on the source and target of
these arrows.

The strict 2-category $\twoDiff$ that we are interested in is the 
2-category of 2-spaces, whose objects are 2-spaces, whose morphisms are 
2-maps
and whose 2-morphisms are natural transformations between 2-maps.

This allows us to finally state the definition of a 2-bundle:

\begin{definition}
  \label{2-bundle}
  A {\bf 2-bundle} is a \emph{bundle in} $\twoDiff$, i.e.
  the collection of
  \begin{itemize}
    \item
      a 2-space $\twoP$ (the {\bf total space})
    \item
      a 2-space $\twoB$ (the {\bf base space})
    \item
      a 2-map $p \maps \twoP \to \twoB$ 
      (the {\bf projection})
    \,.
  \end{itemize}
\end{definition} 

For applications in gauge theory, a bundle should be \emph{locally trivializable}.
This is the content of the next subsection.

\subsubsection{Locally trivializable 2-Bundles}

We shall be interested in \emph{locally trivializable} 2-bundles. 
These require the notion of a 2-cover. Ordinarily, a locally trivializable
bundle is a bundle $E \stackto{p} B$
together with a cover $U = \bigdisunion_{i\in I} U_i$ being the disjoint
union of open
contractible subsets $U_i \subset B$ with $\bigcup\limits_{i\in I} U_i = B$,
such that the restriction to any of the $U_i$ yields a bundle isomorphic to
a trivial bundle. In the context of 2-bundles this is implemented as follows:

\paragraph{2-covers.}

\begin{definition}
  \label{sub-2-spaces}
  Given 2-spaces $S$ and $B$, the 2-space $S$ is called a {\bf sub-2-space}
  $S \subset B$ of $B$ if the point and arrow spaces of $S$ are subsets of the
  point and arrow spaces of $B$ and if the source, target and composition maps
  of $S$ are the restriction of these maps on $B$ to these subsets.
\end{definition}

\begin{definition}
  \label{union of 2-spaces}
  Given a 2-space $B$ and a
  set of sub-2-spaces \refdef{sub-2-spaces}
  $\set{U_i}_{i\in I}$, $U_i \subset B$, 
  their {\bf union of 2-spaces}
  denoted by $\bigcup\limits_{i\in I} U_i \subset B$ is defined to be the
  sub-2-space of $B$ \refdef{sub-2-spaces} whose point space is the union of 
  the point spaces of the
  $U_i$ and whose arrow space is the free space of morphisms generated 
  under composition by the
  union of all morphisms in the arrow spaces of the $U_i$.

  Similarly the {\bf intersection of 2-spaces} $\bigcap\limits_{i\in I} U_i \subset B$
  is defined to be the 2-space 
  $\left(\bigcap\limits_{i\in I} U_i^1, \bigcap\limits_{i\in I} U_i^2\right)$
  whose point space is the intersection of all the
  $U_i^1$ and whose arrow space is the intersection of all the $U_i^2$
\end{definition}

Note that the notion of union of sub-2-spaces depends on the total 2-space
$B$, from which the union inherits its source, target and composition map.

\begin{definition}
  \label{2-cover}
  A {\bf 2-cover} $U$ of a 2-space B is a \emph{disjoint union}
  of sub-2-spaces 
  $\bigdisunion\limits_{i\in I} U_i$
  \refdef{sub-2-spaces} such that the point and arrow space
  of each $U_i$ is open and contractible and such that the
  union \refdef{union of 2-spaces} of all $U_i$ is $B$, 
  $\bigcup\limits_{i\in I} U_i = B$.
\end{definition}
Every 2-cover is equippend with a 2-map
\begin{eqnarray}
 \label{cover injection map}
  U \stackto{j} B
\end{eqnarray}
that restricts on each $U_i$ to the inverse of the restriction map of $B$ to $U_i$.

\begin{definition}
  Given a manifold $\manifold$ 
  the 2-space $(\manifold,\loops\manifold)$ whose point space is $M$ and
  whose arrows space is $\loops\manifold$, the space of free loops over 
  $\manifold$ with the obvious source, target and composition maps, is called
  the {\bf free loop 2-space over $\manifold$}.

  Given any ordinary cover $\set{U_i^1}_{i\in I}$ of $\manifold$ 
  by 1-spaces $U_i^1$, the 2-space obtained as the union 
  \refdef{union of 2-spaces}
  with respect to
  $(\manifold,\loops\manifold)$ of all free 
  loop 2-spaces over the $U_i^1$ is called a {\bf local free loop 2-space}
  over $\manifold$ with respect to the cover $\set{U_i^1}_{i\in I}$:
  \[
    \bigcup\limits_{i \in I} (U_i^1, U_i^2 \defas\loops U_i^1) \;\subset (\manifold,\loops\manifold)
    \,.
  \]
\end{definition}

\begin{definition}
Given a 2-cover $U$, one will often need
the spaces of 
{\bf multiple intersections} of the $U_i$. We denote by $U^{[n]}$ the
2-space that is the disjoint union of the $n$-fold 2-space intersections
\refdef{union of 2-spaces} of $U_i$:
\[
  U^{[n]}
  =
  \bigsqcup\limits_{i_1, i_2,\dots,i_n \in I}
    U_{i_1} \cap U_{i_2} \cap \dots \cap U_{i_n}
  \,.
\]
\end{definition}
Each 2-space of multiple intersections comes with 2-maps
\begin{eqnarray}
  \label{cover projection maps}
  U^{[n]} & \stackto{j_{01\cdots(k-1)(k+1)\cdots n}} & U^{[n-1]}
  \nonumber\\
  (i_1,i_2,\dots,i_n,x \stackto{\gamma} y)
  &\mapsto&
  (i_1,\dots,i_{k-1},i_{k+1},\dots,i_n,x \stackto{\gamma} y)  
\end{eqnarray}
that forget about the $k$-th member of the multiple intersection.

\paragraph{Locally trivalizable 2-Bundles.}

With the notion 2-cover in hand, we can now state the definition of a
locally trivializable 2-bundle:

\begin{definition}
 \label{2-bundle with local trivialization}
A {\bf locally trivializable 2-bundle} is a 2-bundle \refdef{2-bundle} 
together with a 2-space $F$ called the {\bf local fiber} such that there
is a 2-cover \refdef{2-cover} $\set{U_i}_{i\in I}$ of the base 2-space $B$ 
for which there exist smooth equivalences
  \[
    p^{-1}\twoU_i \stackto{t_i} \twoU_i \times \twoF
  \]
  such that the diagram
\[ 
\xymatrix @!0
{ p^{-1} U_i
 \ar [dddrr]_{p|_{p^{-1}U_i}}
  \ar[rrrr]^{t_i} 
  & &  & &
  U_i \times F
  \ar[dddll]^{}
  \\ \\ &  \\ & &
   U_i 
 }
\]
commutes for all $i\in I$ up to natural isomorphism.
\end{definition}
Note that if $\set{t_i}_{i\in I}$ is a local trivialization, then so is
$\set{t_i^\prime}_{i\in I}$ with $t^\prime_i$ naturally isomorphic to $t_i$.
Denote by $[t_i]$ the equivalence class of $t_i$ under natural isomorphisms.
\begin{definition}
  \label{local trivialization}
  The 2-cover $\set{U_i}_{i\in I}$ together with 
 the set $\set{[t_i]}_{i\in I}$ is called a {\bf local trivialization}. 
\end{definition}

This definition is concise and elegant, but rather abstract. In 
\S\fullref{section: 2-Transitions in Terms of local Data}
we translate its meaning into transition laws for local data
specifying the 2-bundle. In order to do so, first \emph{transition 
functions} need to be extracted from a local trivialization:

By composing the local trivializations and their weak inverses 
on double intersections $U_{ij}$
one gets autoequivalences of $U_{ij}\times F$ 
of the form
\[
  U_{ij}\times \twoF \stackto{\bar t_i \circ t_j} U_{ij}\times \twoF
\]
and similarly for other index combinations.

\begin{definition}
\label{principal 2-bundle}
A locally trivializable 2-bundle \refdef{2-bundle with local trivialization}
with a local trivialization \refdef{local trivialization}
\begin{itemize}
\item 
where all 
autoequivalences $U_{ij} \times F \stackto{\bar t_i \circ t_j} U_{ij} \times F$ 
act trivially on the $U_{ij}$ factor, so that
\[
  \bar t_i \circ t_j = \mathrm{id}_{U_{ij}} \times g_{ij}
  \,,
\]
\item
where
$F$ is a 2-group $\twogroup$,
\item
and where the $g_{ij}$ act by left horizontal 2-group multiplication on $\twoF$
\end{itemize}
we say that our 2-bundle is a {\bf principal $\twogroup$-2-bundle} and
that 
\begin{eqnarray}
  U^{[2]} & \stackto{g} & \twogroup
  \nonumber\\
  U_{ij} &\mapsto& g_{ij}
  \nonumber 
\end{eqnarray}
is the {\bf transition function}. 
\end{definition}

Note that according to def \ref{local trivialization} each $g_{ij}$ involves 
\emph{choosing}
maps $t_i$, $t_j$ from the equivalence classes $[t_i], [t_j]$ of the local
trivialization. This additional freedom gives rise to the 
modification of the transition law in a principal $\twogroup$-2-bundle as
compared to that of an ordinary principal bundle. This is discussed in the next
section.

\subsubsection{2-Transitions in Terms of local Data}
\label{section: 2-Transitions in Terms of local Data}

Consider a triple intersection $U_{ijk} = U_i \cap U_j \cap U_k$
in a principal $\twogroup$-2-bundle \refdef{principal 2-bundle}.
The existence of the local trivialization implies that the following diagram
2-commutes (all morphisms here are 2-maps and all 
2-morphisms are natural isomorphisms between these):
\begin{center}
\begin{picture}(300,340)
\includegraphics{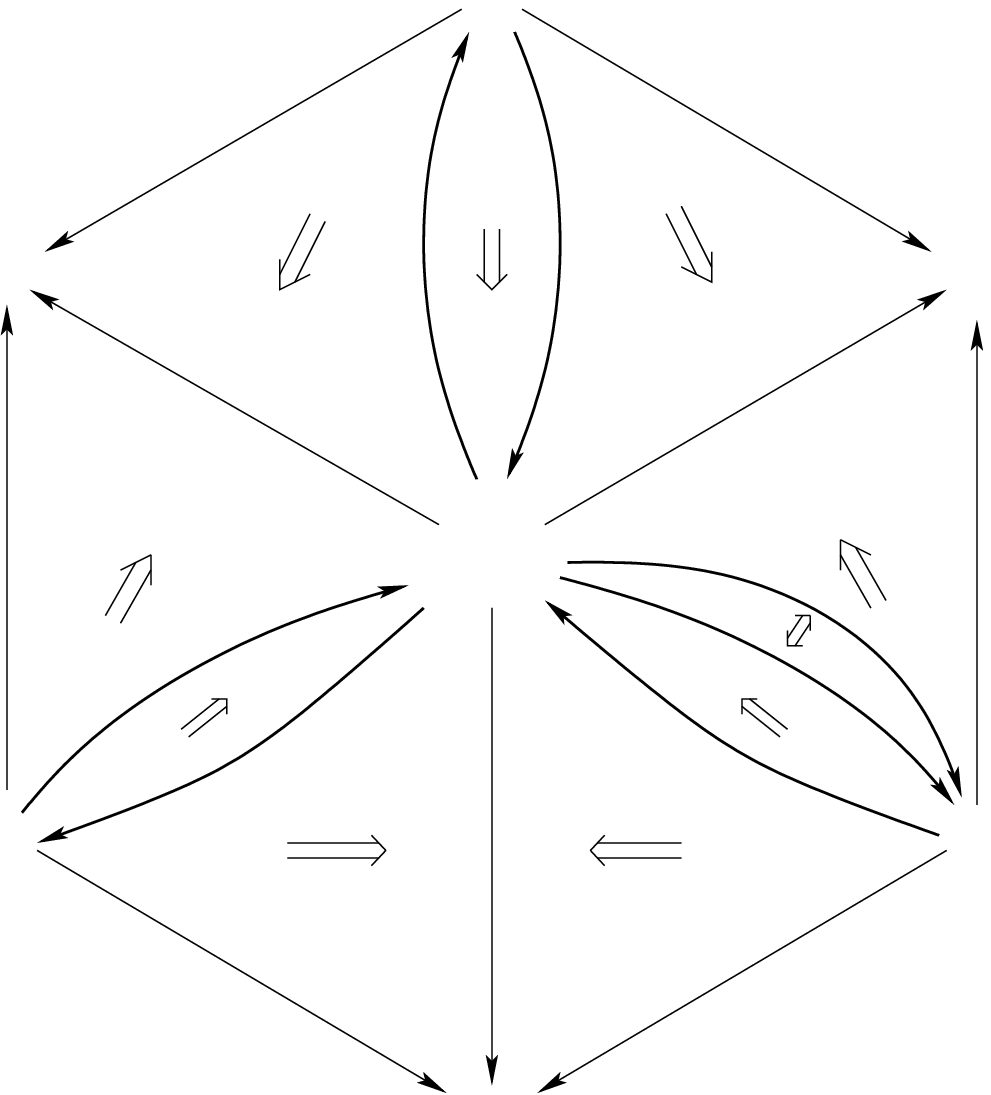}
\put(-162,152){$p^{-1}{U_{ijk}}$}
\put(-155,80){$p$}
\put(-65,190){$p$}
\put(-224,190){$p$}
\put(-150,-7){$U_{ijk}$}
\put(-10,235){$U_{ijk}$}
\put(-293,236){$U_{ijk}$}
\put(-160,320){$U_{ijk}\times F$}
\put(-5,70){$U_{ijk} \times F$}
\put(-320,70){$U_{ijk} \times F$}
\put(-80,87){$\bar t_i$}
\put(-45,120){$t_i$}
\put(-18,122){$t^\prime_i$}
\put(-215,87){$t_j$}
\put(-240,128){$\bar t_j$}
\put(-176,245){$t_k$}
\put(-120,245){$\bar t_k$}
\end{picture}
\end{center}

\vskip 2em

Compared to the analogous diagram for an uncategorified bundle two important new
aspects are
that the barred morphisms are inverses-up-to-isomorphism of the local
trivializations and that the local trivialization itself is unique only up to
natural isomorphisms \refdef{local trivialization}. 
The latter is indicated by the presence of an arrow
denoting a trivialization
$t^\prime_i$ naturally isomorphic to $t_i$.

From the diagram it is clear that the usual transition law 
$g_{ij}g_{jk} = g_{ik}$
here becomes a natural isomorphism called a 2-transition, 
which was first
considered in \cite{Bartels:2004} for the special case of
trivial base 2-spaces \refdef{simple 2-space}, but which directly
generalizes to arbitrary base 2-spaces:

\begin{definition}
  \label{2-transition}
  Given a base 2-space $B$ with cover $U \stackrel{j}{\longrightarrow} B$
  a {\bf 2-transition} is 
  \begin{itemize}
  \item
 a 2-map
\[
  U^{[2]}
  \stackrel{g}{\longrightarrow}
  \twogroup
\]
called the {\bf transition function}, 
\item
and a natural isomorphism
  $f$ 
 \begin{eqnarray}
    &&U^{[3]} 
    \stackrel{j_{02}}{\longrightarrow}
    U^{[2]}
    \stackrel{g}{\longrightarrow}
    \twogroup
    \nonumber\\
    & \stackTo{f} &
    \nonumber\\
    &&
    U^{[3]} \stackrel{\stackrel{\vee}{U}{}^{[3]}}{\longrightarrow}
    U^{[3]} \times U^{[3]}
    \stackrel{j_{01}\times j_{12}}{\longrightarrow}
    U^{[2]} \times U^{[2]}
    \stackrel{g \times g}{\longrightarrow}
    \twogroup \times \twogroup
    \stackrel{m}{\longrightarrow}
    \twogroup
    \,,
    \nonumber
 \end{eqnarray}
(which expressed the categorification of the ordinary transition law
$g_{ij}g_{jk} = g_{ik}$),
  together with the coherence law for $f$ enforcing the
  associativity of the product $g_{ij}g_{jk}g_{kl}$,
  \item
  and a natural isomorphism
  \begin{eqnarray}
    && U \stackto{j_{00}} U^{[2]} \stackto{g} \twogroup
    \nonumber\\
    & \stackTo{\eta} &
    \nonumber\\
    &&
    U \stackto{\hat U} 1 \stackto{i} \twogroup \,.    
    \nonumber
  \end{eqnarray}
  (expressing the categorification of the ordinary $g_{ii} = 1$) together
  with its coherence laws.
\end{itemize}

\end{definition}

In the above $U^{[3]} \stackto{\stackrel{\vee}{U}{}^{[3]}}  U^{[3]}\times U^{[3]}$ 
denotes the
diagonal embedding of $U^{[3]}$ in its second tensor power and $m$ denotes the
horizontal multiplication (functor) in the 2-group $\twogroup$. The maps
$j_{\cdots}$ have been defined in \refer{cover projection maps}.

In terms of local functions this means the following:

\begin{proposition}
  \label{proposition on local meaning of 2-transition}
  A 2-transition 
  \refdef{2-transition}
 on a $\twogroup$-2-bundle with base 2-space being 
a simple 2-space
\refdef{simple 2-space}
and $\twogroup$ a strict 2-group induces the
transition law \refer{gerbe transition law for the transition functions}
of a nonabelian gerbe.
\end{proposition}

\Proof

The existence of the natural isomorphism
means that there is a map
\begin{eqnarray}
  (U^{[3]})^1 &\stackrel{f}{\longrightarrow}& \twogroup^2
  \nonumber\\
  (x,i,j,k) &\mapsto& f_{ijk}\of{x}
  \,,
  \nonumber
\end{eqnarray}
with the property
\begin{eqnarray}
   \label{a relation in a 2-transition}
   g^2_{ik}\of{x} \circ f_{ijk}\of{x}
   &=&
   f_{ijk}\of{x}
   \circ
   (g^2_{ij}\of{x} \cdot g^2_{jk}\of{x})
   \,,
   \hspace{1cm}
   \forall\, (x,i,j) \in U^{[2]}
   \,.
\end{eqnarray}
(Here $\circ$ denotes the vertical and $\cdot$ the horizontal product in the 2-group,
see prop. \ref{strict 2-group coming from crossed module})

For strict $\twogroup$ the source/target matching condition implies that
(again prop. \ref{strict 2-group coming from crossed module})
\begin{eqnarray}
  \label{source/target matching in proof for 2-transition}
  t\of{f^2_{ijk}} g^1_{ik}
  &=&
  g^1_{ij}g^1_{jk}
  \,,
\end{eqnarray}
where we have decomposed the 2-group element
\[
  f_{ijk}\of{x} = (f^1_{ijk}\of{x},f^2_{ijk}\of{x})
\]
into its source label $f^1_{ijk}\of{x} \in G$ 
and its morphism label $f^2_{ijk}\of{x}\in H$.

Identifying $g_{ij} = \phi_{ij}$ this is the gerbe transition law 
\refer{gerbe transition law for the transition functions}.
\endofproof

Note that the assumption that the base 2-space is simple
is crucial for this argument. For more general base 2-spaces the matching
condition would instead read
\begin{eqnarray}
  \label{most general matching condition for 2-transition}
  t(f^2_{ijk})t\of{g^2_{ik}}g^1_{ik}
  &=&
  t\of{g^2_{ij}}g^1_{ij}t\of{g^2_{jk}}g^1_{jk}
  \,.
\end{eqnarray}
Only when the simplicity of the base 2-space forces all $g^2_{ij}$ to take
values in $\ker\of{t}$ does this reduce to the transition law for
a nonabelian gerbe.
But 2-bundles of course exist also for the more general case.

\paragraph{The coherence law for the 2-transition.}

The natural transformation $f$ which weakens the ordinary transition law
$g_{ij}g_{jk} = g_{ik}$ has to satisfy a coherence law which makes its application
on multiple products $g_{ij}g_{jk}g_{kl}$ well defined. 

Note that first of all the result \refer{strict 2-group coming from crossed module} 
implies a certain relation among the $f_{ijk}$:
By using the relation $g^1_{ij}g^1_{jk} = t\of{f_{ijk}}g^1_{ik}$ 
in the expression $g_{ij}g_{jk}g_{kl}$ in two
different ways one obtains
\[
  t\of{f_{ijk}}
  t\of{f_{ikl}}
  =
  g_{ij}t\of{f_{jkl}}g_{ij}^{-1}
  t\of{f_{ijl}}
  \,.
\]
This equation implies that
\begin{eqnarray}
  \label{2-transition coherence law}
  f^{-1}_{ikl}f^{-1}_{ijk}\alpha\of{g_{ij}}\of{f_{jkl}}f_{ijl}
  =
  \lambda_{ijkl}
\end{eqnarray}
with 
\[
  \lambda_{ijkl} \maps U^1_{ijkl} \to  \ker\of{t} \subset H
  \,.
\]

This is the gerbe transition law 
\refer{gerbe coherence law for transformators of transition functions}.
The function $\lambda_{ijkl}$ is the `twist' 0-form 
\refer{gerbe twist p-forms}.

From the perspective of 2-bundles
the twist can be understood as coming from a nontrivial natural
transformation between 2-maps from $U^{[4]}$ to $U^{[2]}$:

First assume that the natural transformation
\begin{eqnarray}
  \label{natural transformation between j023j02 and j013j02}
  &&U^{[4]} \stackto{j_{023} \circ j_{02}} U^{[2]}
  \nonumber\\
  &\stackrel{\omega_{03}}{\Rightarrow}&
  \nonumber\\
  &&
  U^{[4]} \stackto{
  j_{013} \circ j_{02}
  }
  U^{[2]}
  \,.
\end{eqnarray}
is trivial, which means 
that sending a based loop $\gamma_{(x,i,j,k,l)}$ in $(U^{[4]})^2$
first to the based loop $\gamma_{(x,i,k,l)}$ in $(U^{[3]})^2$
and then to $\gamma_{(x,i,l)}$ in $(U^{[2]})^2$
yields the same result as first sending it to
$\gamma_{(x,i,j,l)}$ in $(U^{[3]})^2$ and then to 
$\gamma_{(x,i,l)}$ in $(U^{[2]})^2$.

Using \refer{a relation in a 2-transition} we have

\hspace{-3cm}\parbox{20cm}{
\begin{eqnarray}
  \label{coherence law for transition transformation}
  &&
  (g_{ij}^2 \cdot g^2_{jk}) \cdot g^2_{kl} = 
  g^2_{ij} \cdot (g^2_{jk} \cdot g^2_{kl})
  \nonumber\\
  &\stackrel{\refer{a relation in a 2-transition}}{\Leftrightarrow}&
  \left(
    (f_{ijk})^{r} \circ g^2_{ik} \circ f_{ijk}
  \right) 
  \cdot 
  \left(
    1_{g^1_{kl}} \circ g^2_{kl} \circ 1_{g^1_{kl}}
  \right)
  =
  \left(
    1_{g^1_{ij}}
    \circ
    g^2_{ij}
    \circ
    1_{g^1_{ij}}
  \right)
  \cdot
  \left(
    (f_{jkl})^{r} \circ g^2_{jl} \circ f_{jkl}
  \right)
  \nonumber\\
  &\stackrel{\refer{the exchange law}}{\Leftrightarrow}&
  \left(
    (f_{ijk})^{r} \cdot 1_{g^1_{kl}}
  \right)
  \circ
  \left(
    g^2_{ik}\cdot g^2_{kl}
  \right)
  \circ
  \left(
    f_{ijk} \cdot 1_{g^1_{kl}}
  \right)
  =
  \left(
    1_{g^1_{ij}} \cdot (f_{jkl})^{r} 
  \right)
  \circ
  \left(
    g_{ij}^2 \cdot g_{jl}^2
  \right)
  \circ
  \left(
    1_{g^1_{ij}} \cdot f_{jkl}
  \right)
  \nonumber\\
  &\stackrel{\refer{a relation in a 2-transition}}{\Leftrightarrow}&
  \left(
    (f_{ijk})^{r} \cdot 1_{g^1_{kl}}
  \right)
  \circ
  \left(
    (f_{ikl})^{r} \circ g^2_{il} \circ f_{ikl}
  \right)
  \circ
  \left(
    f_{ijk} \cdot 1_{g^1_{kl}}
  \right)
  =
  \left(
    1_{g^1_{ij}} \cdot (f_{jkl})^{r} 
  \right)
  \circ
  \left(
    (f_{ijl})^{r} \circ g_{il}^2 \circ f_{ijl}^2
  \right)
  \circ
  \left(
    1_{g^1_{ij}} \cdot f_{jkl}
  \right) 
  \,.
  \nonumber\\
\end{eqnarray}
}
This has the form
\begin{eqnarray}
  A^r \circ g^2_{il} \circ A
  &=&
  B^r \circ g^2_{il} \circ B
  \nonumber
\end{eqnarray}
with
\begin{eqnarray}
  \label{conjugation elements in discussion of transition coherence law}
  A &=& f_{ikl} \circ (f_{ijk} \cdot 1_{g^1_{kl}})
  \nonumber\\
  B &=& f_{ijl} \circ (1_{g^1_{ij}} \cdot f_{jkl})
  \,.
\end{eqnarray}
If we identify both `conjugations' we obtain
\begin{eqnarray}
  \label{equality of conjugation elements in discussion of transition coherence law}
  A = B \;&\Leftrightarrow&\;
  (f^2_{ikl})^{-1}(f^2_{ijk})^{-1}\alpha\of{g^1_{ij}}\of{f^2_{jkl}}f^2_{ijl}
  = 1
  \,.
\end{eqnarray}
This reproduces \refer{2-transition coherence law} without the twist.

Now generalize to nontrivial
natural transformations
\refer{natural transformation between j023j02 and j013j02}.
This implies the existence of a function
\[
  (U^{[4]})^1 
  \stackto{\ell}
  (U^{[2]})^2  
\]
that assigns loops based in double overlaps to points in 
quadruple overlaps. Applying the `transition function' $g$ 
to these loops implies that
\begin{eqnarray}
  \label{natural transformation inducind transition twist}
  g^2\of{\ell} \circ g^2\of{j_{013}\circ j_{02}}
  &=&
  g^2\of{j_{023} \circ j_{02}}\circ g^2\of{\ell}
  \,.
\end{eqnarray}
The 2-group element $g^2\of{\ell}$ is specified by a function
\begin{eqnarray}
  (U^{[4]})^1 \stackto{\lambda} \ker\of{t} \subset H
   \nonumber
\end{eqnarray}
as
\begin{eqnarray}
  (U^{[4]})^1 & \stackto{\ell \circ g} & \twogroup^2
  \nonumber\\
  (x,i,j,k,l) & \mapsto & (g^1_{il}, \lambda_{ijkl}\of{x})
  \,.
  \nonumber
\end{eqnarray}

All this applies to \refer{coherence law for transition transformation}
by noting that there on the left hand side the $g_{il}$
in general is $g\of{j_{023}\circ j_{02}}$ while that on the right hand
is $g\of{j_{013}\circ j_{02}}$.

Hence in the case of nontrivial arrow base space we have to
replace the $g^2_{il}$ in the last line on 
the left with $g^2\of{j_{013}\circ j_{02}}$
and that on the right with
$g^2\of{\ell} \circ g^2\of{j_{013}\circ j_{02}} \circ (g^2\of{\ell})^{-1}$.

When doing so the 2-group elements $A$ and $B$ of 
\refer{conjugation elements in discussion of transition coherence law}
become
\begin{eqnarray}
  A &=& (g_{il}^1,\lambda^{-1}_{ijkl}) \circ f_{ikl} \circ (f_{ijk} \cdot 1_{g^1_{kl}})
  \nonumber\\
  B &=& f_{ijl} \circ (1_{g^1_{ij}} \cdot f_{jkl})
  \,.
  \nonumber
\end{eqnarray}
Equating these generalizes 
\refer{equality of conjugation elements in discussion of transition coherence law}
to
\begin{eqnarray}
  A = B \;&\Leftrightarrow&\;
  (f^2_{ikl})^{-1}(f^2_{ijk})^{-1}\alpha\of{g^1_{ij}}\of{f^2_{jkl}}f^2_{ijl}
  = \lambda_{ijkl}
  \,.
  \nonumber
\end{eqnarray}

\paragraph{Restriction to the case of trivial base 2-space.}

It is instructive to restrict the above general disucssion to the
case where the base 2-space is trivial \refdef{simple 2-space}:

In that case the 2-transition specifies the following data:
\begin{itemize}
  \item
    smooth maps
    \[
       g_{ij} \maps U_i \cap U_j \to \twogroup^1
    \] 
   \item
     smooth maps
    \[
      f_{ijk} \maps U_i \cap U_j \cap U_k \to \twogroup^2
    \]
    with
    \[
       f_{ijk}\of{x} \maps  g_{ik}\of{x} \to g_{ij}\of{x}g_{jk}\of{x}
    \]
   \item
     smooth maps
    \[
      k_i : U_i \to \twogroup^2
    \]
    with
    \[
      k_i \maps g_{ii} \to 1 \in \twogroup
      \,.
    \]
\end{itemize}

The coherence law 
\refer{2-transition coherence law} 
says that on quadruple intersections
$U_i \cap U_j \cap U_k \cap U_l$
the following 2-morphisms in $\twogroup$ are identical:

\begin{center}
\begin{picture}(300,150)
\includegraphics{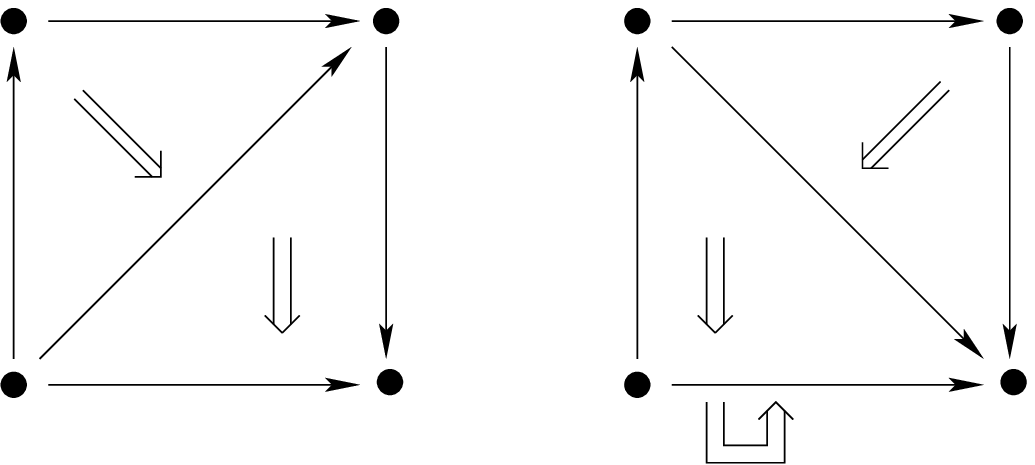}
\put(-5,20){
\begin{picture}(300,150)
\put(-245,115){$g_{jk}$}
\put(-65,115){$g_{jk}$}
\put(-245,-5){$g_{il}$}
\put(-45,-5){$g_{il}$}
\put(-90,-30){$\lambda_{ijkl}$}
\put(-307,55){$g_{ij}$}
\put(-127,55){$g_{ij}$}
\put(-180,55){$g_{kl}$}
\put(-1,55){$g_{kl}$}
\put(-152,55){$=$}
\put(-280,30){$g_{ik}$}
\put(-40,21){$g_{jl}$}
\put(-60,80){$f^{r}_{jkl}$}
\put(-84,30){$f^r_{ijl}$}
\put(-234,30){$f^r_{ikl}$}
\put(-254,77){$f^r_{ijk}$}
\end{picture}
}
\end{picture}
\end{center}

This diagram gives a nice visualization of the different ways to
go from the upper arc $g_{ij}g_{jk}g_{kl}$ of the square to the
bottom edge $g_{il}$.

There are also coherence laws for $k_i$,
the {\bf left unit law} and {\bf right unit law},
which express the relation of $k$
to $f$ when two of the indices of the latter coincide:
\begin{center}
\begin{picture}(230,200)
\includegraphics{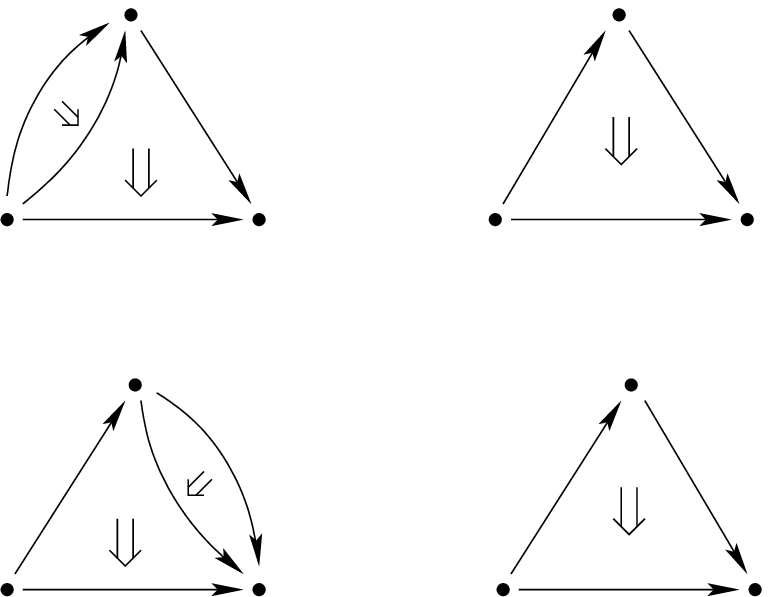}
\put(0,0){X}
\put(-45,-6){$g_{ij}$}
\put(-190,-6){$g_{ij}$}
\put(-45,100){$g_{ij}$}
\put(-190,100){$g_{ij}$}
\put(-20,145){$g_{ij}$}
\put(-161,145){$g_{ij}$}
\put(-71,35){$g_{ij}$}
\put(-215,35){$g_{ij}$}
\put(-75,145){$g_{ii}$}
\put(-224,145){$g_{ii}$}
\put(-200,144){$k_i$}
\put(-193,129){$1$}
\put(-176,122){$1$}
\put(-117,130){$=$}
\put(-55,120){$f^r_{iij}$}
\put(-117,20){$=$}
\put(-16,35){$g_{jj}$}
\put(-149,35){$g_{jj}$}
\put(-172,39){$k_j$}
\put(-175,20){$1$}
\put(-194,15){$1$}
\put(-52,13){$f^r_{ijj}$}
\end{picture}
\end{center}

\vskip 2em

The freedom of having nontrivial $k_i$ is special to 2-bundles and not
known in (nonabelian) gerbe theory. Gerbe cocycles involve {\v C}ech cohomology
and hence \emph{antisymmetry} in indices $i,j,k,\dots$ in the sense that
group valued functions go into their inverse on an odd permutation of their
cover indices.

Whenever we derive nonabelian gerbe cocycles from 2-bundles with 2-connection
we will hence have to restrict to $k_i = 1$ for all $i$.

\vskip 2em

The above diagrams applied to trivial base 2-spaces.
We can also consider nontrivial base 2-spaces, i.e. those
with nontrivial base arrow-space. It turns out however that
the local description of a 2-bundle coincides with the cocycle data
of a nonabelian gerbe only in the limit that the base 2-space morphisms
differ `infinitesimally' from identity morphisms. What this means is
explained in the next subsection.

\subsubsection{2-Bundles on base 2-Spaces of infinitesimal Loops}
\label{2-Bundles on base 2-Spaces of infinitesimal Loops}

The general transition law 
\refer{a relation in a 2-transition}
for the arrow part of the transition function $g$ does
not  seem to have any known counterpart in the theory of nonabelian gerbes
(compare the comment on \refer{most general matching condition for 2-transition}).
But if we let the morphisms of our base 2-space `tend to zero' in
a certain way, the transition law for the arrow part of the $g_{ij}$
reproduces that of the `curving transformation 2-form' 
of a nonabelian gerbe
\refer{curving transformation 2-forms of nonabelian gerbe}.

In order to motivate the definition of `infinitesimal loops'
consider some $\h$-valued 1-form on $\manifold$ and 
some loop $\gamma_x$ in $\manifold$, based at $x$. When the loop 
is very small, the holonomy of $a$ along $\gamma$ is 
approximately given by
\[
  W_a[\gamma_x] \approx \exp\of{\int_\Sigma F_a}
  \,.
\]
Where $F_a = da + a \wedge a$ 
is the curvature of $a$
and $\Sigma$ is some surface in $\manifold$ with
boundary $\partial \Sigma = \gamma$.

For this reason it makes sense to \emph{define} an \emph{infinitesimal} loop 
based at $x$ to be a `tangent parallelogram' at $x$, i.e. an element of the 
dual space $\Omega^2_{\ast x}\of{\manifold}$ of 
the cotangent space $\Omega^2_{x}\of{\manifold}$,
which is the space of antisymmetric rank $(2,0)$ tensors 
(not tensor fields!)
over $\manifold$. The space of all infinitesimal
loops naturally has the structure of a 2-space:

\begin{definition}
  \label{infinitesimal loop 2-space}
  The {\bf free 2-space of infinitesimal loops} 
  over $\manifold$, denoted $d\loops\of{\manifold}$, has
  \begin{itemize}
    \item
      point space 
      \[
        (d\loops\of{\manifold})^1 = \manifold
      \]
    \item
      arrow space 
      \[
        (d\loops\of{\manifold})^2 
        = 
        \bigcup\limits_{x\in \manifold}
        \Omega^2_{\ast x}\of{\manifold}
      \]
    \item
     source and target maps
     \begin{eqnarray}
        d_0,d_1 : (d\loops\of{\manifold})^2 &\to& \manifold
        \nonumber\\
        T_x &\mapsto& x
        \nonumber
     \end{eqnarray}
     \item
      composition of arrows
      \begin{eqnarray}
        \label{composition of infinitesimal loops}
        T_x \circ S_x \defas T_x + S_x
        \,.
        \nonumber
      \end{eqnarray}
  \end{itemize}
\end{definition}

Now consider a principal $\twogroup$-2-bundle over $\manifold$ with base 2-space 
given by $d\loops\of{\manifold}$. The transition functions 
$g^2_{ij}$ must map these infinitesimal loops to closed
2-morphisms in the structure 2-group $\twogroup$
while respecting composition. In other words,
the $g^2_{ij}$ are given by 2-forms 
\[
  d_{ij} \in \Omega^2\of{U_{ij},\ker\of{dt} \subset \h}
  \,,
\] 
such that
\begin{eqnarray}
  \label{action of g^2 on infinitesimal loops}
  g^2_{ij}\of{T_x} \defas \left(g^1_{ij}\of{x}, \exp\of{d_{ij}\of{x}\of{T_x}}\right)
  \,,
\end{eqnarray}
where on the right the parentheses denote the evaluation of a 2-form 
$d_{ij}\of{x}$ on an
antisymmetric rank $(2,0)$ tensor $T_x$.

Since the $d_{ij}$ take values in an abelian subalgebra this action
of $g^2_{ij}$ does respect composition of morphisms as it should:
\begin{eqnarray}
  g^2_{ij}\of{T_x \circ S_x}
  &\equalby{composition of infinitesimal loops}&
  g^2_{ij}\of{T_x + S_x}
  \nonumber\\
  &=&
  \left(
    g^1_{ij}\of{x},
    \exp\of{d_{ij}\of{T_x}}\exp\of{d_{ij}\of{S_x}}
  \right)
  \nonumber\\
  &=&
  g^2_{ij}\of{T_x} \circ g^2_{ij}\of{S_x}
  \,.
  \nonumber
\end{eqnarray}

Given this definition of spaces of `infinitesimal loops' we can now
prove that for such base 2-spaces the 2-transition of a 2-bundle
induces on the arrow-part of $g_{ij}$ a law known from nonabelian gerbe
cocycles:

\begin{proposition}

For a principal $\twogroup$-2-bundle with strict structure 2-group over a base 2-space 
of free infinitesimal loops \refdef{infinitesimal loop 2-space}
the 2-transition
encodes curving transformation 2-forms
  $d_{ij}$ \refer{curving transformation 2-forms of nonabelian gerbe}
 of a nonabelian gerbe 
 in that the arrow part of the 2-transition \refdef{2-transition}
 is equivalent to \refer{gerbe trans law for curving transformation 2-forms}
  for the special case
  \begin{eqnarray}
    &&dt\of{B_i} + F_{A_i} = 0
    \nonumber\\
    &&d_{ij} \in \Omega^2\of{U_{ij},\ker\of{dt}\subset \h}
    \,.
    \nonumber
  \end{eqnarray}
\end{proposition} 

\Proof 
The arrow part of the 2-transition law \refer{a relation in a 2-transition}
is equivalent to
\begin{eqnarray}
  (g^1_{ik}, g^2_{ik})
  \circ
  (g^1_{ik}, f^2_{ijk})
  &=&
  \left(
    g^1_{ik},
    f^2_{ijk}
  \right)
  \circ
  \left(
    g^1_{ij}g^1_{jk},
    g^2_{ij}\, \alpha\of{g^1_{ij}}\of{g^2_{jk}}
  \right)
  \nonumber\\
  \Leftrightarrow
    f^2_{ijk}g^2_{ik}(f^2_{ijk})^{-1}
  &=&
     g^2_{ij}\, \alpha\of{g^1_{ij}}\of{g^2_{jk}} 
  \,.
  \nonumber
\end{eqnarray}

Inserting here the expression \refer{action of g^2 on infinitesimal loops}
for $g_{ij}^2$,
the above becomes
\begin{eqnarray}
  \exp\of{
    f_{ijk}\,d_{ik}\of{T_x} \,f_{ijk}^{-1}
  }
  &=&
  \exp\of{
   d_{ij}\of{T_x}
  }
  \exp\of{
   \alpha\of{g^1_{ij}}\of{d_{jk}\of{T_x}}
  }
  \nonumber\\
  &=&
  \exp\of{
   d_{ij}\of{T_x}
   +
   \alpha\of{g^1_{ij}}\of{d_{jk}\of{T_x}}
  }
  \,.
  \nonumber
\end{eqnarray}
Requiring this for all $T_x$implies
\begin{eqnarray}
  f_{ijk}\, d_{ik} \,f_{ijk}^{-1}
  &=&
  d_{ij}
  +
  \alpha\of{g^1_{ij}}\of{d_{jk}}
  \,.
\end{eqnarray}
This is indeed the gerbe law 
\refer{gerbe trans law for curving transformation 2-forms}
for the given special case.
\endofproof

\paragraph{Summary.}

By internalizing the concept of an ordinary bundle in the 2-category of
2-spaces (which again are categories internalized in $\Diff$, the category
of smooth spaces) one obtains a categorified notion
of the fiber bundle concept, called 2-bundle, which differs from an ordinary bundle 
essentially in that what used to be ordinary maps between sets
(like the projection map of the bundle)
now become (smooth) functors between categories. This adds to the original bundle 
(at the `point level' of the 2-space) a dimensional generalization 
(at the `arrow level' of the 2-space)
of all concepts involved. In addition to providing new `degrees of freedom' the
categorification weakens former notions of equality. 

By re-expressing the abstract arrow-theoretic construction of a 2-bundle in terms
of concrete local group-valued and algebra-valued $p$-forms and relations between them,
we find a generalization of the ordinary transition laws for such local data
in an ordinary bundle. Under certain conditions these generalized transition
laws coincide with the cocycle data of nonabelian gerbes.

So far all of this pertained to 2-bundles (and nonabelian gerbes) without
a notion of connection. For constructing a categorified connection and
hence a notion of nonabelian surface holonomy, it is helpful to 
first consider ordinary connections on spaces of paths in a manifold.
This is the content of the next section.

\newpage

\subsection{Path Space}
\label{section: Path space}

The space of all `paths' in a manifold constitutes an infinite-dimensional
manifold by itself. In the context of Frechet spaces one can study
differential geometry on such infinite dimensional spaces
(e.g. \cite{GetzlerJonesPetrack:1991}). In particular,
we study the notion of holonomy of curves in path space.
A curve in path space over $\manifold$ maps to a (possibly degenerate) 
surface in $\manifold$ and hence its path space holonomy gives rise to
a notion of surface holonomy in $\manifold$.

In this section we first discuss
basic concepts of differential geometry on path spaces
and then apply them to define path space holonomy.
Using that,
a 2-functor $\hol$ from the 2-groupoid of bigons in $\manifold$
(to be defined below)
to a strict structure 2-group is defined and shown to be consistent.
This functor gives us a notion of \emph{local} 2-holonomy which is used
in the subsequent section 
\S\fullref{section 2-bundles with 2-connection} to define
a \emph{global} 2-holonomy by means of 2-transitions.

Throughout the following, various $p$-forms taking values in 
Lie algebras $\g$ and $\h$ are used, where $\g$ and $\h$ are
equipped with the structure of a differential crossed module
$\crossedmodule$ \refdef{differential crossed module}.

Elements of a basis of $\g$ will be denoted by $T_a$
with $a \in (1,\dots,\mathrm{dim}\of{\g})$
and those of a basis of $\h$ by 
$S_a$ with $a \in (1,\dots,\mathrm{dim}\of{\h})$. Arbitrary
elements will be expanded as $A = A^a T_a$.	

  Given a $\g$-valued 1-form $A$ its {\bf gauge covariant exterior derivative}
  is
  \begin{eqnarray}
    \extd_A \omega &\defas& 
    \commutator{\extd + A}{\omega}
    \nonumber\\
    &\defas&
    \extd \omega + A^a \wedge d\alpha\of{T_a}\of{\omega}
    \nonumber
  \end{eqnarray}
  and its {\bf curvature} is
  \begin{eqnarray}
    F_A 
    &\defas&
    (\extd + A)^2
    \nonumber\\
    &\defas&
    \extd A + \frac{1}{2}A^a \wedge A^b \commutator{T_a}{T_b}
    \,.
    \nonumber
  \end{eqnarray}
  By
  a {\bf $\crossedmodule$-valued (1,2)-form} on a manifold 
  $\manifold$ we shall mean a pair $(A,B)$ with
  \begin{eqnarray}
    \label{definition 1,2 form}
    A &\in&  \Omega^1\of{\manifold,\g}
    \nonumber\\
    B &\in&  \Omega^2\of{\manifold,\h}  
   \,.
  \end{eqnarray}

\subsubsection{Path Space Differential Calculus}
\label{path space differential calculus}
	
Differential calculus on spaces of \emph{parametrized} paths can relatively
easily be handled. We start by establishing some basic facts on
parametrized paths and then define the \emph{groupoid of paths} by
considering thin homotopy equivalence classes of parametrized paths. 

\begin{definition}
  \label{based path space}
  Given a manifold $\manifold$,
  the {\bf based parametrized path space}
  $\paths_s^t\of{\manifold}$
  over $\manifold$ with source $s \in \manifold$ and target $t \in \manifold$
is the space of smooth maps  
\begin{eqnarray}
  X : [0,1] &\to& \manifold
  \nonumber\\
  \sigma &\mapsto& X\of{\sigma}
\end{eqnarray} 
which are constant in a neighborhood of $\sigma = 0$ and in a neighborhood of
 $\sigma = 1$.
  When source and target coincide 
  \[
  \Omega_x\of{\manifold}
  \defas
  \paths_x^x\of{\manifold} 
  \]
  is the {\bf based loop space} over $\manifold$ based at $x$.
\end{definition}

The constancy condition at the boundary is known as the property of having
{\bf sitting instant}, compare for instance \cite{CaetanoPicken:1993}.
It serves in def. \ref{groupoid of paths} to ensure that the
composition of two smooth parametrized paths is again a smooth
parametrized path.

We denote a generic path by $X : [0,1] \to \manifold$ 
or by $\gamma: [0,1] \to \manifold$ depending on whether we want to emphasize
that it specifies a point in $\paths_s^t\of{\manifold}$ or a curve 
in $\manifold$, respectively.


  In the study of differential forms on 
  parametrized path space the following notions play an
  important role (\cf \cite{GetzlerJonesPetrack:1991}, section 2):

\begin{definition}
  \label{notions on path space}
  \begin{enumerate}
  \item
  Given any path space $\paths_s^t\of{\manifold}$ \refdef{based path space}, 
  the 1-parameter family of maps
  \begin{eqnarray}
    e_\sigma : \paths_s^t\of{\manifold} &\to& \manifold
    \hspace{1cm}(\sigma \in (0,1))
    \nonumber\\
     X &\mapsto& X\of{\sigma}
     \nonumber
  \end{eqnarray}
  maps each path to its position in $\manifold$ at parameter value $\sigma$.

  \item
  Given any differential $p$-form $\omega \in \Omega^p\of{\manifold}$
  the pullback to $\paths_s^t\of{\manifold}$ by $e_\sigma$ shall be denoted simply  by
  \begin{eqnarray}
    \omega\of{\sigma}
    &\defas&
    e^\ast_\sigma\of{\omega}
    \,.
    \nonumber
  \end{eqnarray}

  \item
  With respect to a local coordinate patch on $\manifold$ 
  the differential forms on $\paths_s^t\of{\manifold}$ are generated by
  the 1-forms $\set{dX^\mu\of{\sigma} | \mu \in \set{1,\dots, \mathrm{dim}\of{\manifold}},
  \sigma\in (0,1)}$, which we equivalently write as
  \begin{eqnarray}
    dX^{(\mu,\sigma)}
    &\defas&
    dX^\mu\of{\sigma}
    \,.
    \nonumber
  \end{eqnarray}
  Integration over $\sigma$ will be abbreviated as implicit index contraction, as in
  \begin{eqnarray}
    f_{(\mu,\sigma)}dX^{(\mu,\sigma)}
    &\defas&
    \sum\limits_{\mu=1}^{\mathrm{dim}\of{\manifold}}\int\limits_0^1 d\sigma\; 
      f_\mu\of{\sigma}dX^\mu\of{\sigma}
   \,. 
   \nonumber
 \end{eqnarray}

  \item
  In local coordinates the pullback of any form via $e_\sigma$ reads 
  \begin{eqnarray}
    \omega\of{\sigma}
    &=&
    \omega_{\mu_1\dots\mu_p}
    \underbrace{
      \frac{\partial X^{(\mu_1,\sigma)}}{\partial X^{(\nu_1,\rho_1)}}
    }_{= \delta^{\mu_1}_{\nu_1}\delta\of{\sigma,\rho_1}}
    \cdots
    \underbrace{
      \frac{\partial X^{(\mu_p,\sigma)}}{\partial X^{(\nu_p,\rho_p)}}
    }_{= \delta^{\mu_p}_{\nu_p}\delta\of{\sigma-\rho_p}}
    \of{X\of{\sigma}}
    \,
    dX^{(\nu_1,\rho_1)}
    \wedge
    \cdots
    \wedge
    dX^{(\nu_p,\rho_p)}
    \nonumber\\
    &=&
    \omega_{\mu_1\dots\mu_p}\of{X\of{\sigma}}
    dX^{(\mu_1,\sigma)}
    \wedge \cdots \wedge
    dX^{(\mu_p,\sigma)}    
    \,.
    \nonumber
  \end{eqnarray}

  \item
  The vector field
  \begin{eqnarray}
    K(X) &\defas& \frac{d}{d\sigma}X
    \nonumber\\
    &\defas&
    X^\prime
    \nonumber
  \end{eqnarray}
  on path space generates rigid reparameterizations. Its push-forward $e_{\sigma\ast}\of{K}(X)$
  to $\manifold$ is the tangent to the path $X$ at $X\of{\sigma}$.   

  \item
  The contraction of $\omega\of{\sigma}$ with $K$ is denoted by
  $\iota_K \omega\of{\sigma}$.
 
  \item
   The exterior differential on path space reads in local coordinates
  \begin{eqnarray}
     \label{local version of path space exterior derivative}
     \extd &=& dX^{(\mu,\sigma)}\wedge \frac{\delta}{\delta X^{(\mu,\sigma)}}
     \,.
  \end{eqnarray}

 \end{enumerate}
\end{definition}

A special class of differential forms on path space play a major role:

\begin{definition}
  \label{Chen forms}
  Given a familiy $\set{\omega_i}_{i=1}^N$ of differential forms on a manifold $\manifold$
  with degree
  \begin{eqnarray}
    \mathrm{deg}\of{\omega_i} &\defas& p_i +1
    \nonumber
  \end{eqnarray}
  one gets a differential form (\cf \refdef{notions on path space})
  \begin{eqnarray}
    \Omega_{\set{\omega_i},(\alpha,\beta)}\of{X}
    &\defas&
    \oint\limits_{X|_{\alpha}^{\beta}} (\omega_1, \dots, \omega_n)
    \;\defas\;
    \int\limits_{\alpha < \sigma_i < \sigma_{i+1} < \beta}
    \!\!\!\!\!\!
    \iota_K \omega_1\of{\sigma^1}
    \wedge
    \cdots
    \wedge
    \iota_K \omega_N\of{\sigma^N}
    \nonumber
  \end{eqnarray}
  of degree
  \begin{eqnarray}
    \mathrm{deg}\of{\Omega_{\set{\omega_i}}} &=& \sum\limits_{i=1}^N p_i
    \,,
    \nonumber
  \end{eqnarray}
  on any based parametrized path space $P_s^t\of{\manifold}$
  \refdef{based path space}.

  For $\alpha=0$ and $\beta = 1$ we write
  \begin{eqnarray}
    \Omega_{\set{\omega_i}}
    &\defas&
    \Omega_{\set{\omega_i},(0,1)}
    \,.
    \nonumber
  \end{eqnarray}
  These path space forms are known as
  {\bf multi integrals} or {\bf iterated integrals}
  or {\bf Chen forms} (\cf \cite{GetzlerJonesPetrack:1991,Hofman:2002}).
\end{definition}

It turns out that the exterior derivative on path space maps
Chen forms \refdef{Chen forms} to Chen forms in a nice way:

\begin{proposition}
  \label{action of extd on Chen forms} 
  The action of the path space exterior derivative 
  \refer{local version of path space exterior derivative}
  on Chen forms \refdef{Chen forms} is
\begin{eqnarray}
  \label{extd on path space Chen forms}
  \extd \oint (\omega_1, \cdots,\omega_n)
  &=&
  (\tilde \extd + \tilde M)
  \oint (\omega_1, \cdots,\omega_n)
  \,,
\end{eqnarray}
where
\begin{eqnarray}
  \tilde \extd \oint (\omega_1, \cdots,\omega_n)
  &\defas&
  -\sum\limits_k (-1)^{\sum\limits_{i < k}p_i}
    \oint (\omega_1, \cdots, \extd\omega_k, \cdots, \omega_n)
  \nonumber\\
  \tilde M \oint (\omega_1, \cdots,\omega_n)
  &\defas&
  -\sum\limits_k (-1)^{\sum\limits_{i < k}p_i}
    \oint (\omega_1,\cdots, \omega_{k-1}\wedge \omega_k,\cdots,\omega_n)
  \,,
  \nonumber
\end{eqnarray}
satisfying
\begin{eqnarray}
  \tilde \extd^2 &=& 0
  \nonumber\\
  \tilde M^2 &=& 0
  \nonumber\\
  \antiCommutator{\tilde \extd^2}{\tilde M^2} &=& 0
  \,. 
\end{eqnarray}
\end{proposition}
(\cf \cite{GetzlerJonesPetrack:1991,Hofman:2002})

Before using these facts for the investigation of path space holonomy
let us conclude by mentioning 
\paragraph{The groupoid of paths.}

\begin{definition}
  \label{groupoid of paths}
  The {\bf groupoid of paths} $\paths_1\of{\manifold}$ 
  in a manifold $M$ is the groupoid for which
  \begin{itemize}
    \item
      objects are points $x \in \manifold$
    \item
      morphisms with source $s \in \manifold$ and target $t\in \manifold$
      are thin homotopy equivalence classes $[\gamma]$
      of parametrized paths $\gamma \in \paths_s^t\of{\manifold}$
      \refdef{based path space}
     \[
       \xymatrix{
         x \ar@/^1pc/[rr]^{[\gamma]}
         && y
       }
     \]
    \item
      composition is given by
   \[
     \xymatrix{
     x \ar@/^1pc/[rr]^{[\gamma_1]}
     && y \ar@/^1pc/[rr]^{[\gamma_2]}
     && z
     }
     =
     \xymatrix{
       x \ar@/^1pc/[rr]^{[\gamma_1 \circ \gamma_2]}
       && z
      }
   \]
    where
    \begin{eqnarray}
       \circ \maps \paths_x^y\of{\manifold} \times \paths_y^z\of{\manifold}
       &\to&
       \paths_x^z\of{\manifold}
       \nonumber\\
       (\gamma_1,\gamma_2) &\mapsto&
       \gamma_{1,2}
       \nonumber
    \end{eqnarray}
    with
    \begin{eqnarray}
      \gamma_{1,2}\of{\sigma}
      &\defas&
      \left\lbrace
        \begin{array}{cc}
           \gamma_1\of{2\sigma} & \mbox{for $0 \leq \sigma \leq 1/2$} \\
           \gamma_2\of{2\sigma-1} & \mbox{for $1/2 \leq \sigma \leq 1$}
        \end{array}
      \right.
      \,.
      \nonumber
    \end{eqnarray}
  \end{itemize}
\end{definition}

Note that taking thin homotopy equivalence classes makes this composition
asscociative and invertible.

In a similar manner we define the 2-groupoid of bigons in 
def. \ref{2-groupoid of bigons} below.

\subsubsection{The Standard Connection 1-Form on Path Space}
\label{section: The Standard Connection 1-Form on Path Space}

There are many 1-forms on path space that one could consider as
local connection 1-forms in order to define a local holonomy on
path space. Here we restrict attention to a special class, to be called
the \emph{standard connection 1-forms} 
\refdef{standard path space 1-form}, because, as is shown 
in \S\fullref{section: Local 2-Holonomy from local Path Space Holonomy},
these turn out to be the ones which compute local 2-group holonomy.
(This same `standard connection 1-form' can however also be motivated
from other points of view, as done in 
\cite{AlvarezFerreiraSanchezGuillen:1998,Schreiber:2004e}.)

\paragraph{Motivation of the standard path space connection 1-form.}

In order to roughly see how 2-group holonomy gives rise to a
connection on path space, consider an iterated horizontal product
of 2-group morphisms labeling
 a row of small
`surface elements' as follows:
\[
\xymatrix{
   \bullet \ar@/^1pc/[rr]^{g_1}_{}="0"
           \ar@/_1pc/[rr]_{}="1"
           \ar@{=>}"0";"1"^{h_1}
&& \bullet \ar@/^1pc/[rr]^{g_2}_{}="2"
           \ar@/_1pc/[rr]_{}="3"
           \ar@{=>}"2";"3"^{h_2}
&& \bullet \ar@/^1pc/[rr]^{g_3}_{}="2"
           \ar@/_1pc/[rr]_{}="3"
           \ar@{=>}"2";"3"^{h_3}
&& \bullet
}
 \cdots
\xymatrix{
   \bullet \ar@/^1pc/[rr]^{g_N}_{}="0"
           \ar@/_1pc/[rr]_{}="1"
           \ar@{=>}"0";"1"^{h_N}
&& \bullet 
}
 =
  \xymatrix{
   \bullet \ar@/^1pc/[rr]^{g^\mathrm{tot}}_{}="0"
           \ar@/_1pc/[rr]_{}="1"
           \ar@{=>}"0";"1"^{h^\mathrm{tot}} 
   && \bullet
   }
\]
Here the $j$-th morphism is supposed to be given by $(g_j,h_j) \in \twogroup$
with $g \in G$ and $h\in H$.
By the rules of 2-group multiplication 
(prop. \ref{strict 2-group coming from crossed module})
the total horizontal product 
$(g^\mathrm{tot},h^{\mathrm{tot}})$ is given by
\begin{eqnarray}
  g^\mathrm{tot} &=& g_1\, g_2\, g_3 \cdots g_N
  \nonumber\\
  h^\mathrm{tot} &=& h_1 \,\alpha\of{g_1}\of{h_2}\, \alpha\of{g_1g_2}\of{h_3}
  \cdots \alpha\of{g_1g_2g_3\cdots g_{N-1}}\of{h_N}
  \,.
  \nonumber
\end{eqnarray}
The products of the $g_j$ can be addressed as a \emph{holonomy} along the 
upper edges, which, for reasons to become clear shortly, we shall write as
\[
  g_1 \, g_2 \cdots g_{j} \defas (W_{j+1})^{-1}
  \,.
\]
Now suppose the group elements come from algebra elements $A_j \in \g$ and $B_j \in \h$
as
\begin{eqnarray}
  g_j &\defas& \exp\of{\epsilon A_j}
  \nonumber\\
  h_j &\defas& \exp\of{\epsilon^2 B_j}
  \nonumber
\end{eqnarray}
where
\[
  \epsilon \defas 1/N
  \,,
\]
then
\[
  h^\mathrm{tot}
  =
  1
  +
  \epsilon^2
  \sum\limits_{j=1}^N
  \alpha\of{W^{-1}_j}\of{B_j}
  +
  \order{\epsilon^4}
  \,.
\]
Using the notation
\begin{eqnarray}
  W_j &\defas& W\of{1-j\epsilon,1}
  \nonumber\\
  B_j &\defas& B\of{1-\epsilon j}
  \nonumber
\end{eqnarray}
we have
\begin{eqnarray}
  h^\mathrm{tot}
  &=&
  1+
  \epsilon \,
  \int\limits_0^1 d\sigma\;
  \alpha\of{W^{-1}\of{\sigma,1}}\of{B\of{\sigma}}
  + 
  \order{\epsilon^3}
  \,.
  \nonumber
\end{eqnarray}
Finally, imagine that the $\twogroup$-labels $h^{\mathrm{tot}}_k$ of 
many such thin horizontal rows of `surface elements' 
are composed \emph{vertically}. Each of them comes from algebra elements
\[
  B_k\of{\sigma} \defas B\of{\sigma,k\epsilon}
\]
and holonomies
\[
  W_k\of{\sigma,1} \defas W_{k\epsilon}\of{\sigma,1}
\]
as
\[
  h_k^\mathrm{tot}
  \defas
  1+
  \epsilon \,
  \int\limits_0^1 d\sigma\;
  \alpha\of{W_{k\epsilon}^{-1}\of{\sigma,1}}\of{B\of{\sigma,k\epsilon}}
  + 
  \order{\epsilon^3}  
  \,.
\]
In the limit of vanishing $\epsilon$ their total vertical product is
\begin{eqnarray}
  \lim\limits_{\epsilon = 1/N \to 0}
  h^\mathrm{tot}_0 h^\mathrm{tot}_\epsilon h^\mathrm{tot}_{2\epsilon}
  \cdots h^\mathrm{tot}_1
  &=&
  \mathrm{P}
  \exp\of{
    \int\limits_0^1
      d\tau\;
      \mathcal{A}\of{\tau}
  }
  \nonumber
\end{eqnarray}
for
\begin{eqnarray}
  \label{motivation for standard path space connection}
  \mathcal{A}\of{\tau} 
  &=& 
  \int\limits_0^1 d\sigma\; \alpha\of{W^{-1}_\tau\of{\sigma,1}}\of{B\of{\sigma,\tau}}
  \,,
\end{eqnarray}
where $\mathrm{P}$ denotes path ordering with respect to $\tau$.

Thinking of each of these vertical rows of surface elements as 
paths (in the limit $\epsilon \to 0$), this shows roughly how the computation
of total 2-group elements from vertical and horizontal products of many
`small' 2-group elements can be reformulated as the holonomy of a connection
on path space of the form \refer{motivation for standard path space connection}. 

What is missing in the above
discussion is the precise identification of path space differential forms.
In the following a path space 1-form having the structure
\refer{motivation for standard path space connection} is defined and
it is shown that indeed its holonomy defines a functor from a 2-groupoid
of `surface elements' (`bigons') to the 2-group $\twogroup$, thus making
the above discussion precise.

In order to get there, we first need to deal with some basic issues of
holonomies and parallel transport:

\paragraph{Holonomy and parallel transport.}

In order to set up some notation and conventions and for later references, 
the following gives a list of well-known 
definitions and facts that are crucial for the further developments:

\begin{definition}
  \label{line holonomy and parallel transport}
  Given a path space $\paths_s^t\of{\manifold}$ \refdef{based path space}
  and a 
  $\crossedmodule$-valued (1,2)-form $(A,B)$ 
  \refer{definition 1,2 form}
  on $\manifold$, the following objects are of interest:
  \begin{enumerate}
    \item
      The {\bf line holonomy} of $A$ along a given path $X$ is denoted by
      \begin{eqnarray}
        \label{def line holonomy}
        W_A[X]{\of{\sigma^1,\sigma^2}} 
        &\defas&
        \mathrm{P}\exp\of{\int\limits_{X|_{\sigma^1}^{\sigma^2}} A}
        \nonumber\\
        &\defas&
        \sum\limits_{n=0}^{\infty}
        \oint\limits_{X|_{\sigma^1}^{\sigma^2}}
        (A^{a_1},\dots ,A^{a_n})
        T_{a_1}\cdots T_{a_n}
        \,.
      \end{eqnarray}

   \item
    The {\bf parallel transport} of elements in $T \in \g$ and 
    $S\in \h$
   is written
\begin{eqnarray}
  \label{notions of parallel tranport}
  W_A[X]\of{\sigma,1}\of{T\of{\sigma}}
  &\defas&
  T^{W_A[X]}\of{\sigma}
  \nonumber\\
  &\defas&
  W^{-1}_A[X]\of{\sigma,1}T\of{\sigma}W_A[X]\of{\sigma,1}
  \nonumber\\
  &=&
  \sum_{n=0}^\infty
   \oint_{X|_\sigma^1} (-A^{a_1},\cdots,-A^{a_n})
   \;
   \commutator{T_{a_n}}{\cdots \commutator{T_{a_1}}{T\of{\sigma}}\cdots}     
  \,,
  \nonumber\\
  W_A[X]\of{\sigma,1}\of{S\of{\sigma}}
  &\defas& S^{W_A[X]}\of{\sigma}
  \nonumber\\
  &\defas&
  \alpha\of{W^{-1}_A[X]\of{\sigma,1}}\of{S\of{\sigma}}
  \nonumber\\
  &\defas&
  \sum_{n=0}^\infty
   \oint_{X|_\sigma^1} (-A^{a_1},\cdots,-A^{a_n})
   \;
   d\alpha(T_{a_n})\circ \cdots \circ d\alpha\of{T_{a_1}}\of{S\of{\sigma}}    
  \,.
  \nonumber\\
\end{eqnarray}
  \end{enumerate}
\end{definition}

For convenience the dependency $[X]$ on the path $X$ will often be omitted.

\begin{proposition}
  \label{properties of parallel transport}
  Parallel transport \refdef{line holonomy and parallel transport} has the following
  properties:
  \begin{enumerate}
   \item 
   Let $\sigma_1 \leq \sigma_2 \leq \sigma_3$
   then
   \begin{eqnarray}
     W_A[X]\of{\sigma_1,\sigma_2} \circ W_A[X]\of{\sigma_2,\sigma_3}  
     &=&
     W_A[X]\of{\sigma_1,\sigma_3}
     \,.
     \nonumber
   \end{eqnarray}
   \item
  Conjugation of elements in $\g$ with parallel tranport of elements
in $\h$ yields
\begin{eqnarray}
  \label{conjugation of g by holonomies}
  W_A\of{\sigma,1}\of{
    d\alpha\of{T}\of{\sigma}
    \of{
      W_A^{-1}\of{\sigma,1}
      \of{
        S
      }
    }
  }
  &=&
  d\alpha\of{T^{W_A}\of{\sigma}}\of{S}
  \,.
\end{eqnarray}

 \item
  Given a $G$-valued 0-form
  $
    g \in \Omega^0\of{\manifold,G} 
  $
  and a path $X \in \paths_x^y\of{\manifold}$
  we have
  \begin{eqnarray}
    \label{gauge trafo of parallel transport}
    g\of{x}W_A[X](g\of{y})^{-1}
    &=&
    W_{(gAg^{-1} + g^{-1}\extd g)}[X]
    \,.
  \end{eqnarray}

  \item
  Given a $G$-valued 0-form
  $
    g \in  \Omega^0\of{\manifold,G} 
  $
  and a based loop $X \in \paths_x^x\of{\manifold}$
  we have
  \begin{eqnarray}
     \label{action of phi on parallel transport}
     \alpha\of{\phi\of{x}}\of{W_A[X]\of{\sigma,1}\of{S\of{\sigma}}}
     &=&
     W_{A^\prime}[X]\of{\sigma,1}\of{\alpha\of{\phi\of{X\of{\sigma}}}\of{S\of{\sigma}}}
  \end{eqnarray}
  with
  \begin{eqnarray}
    A^\prime &\defas& \phi A \phi^{-1} + \phi (d\phi^{-1})
    \,.
    \nonumber
  \end{eqnarray}
  \end{enumerate}
\end{proposition}

\Proof
\begin{enumerate}
  \item
    This follows by looking at infinitesimal parallel transport.
  \item
  Integrate up the infinitesimal relation
  \begin{eqnarray}
    d\alpha\left(
      1- \epsilon X^\prime \inner A
    \right)
    \of{
      d\alpha\of{T}\of{
        d\alpha\left(
           1 + \epsilon X^\prime \inner A
        \right)
        \of{
          S
        }
      }
    }
   &=&
   d\alpha\of{T}\of{S}
   -
   \epsilon d\alpha\of{\commutator{X^\prime\inner A}{T}}\of{S}
   +
   \order{\epsilon^2}
   \,.
   \nonumber\\
  \end{eqnarray}  

  \item
   Using infinitesimal steps one finds
    
   \hspace{-3cm}\parbox{20cm}{
   \begin{eqnarray}
    &&g\of{x}W_A[X](g\of{y})^{-1}
    \nonumber\\
    &=&
    \lim\limits_{\epsilon\to 0}
    g\of{x}
      (1 + \epsilon X^\prime \inner A\of{x})
      (1 + \epsilon X^\prime \inner A\of{x+\epsilon X^\prime})
      \cdots
      (1 + \epsilon X^\prime \inner A\of{y})
   (g\of{y})^{-1}
   \nonumber\\
   &=&
    \lim\limits_{\epsilon\to 0}
    g\of{x}
      (1 + \epsilon X^\prime \inner A\of{x})
    (g\of{x+\epsilon})^{-1}g\of{x+\epsilon}
      (1 + \epsilon X^\prime \inner A\of{x+\epsilon X^\prime})
      \cdots
      (1 + \epsilon X^\prime \inner A\of{y})
   (g\of{y})^{-1}
   \nonumber\\
   &=&
    \lim\limits_{\epsilon\to 0}
    g\of{x}
      (1 + \epsilon X^\prime \inner A\of{x})
    (g\of{x})^{-1}(1+ \epsilon X^\prime\inner  (g\of{x}\extd (g\of{x})^{-1}))g\of{x+\epsilon}
      (1 + \epsilon X^\prime \inner A\of{x+\epsilon X^\prime})
      \cdots
      (1 + \epsilon X^\prime \inner A\of{y})
   (g\of{y})^{-1}
   \nonumber\\
   &=& \cdots
   \nonumber\\
   &=&
   W_{gAg^{-1} + g\extd g^{-1}}[X]
   \,.
  \end{eqnarray}
  }

  \item
  Again consider infinitesimal steps to obtain
  \begin{eqnarray}
  &&
  \phi\of{x}\of{
    W_A[X]\of{\sigma_1,1}
    \of{
      S
    }  
  } 
  \nonumber\\
  &=&
  \lim\limits_{\epsilon \to 0}
  \phi\of{x}\of{
      W_A\of{1-\epsilon,1}
    \of{
      W_A\of{1-2\epsilon,1-\epsilon}
    \of{
      \cdots
    W_A\of{\sigma,\sigma+\epsilon}
      \of{
      S
      \cdots
    }  
  }   
  }
  }
  \nonumber\\
  &=&
  \lim\limits_{\epsilon \to 0}
  \phi\of{x}
   \of{
    W_A\of{1-\epsilon,1}
   \of{
     \phi^{-1}\of{x-\epsilon X^\prime\of{x}}
    \of{
      \phi\of{x-\epsilon X^\prime\of{x}}
    \of{
      \cdots
    }}}}
  \nonumber\\
  &=&
  \lim\limits_{\epsilon \to 0}
  \phi\of{x}
   \of{
    W_A\of{1-\epsilon,1}
   \of{
     \phi^{-1}\of{x}
     (1 - \phi\of{x}\epsilon X^\prime \inner (d\phi^{-1}\of{x}))
    \of{
      \phi\of{x+\epsilon X^\prime\of{x}}
    \of{
      \cdots
    }}}}  
   \nonumber\\
   &=&
   \dots
   \nonumber\\
   &=&
   W_{A^\prime}[X]\of{\sigma,1}\of{\phi\of{X\of{\sigma}}\of{S}}
   \,.
  \end{eqnarray}
\end{enumerate}
\endofproof

As the discussion at the beginning of this section showed, 
integrals over $p$-forms pulled back to a path and 
parallel tranported to some base point
play an important role for path space holonomy. Following
\cite{Hofman:2002,Schreiber:2004e} 
we introduce special notation to take care
of that automatically:

\begin{definition}
  \label{notation Chen form with parallel transport}
  A natural addition to the notation \refer{Chen forms}
  for iterated integrals in the presence of a $\g$-valued
  1-form $A$ is the abbreviation
  \begin{eqnarray}
    \oint_A\of{
      \omega_1,\dots\omega_N
    }
    &\defas&
    \oint\of{
      \omega_1^{W_A},\dots \omega_N^{W_A}
    }
    \,,
    \nonumber
  \end{eqnarray}
  where $(\cdot)^{W_A}$ is defined in def \ref{line holonomy and parallel transport}.
  When Lie algebra indices are displayed on the left they are defined to 
pertain to the parallel tranported object:
  \begin{eqnarray}
    \label{special notation in Chen form with parallel transport}
    \oint_A (\dots,\omega^a,\dots)
    &\defas&
    \oint (\dots,(\omega^{W_A})^a,\dots)
    \,.
  \end{eqnarray}
\end{definition}

Using this notation first of all the following fact can be conveniently 
stated, which plays a central role 
in the analysis of the 
transition law for the 2-holonomy in 
\S\fullref{section 2-bundles with 2-connection}:

\begin{proposition}
  \label{difference in holonomies wrt different connections}
  The difference in line holonomy \refdef{line holonomy and parallel transport}
  along a given loop with respect to two different 1-forms $A$ and $A^\prime$
  can be expressed as
  
  \hspace{-3cm}\parbox{20cm}{
  \begin{eqnarray}
    (W_{A}[X])^{-1}W_{A^\prime}[X] 
    &=&
    \lim\limits_{\epsilon = 1/N \to 0}
    \left(
      1 + \epsilon\oint\limits_{A} (A^\prime - A)
    \right)
    \left(
      1 + \epsilon\oint\limits_{A+\epsilon(A^\prime-A)} (A^\prime - A)
    \right)
    \cdots
    \left(
      1 + \epsilon\oint\limits_{A^\prime- \epsilon (A^\prime-A)} (A^\prime - A)
    \right)_X
    \,.
    \nonumber
  \end{eqnarray}
}

\end{proposition}
\Proof

First note that from def. \ref{line holonomy and parallel transport} 
it follows that
\begin{eqnarray}
  \oint\limits_A \of{A^\prime - A}
  &=&
  \int\limits_{0}^1
  d\sigma
  (W_A[X](\sigma,1))^{-1} \iota_K (A^\prime - A)\of{\sigma}
  W_A[X](\sigma,1)
  \,.
  \nonumber
\end{eqnarray}
This implies that
\begin{eqnarray}
  W_A[X]\left(1 + \epsilon \oint_{A}\of{A^\prime - A}\right)_X
  &=&
  W_{A+\epsilon (A^\prime-A)}[X] + \order{\epsilon^2}
  \,.
  \nonumber
\end{eqnarray}
The proposition follows by iterating this.
\endofproof

\paragraph{Exterior derivative and curvature for Chen forms.}

The exterior derivative on path space maps Chen forms to Chen forms
(prop. \ref{action of extd on Chen forms}). Since, for
reasons explained at the beginning of this section, we shall be interested 
in Chen forms involving parallel transport 
\refdef{notation Chen form with parallel transport}, it is important
to know also the particular action of the exterior derivative on these:

\begin{proposition}
  \label{extd on A-Chen forms of one form}
  The action of the path space exterior derivative on
  $\oint_A\of{\omega}$ is
  \begin{eqnarray}
    \label{path space extd on Chen forms with parallel transport}
    \extd \oint_A (\omega)
    &=&
    -\oint_A (\extd_A \omega)
    -
    (-1)^{\mathrm{deg}\of{\omega}}
    \oint_A\of{
      d\alpha\of{T_a}\of{\omega}, F_A^a
    }
    \,.
  \end{eqnarray}
\end{proposition}
(Recall the convention \refer{special notation in Chen form with parallel transport}).

\Proof

Using the identities
\begin{eqnarray}
  \label{defintion dA B}
  \extd_A \omega
  &=&
  \extd \omega + A^{a_1}\wedge d\alpha\of{T_{a_1}}\of{\omega}
\end{eqnarray}
and
\begin{eqnarray}
 \label{equation involving field strength}
 (\extd A^{a}) \; d\alpha\of{T_a}
 +
 (A^a \wedge A^b)
  \;
 d\alpha\of{T_a}\circ d\alpha\of{T_b}
  &=&
 (\extd A^{a}) \; d\alpha\of{T_a} 
 +
 (A^a \wedge A^b)
 \frac{1}{2}
 d\alpha\of{\commutator{T_a}{T_b}}\of{S}
 \nonumber\\
 &=&
 F_A^{a} \;d\alpha\of{T_a}
 \,.
\end{eqnarray}
the exterior derivative for the case $\mathrm{deg}(C) = \mathrm{odd}$ yields:
\begin{eqnarray}
  &&
  \extd \oint (\omega^{W_A})
  \nonumber\\
  &=&
    \sum_{n=0}^\infty
    \extd\oint (d\alpha(T_{a_n})\circ \cdots \circ d\alpha\of{T_{a_1}}\of{\omega},-A^{a_1},\cdots,-A^{a_n})    
   \nonumber\\
  &\equalby{extd on path space Chen forms}&
    -
    \sum_{n=0}^\infty
      \oint (d\alpha(T_{a_n})\circ \cdots \circ d\alpha\of{T_{a_1}}\of{\extd \omega},-A^{a_1},\cdots,-A^{a_n})  
    \nonumber\\
    &&
      -
      \oint (d\alpha(T_{a_n})\circ \cdots \circ d\alpha\of{T_{a_1}}\of{\omega}\wedge(-A^{a_1}) ,-A^{a_2}\cdots,-A^{a_n})        
   \nonumber\\
    &&
      -
    \sum_{n=0}^\infty
    \sum_{1 \leq k \leq n}
    \oint (d\alpha(T_{a_n})\circ \cdots \circ d\alpha\of{T_{a_1}}\of{\omega},iA^{a_1},\cdots,
    -A^{a_{k-1}},\extd (-A^{a_k}),
    -A^{a_{k+1}},
    \cdots,
    -A^{a_n})    
    \nonumber\\
    &&
      -
    \sum_{n=0}^\infty
    \sum_{1 < k \leq n}
    \oint (d\alpha(T_{a_n})\circ \cdots \circ d\alpha\of{T_{a_1}}\of{\omega},-A^{a_1},\cdots,
    (-A^{a_{k-1}}) \wedge (-A^{a_k}),
    -A^{a_{k+1}},
    \cdots,
    -A^{a_n})    
  \nonumber\\
  &\stackrel{\refer{defintion dA B}\refer{equation involving field strength}}{=}&
    -
    \sum_{n=0}^\infty
    \oint (d\alpha(T_{a_n})\circ \cdots \circ d\alpha\of{T_{a_1}}\of{\extd_A \omega},-A^{a_1},\cdots,-A^{a_n})
    \nonumber\\
    &&
    +
    \sum_{n=0}^\infty
    \sum_{1 \leq k \leq n}
    \oint (d\alpha(T_{a_n})\circ \cdots \circ d\alpha\of{T_{a_1}}\of{\omega},-A^{a_1},\cdots,
    -A^{a_{k-1}},iF_A^{a_k},
    -A^{a_{k+1}},
    \cdots,
    -A^{a_n})    
  \nonumber\\
  &=&
  - \oint \left((\extd_A \omega)^{W_A}\right)
  \nonumber\\
  &&
    +
    \sum_{n=0}^\infty
    \int\limits_0^{1}d\sigma_2\;
    \int\limits_{0}^{\sigma_2}d\sigma_1\;
    d\alpha(T_{a_n})\circ \cdots \circ d\alpha\of{T_{a_1}}\of{
      d\alpha\of{T_a}\of{W_A\of{\sigma_1,\sigma_2}\of{\iota_K \omega\of{\sigma_1}}}}
    \times
    \nonumber\\
    &&
    \hspace{5cm}
    \times
    \oint\limits_{\sigma_2}^1 
    (
    F_A^{a},
    -A^{a_{1}},
    \cdots,
    -A^{a_n})
  \nonumber\\
  &=&
  - \oint \left((\extd_A \omega)^{W_A}\right)
  \nonumber\\
  &&
    +
    \int\limits_0^{1}d\sigma_2\;
    \int\limits_{0}^{\sigma_2}
    d\sigma_1\;
    W_A\of{\sigma_2,1}\of{
      d\alpha\of{T_a}\of{W_A\of{\sigma_1,\sigma_2}\of{\iota_K \omega\of{\sigma_1}}}}
    \iota_K F_A^{a}\of{\sigma_2}
  \nonumber\\
  &=&
  - \oint \left((\extd_A \omega)^{W_A}\right)
  \nonumber\\
  &&
    +
    \int\limits_0^{1}d\sigma_2\;
    \int\limits_{0}^{\sigma_2}
    d\sigma_1\;
    W_A\of{\sigma_2,1}\of{
      d\alpha\of{T_a}\of{W_A^{-1}\of{\sigma_2,1}\of{W_A\of{\sigma_1,1}\of{\iota_K \omega\of{\sigma_1}}}}}
    \iota_K F_A^{a}\of{\sigma_2}
  \nonumber\\
  &\equalby{conjugation of g by holonomies}&
  - \oint \left((\extd_A \omega)^{W_A}\right)
   +
    \oint
    \of{
      d\alpha\of{T_a}\of{\omega^{W_A}},
      (F^{W_A})^a
    }
  \,.
  \nonumber
\end{eqnarray}
The case $\mathrm{deg}\of{\omega} = \mathrm{even}$ is completely analogous.

\endofproof

We have restricted attention here to just a single insertion, i.e.
$\oint_A\of{\omega}$ instead of $\oint_A \of{\omega_1,\dots,\omega_n}$,
because this is the form that the
\emph{standard connection 1-form} has:

\begin{definition}
  \label{standard path space 1-form}
  Given a $\crossedmodule$-valued $(1,2)$-form \refer{definition 1,2 form} 
  the path space 1-form
\begin{eqnarray}
   \Omega^1\of{\paths_s^t\of{\manifold},\h}
  \ni
  \;
  \mathcal{A}_{(A,B)}
  &\defas&
  \oint_A (B)
  \,.
  \nonumber
\end{eqnarray}
  is here called the
  {\bf standard local connection 1-form on path space}.
\end{definition}
(\cf \cite{AlvarezFerreiraSanchezGuillen:1998,Chepelev:2002,Schreiber:2004e})

Given a connection, one wants to know its curvature:

\begin{corollary}
  \label{corollary: curvature of standard path space 1-form}
  The curvature of the standard path space 1-form $\mathcal{A}_{(A,B)}$
  \refdef{standard path space 1-form}
  is
  \begin{eqnarray}
    \label{curvature of standard path space 1-form}
    \mathcal{F}_\mathcal{A}
   &=&
    -
    \oint_A \of{\extd_A B}
    -
    \oint_A\of{d\alpha\of{T_a}\of{B},(F_A+ dt\of{B})^a}
    \,.
  \end{eqnarray}
\end{corollary}

\Proof 
 Use \ref{standard path space 1-form} in \ref{extd on A-Chen forms of one form}
 to write
  \begin{eqnarray}
    \mathcal{F}_\mathcal{A}
    &\defas&
    (\extd_\mathcal{A})^2
    \nonumber\\
    &\defas&
    \extd \mathcal{A}
    +
    \mathcal{A} \wedge \mathcal{A}
    \nonumber\\
    &=&
    -
    \oint_A \of{\extd_A B}
    -
    \oint_A\of{d\alpha\of{T_a}\of{B},F_A^a}
    +
    \oint_A (B) \wedge \oint_A\of{B}
    \nonumber\\
   &\equalby{notation for action of h on h'}&
    -
    \oint_A \of{\extd_A B}
    -
    \oint_A\of{d\alpha\of{T_a}\of{B},(F_A+ dt\of{B})^a}
    \,.
  \end{eqnarray}
\endofproof

All this is `local' in the sense that it makes sense only on some contractible
open patch. Suitably generalizing this connection to a globally defined connection
is the content of \S\fullref{section 2-bundles with 2-connection}.
In the remainder of the present section the local character of the constructions
is mostly left implicit. 

The form of the curvature of the standard path space connection 1-form
suggests to identify the following two objects:

\begin{definition}
  \label{curvature and fake curvature}
  Given a standard path space connection 1-form $\mathcal{A}_{(A,B)}$
  \refdef{standard path space 1-form} coming from a $\g$-valued
  1-form $A$ and an $\h$-valued 2-form $B$
\begin{itemize}
  \item
     the 3-form 
     \begin{eqnarray}
        \label{definition curvature 3-form}
        H \defas \extd_A B
     \end{eqnarray}
     is called the {\bf curvature 3-form},
   \item
     the 2-form
     \begin{eqnarray}
       \label{definition fake curvature}
       \tilde F \defas F_A + dt\of{B}
     \end{eqnarray}
     is called the {\bf fake curvature 2-form}.
\end{itemize}
\end{definition}
The term `fake curvature' has been introduced in \cite{BreenMessing:2001}.
The notation $\tilde F$ follows \cite{GirelliPfeiffer:2004}. The 
curvature 3-form was used in \cite{Baez:2002}.

Using this notation the local path space curvature reads
\begin{eqnarray}
  \label{path space curvature in fake flat case}
  \mathcal{F}_\mathcal{A}
  &=&
  - \oint_A \of{H} - \oint_A \of{ d\alpha(T_a)\of{B},\tilde F^a}
  \,.
\end{eqnarray}

\subsubsection{Path Space Line Holonomy and Gauge Transformations}

With the usual tools of differential geometry available for
path space (as discussed in \S\fullref{path space differential calculus})
the holonomy on path space is defined as usual:

\begin{definition}
  \label{def path space holonomy}
  Given a path space 1-form $\mathcal{A}$ and a curve $\Sigma$ in path space the 
  {\bf path space line holonomy} of $\mathcal{A}$ along $\Sigma$ is
  \begin{eqnarray}
    \mathcal{W}_{\mathcal{A}}\of{\Sigma}
    &\defas&
    \mathrm{P} \exp\of{
      \int_\Sigma \mathcal{A}
    }
    \,.
    \nonumber
  \end{eqnarray}
\end{definition}

Note that by definition P here indicates path ordering with objects at 
higher parameter value to the 
\emph{right} of those with lower parameter value, just as in the definition of
ordinary line holonomy in def. \refer{line holonomy and parallel transport}.

Path space line holonomy has a richer set of gauge transformations
than holonomy on base space. In fact, ordinary gauge transformations
on base space correspond to \emph{constant} (`global') gauge transformations
on path space in the following sense:

\begin{proposition}
  \label{target space gauge trafos on path space holonomy}
  Given a path space line holonomy
  \refdef{def path space holonomy}
  coming from a standard path space connection 1-form \refdef{standard path space 1-form}
  $\mathcal{A}_{(A,B)}$
  in a based loop space $\paths_x^x\of{\manifold}$ 
  as well as a 
  $G$-valued 0-form $\phi \in \Omega^0\of{\manifold,G}$
  we have
  \begin{eqnarray}
    \alpha\of{\phi\of{x}}\of{
      \mathcal{W}_{\mathcal{A}_{(A,B)}}\of{\Sigma}
    }
    &=&
      \mathcal{W}_{\mathcal{A}_{(A^\prime,B^\prime)}}\of{\Sigma}    
    \nonumber
  \end{eqnarray}
  with
  \begin{eqnarray}
    A^\prime &=& \phi A \phi^{-1} + \phi (d\phi^{-1}) 
    \nonumber\\
    B^\prime &=& \alpha\of{\phi}\of{B}
    \,.
    \nonumber
  \end{eqnarray}
\end{proposition}

\Proof
Write out the path space holonomy in infinitesimal steps and apply
\refer{action of phi on parallel transport} on each of them.
\endofproof

The usual notion of gauge transformation is obtained by conjugation:

\begin{definition}
  \label{infinitesimal path space holonomy gauge transformation}
  Given the path space holonomy $\mathcal{W}_{\mathcal{A}_{(A,B)}}\of{\Sigma|_{X_0}^{X_1}}$ 
  \refdef{def path space holonomy}
  of a standard local path space connection 1-form $\mathcal{A}_{(A,B)}$
  \refdef{standard path space 1-form} along a curve $\Sigma$ in
  $\paths_s^t\of{\manifold}$ with endpaths
  $X_0$ and $X_1$,
  an {\bf infinitesimal path space holonomy gauge transformation}
 is the map
  \begin{eqnarray}
    \mathcal{W}_{\mathcal{A}_{(A,B)}}\of{\Sigma|_{X_0}^{X_1}}
    &\mapsto&
    \left(
      1 - \epsilon \oint_A \of{a}
    \right)_{X_0}
    \mathcal{W}_{\mathcal{A}_{(A,B)}}\of{\Sigma|_{X_0}^{X_1}}
    \left(
      1 + \epsilon \oint_A \of{a}
    \right)_{X_1}
    \,,
    \nonumber
  \end{eqnarray}  
  for any 1-form 
  \begin{eqnarray}
    a &\in& \Omega^1\of{\manifold,\h}
    \,.
    \nonumber
  \end{eqnarray}
\end{definition}

This 
yields a new sort of gauge transformation in terms of the 
target space (1,2) form $(A,B)$:

\begin{proposition}
  \label{effect of infinitesimal gauge transformations on path space}
  Infinitesimal path space holonomy gauge transformations
  \refdef{infinitesimal path space holonomy gauge transformation}
  for the holonomy of a standard path space connection 1-form
  $\mathcal{A}_{(A,B)}$ and arbitrary transformation parameter $a$
  yields to first order in the parameter $\epsilon$ the path space holonomy
  of a transformed standard path space connection 1-form
  $\mathcal{A}_{(A^\prime,B^\prime)}$ with
  \begin{eqnarray}
    \label{liner transformation of A and B under path space gauge trafos}
    A^\prime &=& A + dt\of{a}
    \nonumber\\
    B^\prime &=& B - \extd_A a
  \end{eqnarray}
  if and only if $\mathcal{A}_{(A,B)}$ is strictly r-flat 
  \refdef{strict r-flatness}. Otherwise the result of the infinitesimal
   gauge transformation is not (to any non-vanishing order in $\epsilon$) 
  the holonomy of any standard path space
  connection 1-form at all for arbitrary $a$.
\end{proposition}
(This was originally considered in \cite{Schreiber:2004e} for the special case $G=H$, 
$t = \mathrm{id},\, \alpha = \mathrm{Ad}$.)

\Proof

As for any holonomy the gauge transformation induces a transformation of the
connection 1-form $\mathcal{A} \to \mathcal{A}^\prime$ given by
\begin{eqnarray}
  \mathcal{A}^\prime
  &=&
  \left(
    1- \epsilon\oint_A \of{a}
  \right)
  \left(\extd + \mathcal{A}\right)
  \left(
    1 + \epsilon\oint_A \of{a}
  \right)
  \nonumber\\
  &=&
  \mathcal{A} 
  + 
  \epsilon\; \extd_\mathcal{A}\oint_A\of{a}
  +
  \order{\epsilon^2}
  \,.
\end{eqnarray}
One finds
(using the notation \refer{special notation in Chen form with parallel transport})
\begin{eqnarray}
  \mathcal{A} + 
  \epsilon\;\extd_{\mathcal{A}}\oint_A (a)
  &\equalby{path space extd on Chen forms with parallel transport}&
  \mathcal{A} 
  - \epsilon \oint_A \left(\extd_A a\right)
    +
    \epsilon \oint_A
    \of{
      d\alpha\of{T_a}\of{a},
      F^a
    }
  \nonumber\\
  &&
  +
  \epsilon \left(
  \oint_A 
  ( B^{a_1}, a^{a_2})
  -
  \oint_A 
  ( a^{a_1}, B^{a_2})
  \right)
  \commutator{S_{a_1}}{S_{a_2}}
  \nonumber\\
  &\equalby{notation for action of h on h'}&
  \mathcal{A} 
  - \epsilon\oint_A \left(\extd_A a\right)
    +
    \epsilon\oint_A
    \of{
      d\alpha\of{T_a}\of{a},
      F^a
    }
  \nonumber\\
  &&
  \mathcal{A} 
  -
  \epsilon
   \oint_A 
  ( d\alpha\of{T_a}\of{B}, dt\of{a}^a)
  +
  \epsilon
  \oint_A 
  ( d\alpha\of{T_a}\of{a}, dt\of{B}^{a})
  \nonumber\\
  &=&
  \mathcal{A} 
  - \epsilon\oint_A \left(\extd_A a\right)
  -
  \epsilon
  \oint_A 
  ( d\alpha\of{T_a}\of{B}, dt\of{a}^a)
    +
  \epsilon
    \oint_A
    \of{
      d\alpha\of{T_a}\of{a},
      (dt\of{B} + F)^a
    }
  \nonumber\\
  &\equalby{liner transformation of A and B under path space gauge trafos}&
  \oint_{A^\prime}\of{B^\prime}
    +
  \epsilon
    \oint_A
    \of{
      d\alpha\of{T_a}\of{a},
      (dt\of{B} + F)^a
    }
  + \order{\epsilon^2}
  \,.
  \nonumber
\end{eqnarray}

Since $a$ is by assumption arbitrary, the last line is equal to
a standard connection 1-form to order $\epsilon$  if and only if $dt\of{B} + F = 0$.
\endofproof

The above infinitesimal gauge transformation is easily integrated to a
finite gauge transformation:

\begin{definition}
  \label{finite path space gauge transformations}
  A {\bf finite path space holonomy gauge transformation}
  is the integration of infinitesimal  path space holonomy gauge transformations
  \refdef{infinitesimal path space holonomy gauge transformation}, i.e. it is 
  a map for any $a \in  \Omega^1\of{\manifold,\h}$
  given by
  \begin{eqnarray}
    &&\mathcal{W}_{\mathcal{A}_{(A,B)}}\of{\Sigma|_{X_0}^{X_1}}
    \mapsto
    \nonumber\\
    &&
    U^{-1}\of{A,a}
    \mathcal{W}_{\mathcal{A}_{(A,B)}}\of{\Sigma|_{X_0}^{X_1}}
    U\of{A,a}
    \nonumber\\
    &\defas&
    \lim\limits_{\epsilon = 1/N \to 0}
    \underbrace{
    \left(
      1 - \epsilon \oint_{A + dt\of{a}} \of{a}
    \right)
    \cdots
    \left(
      1 - \epsilon \oint_{A+\epsilon dt\of{a}} \of{a}
    \right)
    \left(
      1 - \epsilon \oint_{A} \of{a}
    \right)_{X_0}}_{\mbox{$N$ factors}} \times
    \nonumber\\
    &&
    \skiph{$\lim\limits_{\epsilon = 1/N \to 0}$}\;
    \times
    \mathcal{W}_{\mathcal{A}_{(A,B)}}\of{\Sigma|_{X_0}^{X_1}}
    \times
    \nonumber\\
    &&
    \skiph{$\lim\limits_{\epsilon = 1/N \to 0}$}\;
    \times
    \underbrace{
    \left(
      1 + \epsilon \oint_{A} \of{a}
    \right)
    \left(
      1 + \epsilon \oint_{A + \epsilon dt\of{a}} \of{a}
    \right)
    \cdots
    \left(
      1 + \epsilon \oint_{A + dt\of{a}} \of{a}
    \right)_{X_1}}_{\mbox{$N$ factors}}
    \,.
   \nonumber
  \end{eqnarray}
\end{definition}

\begin{proposition}
  \label{effect of finite path space gauge transformations}
  A finite path space holonomy gauge transformation \refdef{finite path space gauge transformations}
  of the holonomy of a standard path space connection 1-form $\mathcal{A}_{(A,B)}$
  is equivalent to a transformation
  \begin{eqnarray}
    \mathcal{A}_{(A,B)} &\mapsto& \mathcal{A}_{(A^\prime,B^\prime)}
    \nonumber
  \end{eqnarray}
  where
  \begin{eqnarray}
    \label{finite path space gauge transformation on 1,2 form}
    A &\mapsto& A + dt\of{a}
    \nonumber\\
    B &\mapsto& B - \underbrace{(d_A a + a\wedge a)}_{\defas k_a} 
  \end{eqnarray}
  is the transformed $(1,2)$-form $(A,B)$.
\end{proposition}

\Proof

We have to integrate up \refer{liner transformation of A and B under path space gauge trafos}.
At each step we have
\begin{eqnarray}
  A_{(n)} &\defas& A_{(n-1)} + \epsilon \, dt\of{a}
  \nonumber\\
  B_{(n)} &\defas& B_{(n-1)} - \epsilon \, da - \epsilon\, A_{(n-1)}^a \wedge d\alpha\of{T_a}\of{a}  
  \nonumber
\end{eqnarray}
it follows that
\begin{eqnarray}
  A^\prime = A_{(N)}
  &=&
  A + \epsilon N \, dt\of{a}
  \nonumber\\
  B^\prime = B_{(N)}
  &=&
  B 
  - \epsilon N\, d a 
  - \epsilon (N-1) A^a \wedge d\alpha\of{T_a}\of{a}
  - 
  \epsilon^2
  \frac{N(N-1)}{2}
  dt\of{a}^a \wedge d\alpha\of{T_a}\of{a}
  \,,
  \nonumber
\end{eqnarray}
which in the limit $N = 1/\epsilon \to \infty$ goes to \refer{finite path space gauge transformation on 1,2 form}
(using $\frac{1}{2}dt\of{a}^a \wedge d\alpha\of{T_a}\of{a} \equalby{notation for action of h on h'} a\wedge a$).
\endofproof

  In summary the above yields two different notions of gauge transformations on path space:
  \begin{enumerate}
    \item If the path space in question is a based loop space then 
       according to prop \ref{target space gauge trafos on path space holonomy}
       a gauge transformation on
       target space  
       yields an ordinary gauge transformation of the $(1,2)$-form $(A,B)$:
       \begin{eqnarray}
          \label{first kind trafo}
          A &\mapsto& \phi A \phi^{-1} + \phi (d\phi^{-1})
          \nonumber\\
          B &\mapsto& \alpha\of{\phi}\of{B}
          \,.
          \nonumber 
       \end{eqnarray}
      We shall call this a {\bf 2-gauge transformation of the first kind}.
    \item
     A gauge transformation in path space itself yields,
according to prop. \ref{effect of finite path space gauge transformations},
 a transformation
     \begin{eqnarray}
        \label{second kind trafo}
        A &\mapsto& A + a
        \nonumber\\
        B &\mapsto& B - (d_A a + a\wedge a)
        \,. 
        \nonumber
     \end{eqnarray}
     We shall call this a {\bf 2-gauge transformation of the second kind.}
  \end{enumerate}
  Recall that according to prop \ref{effect of infinitesimal gauge transformations on path space}
  this works precisely when $(A,B)$ defines a standard connection 1-form 
   \refdef{standard path space 1-form} 
   on path space for which the `fake curvature' \refdef{curvature and fake curvature}
  vanishes
  $\tilde F = dt\of{B} + F_A = 0$.

In the context of loop space these two transformations and the conditions on them
were discussed for the special case $G=H$ and $t = \mathrm{id},\, \alpha = \mathrm{Ad}$ 
in \cite{Schreiber:2004e}. In the context of 2-groups and higher lattice gauge theory 
they were
found in section 3.4 of \cite{GirelliPfeiffer:2004}. They also appear in the
transition laws for nonabelian gerbes 
\cite{BreenMessing:2001,AschieriCantiniJurco:2003,AschieriJurco:2004},
as is discussed in detail in \S\fullref{gerbes}.
The same transformation for the  special case where all groups are abelian 
is well known from abelian gerbe theory \cite{Chatterjee:1998} but also
for instance from string theory (e.g. section 8.7 of \cite{Polchinski:1998}).

\vskip 2em

With holonomy on path space understood, it is now possible to use the fact
that every curve in path space maps to a (possibly degenerate) surface in
target space in order to get a notion of (local) surface holonomy. That is
the content of the next subsection.

\subsubsection{Local 2-Holonomy from local Path Space Holonomy}
\label{section: Local 2-Holonomy from local Path Space Holonomy}

Just like ordinary holonomy is a functor from the groupoid of paths
\refdef{groupoid of paths} to an ordinary group, 2-holonomy
is a 2-functor from some 2-groupoid to a 2-group. This 2-groupoid
is roughly that consisting of bounded surfaces in $M$ whose horizontal
and vertical composition corresponds to the ordinary gluing of bounded
surfaces. This heuristic idea is made precise in the following 
by constructing $\paths_2\of{\manifold}$, the {\bf 2-groupoid of bigons}.

\vskip 1em

First of all a bigon is a `surface with two corners'. More precisely:

\begin{definition}
  \label{parametrized bigon}
  Given any manifold $\manifold$ a {\bf parametrized bigon} in $\manifold$
  is a smooth map
  \begin{eqnarray}
    \Sigma \maps [0,1]^2 &\to& \manifold
    \nonumber\\
    (\sigma,\tau) &\mapsto& \Sigma\of{\sigma,\tau}
  \end{eqnarray}
  with
  \begin{eqnarray}
    \Sigma\of{0,\tau} &=& s \in \manifold
    \nonumber\\
    \Sigma\of{1,\tau} &=& t \in \manifold 
    \nonumber  
  \end{eqnarray}
  for given $s,t\in \manifold$,
  which is constant in a neighborhood of 
  $\sigma = 0,1$
  and independent of $\tau$ near $\tau = 0,1$.

  Equivalently, a parametrized bigon is path in path space
  $\paths_s^t\of{\manifold}$ \refdef{based path space}
  \begin{eqnarray}
    \Sigma \maps [0,1] &\to& \paths_s^t\of{\manifold}
   \nonumber\\
      \tau &\mapsto& \Sigma\of{\cdot,\tau}   
    \,,
      \nonumber
  \end{eqnarray} 
  which is constant in a neighborhood of $\tau = 0,1$.
  We call $s$ the {\bf source vertex} of the bigon, $t$ the 
  {\bf target vertex}, $\Sigma\of{\cdot,0}$ the {\bf source edge}
  and $\Sigma\of{\cdot,1}$ the {\bf target edge}.
\end{definition}

As with paths, the parametrization involved here is ultimately
not of interest and should be devided out:

\begin{definition}
  \label{bigon}
  An {\bf unparametrized bigon} or simply a {\bf bigon}
  is a thin homotopy equivalence class $[\Sigma]$ of 
  parametrized bigons
  $\Sigma$ \refdef{parametrized bigon}.
\end{definition}

More in detail, this means (\cf for instance 
\cite{MackaayPicken:2000} p.26 and \cite{BaezLauda:2003} p.50) 
that two parametrized bigons 
$\Sigma_1, \Sigma_2 \maps [0,1]^2 \to \manifold$ 
are taken to be equivalent
\[
  \Sigma_1 \sim \Sigma_2
\]
precisely if there exists a smooth map
\[
  H \maps [0,1]^3 \to \manifold
\]
which takes one bigon smoothly into the other while preserving their boundary,
i.e. such that
\begin{eqnarray}
  H\of{\sigma,\tau,0} &=& \Sigma_1\of{\sigma,\tau}
  \nonumber\\
  H\of{\sigma,\tau,1} &=& \Sigma_2\of{\sigma,\tau}
  \nonumber\\
  H\of{\sigma,0,\nu} &=& \Sigma_1\of{\sigma,0} = \Sigma_2\of{\sigma,0}
  \nonumber\\
  H\of{\sigma,1,\nu} &=& \Sigma_1\of{\sigma,1} = \Sigma_2\of{\sigma,1}
  \nonumber\\
  H\of{0,\tau,\nu} &=& \Sigma_1\of{0,\tau} = \Sigma_2\of{0,\tau}
  \nonumber\\
  H\of{1,\tau,\nu} &=& \Sigma_1\of{1,\tau} = \Sigma_2\of{1,\tau}
  \,,
  \nonumber
\end{eqnarray}
but which does so in a degenerate fashion, meaning that
\[
  \mathrm{rank}\of{dH}\of{\sigma,\nu,\tau} < 3
\]
for all $\sigma,\tau,\nu \in [0,1]$.

These bigons naturally form a coherent 2-groupoid:

\begin{definition}
  \label{2-groupoid of bigons}
  The {\bf coherent 2-groupoid of bigons} 
  $\paths_2\of{\manifold}$
  in $\manifold$ is the groupoid whose
  \begin{itemize}
    \item 
      objects are points $x\in \manifold$
    \item 
      morphisms are 
      paths $\gamma \in \paths_x^y\of{\manifold}$
     \[
      \xymatrix{
       x \ar@/^1pc/[rr]^{\gamma} 
       && y 
      }
      \]
    \item
      2-morphisms       are bigons 
      \refdef{bigon} with source edge $\gamma_1$ and target
      edge $\gamma_2$
 \[
\xymatrix{
   x \ar@/^1pc/[rr]^{\gamma_1}_{}="0"
           \ar@/_1pc/[rr]_{\gamma_2}_{}="1"
           \ar@{=>}"0";"1"^{{}_{[\Sigma_1]}}
&& y
}
\]
  \end{itemize}
  and whose composition operations are defined as
  \begin{itemize}
   \item 
\[
\xymatrix{
   x \ar@/^1pc/[rr]^{\gamma_1}
&& y \ar@/^1pc/[rr]^{\gamma_2}
&& z
}
  =
\xymatrix{
   x \ar@/^1pc/[rr]^{\gamma_1 \circ \gamma_2}
&& z 
}
\]

\item
\[
\xymatrix{
   x \ar@/^2pc/[rr]^{\gamma_1}_{}="0"
           \ar[rr]^<<<<<<{\gamma_2}_{}="1"
           \ar@{=>}"0";"1"^{[\Sigma_1]}
           \ar@/_2pc/[rr]_{\gamma_3}_{}="2"
           \ar@{=>}"1";"2"^{[\Sigma_2]}
&& y
}
=
\xymatrix{
   x \ar@/^1pc/[rr]^{\gamma_1}_{}="0"
           \ar@/_1pc/[rr]_{\gamma_3}_{}="1"
           \ar@{=>}"0";"1"^{{}_{[\Sigma_1\circ \Sigma_2]}}
&& y
}
\]

\item

\[
  \begin{array}{cccccc}
    x & \stackto{\gamma_1} & y & \stackto{\gamma_2}  & z\\
      &   \skiph{\mbox{\small $[\Sigma_1]$}}  \Big\Downarrow \mbox{\small $[\Sigma_1]$} &   
         & \skiph{\mbox{\small $[\Sigma_2]$}}  \Big\Downarrow \mbox{\small $[\Sigma_2]$} & \\
    x & \stackto{\gamma_1^\prime} & y & \stackto{\gamma_2^\prime}  & z\\     
  \end{array}
  \hspace{2em}
  =
  \hspace{2em}
  \begin{array}{cccc}
    x & \stackto{\gamma_1\circ \gamma_2} & z \\
      &   \skiph{\mbox{\small $[\Sigma_1\cdot \Sigma_2]$}}  \Big\Downarrow \mbox{\small $[\Sigma_1\cdot \Sigma_2]$} & \\  
    x & \stackto{\gamma_1^\prime \circ \gamma_2^\prime} & z \\     
  \end{array}
\]

  \end{itemize}
  where
  \begin{eqnarray}
    \left(\gamma_1\circ\gamma_2\right)\of{\sigma}
    &\defas&
    \left\lbrace
      \begin{array}{cl}
         \gamma_1\of{2\sigma} & \mbox{for $0 \leq \sigma \leq 1/2$}\\
         \gamma_2\of{2\sigma-1} & \mbox{for $1/2\leq \sigma \leq 1$}
      \end{array}
    \right.
  \nonumber\\
    \left(\Sigma_1\circ\Sigma_2\right)\of{\sigma,\tau}
    &\defas&
    \left\lbrace
      \begin{array}{cl}
         \Sigma_1\of{\sigma,2\tau} & \mbox{for $0 \leq \tau \leq 1/2$}\\
         \Sigma_2\of{\sigma,2\tau-1} & \mbox{for $1/2\leq \tau \leq 1$}
      \end{array}
    \right.
  \nonumber\\
    \left(\Sigma_1\cdot\Sigma_2\right)\of{\sigma,\tau}
    &\defas&
    \left\lbrace
      \begin{array}{cl}
         \Sigma_1\of{2\sigma,\tau} & \mbox{for $0 \leq \sigma \leq 1/2$}\\
         \Sigma_2\of{2\sigma-1,\tau} & \mbox{for $1/2\leq \sigma \leq 1$}
      \end{array}
    \right.  
    \,.
    \nonumber
  \end{eqnarray}
\end{definition}

Note that in this definition we did \emph{not} divide out by thin
homotopy of parametrized \emph{paths} 
but only by thin homotopy of parametrized bigons. This implies that
the horizontal composition in this 2-groupoid is \emph{not}
associative. But one can check that 
the above indeed is a coherent 2-groupoid where associativity is \emph{weakly} 
preserved in a \emph{coherent} fashion, as described in 
\cite{BaezLauda:2003}.

Namely there are degenerate bigons for which $\mathrm{rank}\of{d\Sigma} \leq 1$,
whose vertical composition with any other bigon 
hence has the effect of applying a thin homotopy to that bigon's source or
target edges. Therefore associativity of horizontal composition of bigons holds
up to vertical composition with such degenerate bigons and hence up to natural
isomorphism.

\vskip 2 em 

With the 2-groupoid of bigons constructed, 2-holonomy can finally 
be defined:

\begin{definition}
  \label{local strict 2-holonomy}
  Given a manifold $\manifold$ and a strict 2-group $\twogroup$
  a {\bf local strict 2-holonomy} 
  (or simply {\bf local 2-holonomy})
  is a strict 2-functor 
  \[
    \hol \maps \paths_2\of{\manifold} \to \twogroup
  \]
  from the coherent 2-groupoid of
  bigons $\paths_2\of{\manifold}$ \refdef{2-groupoid of bigons}
  to the strict 2-group $\twogroup$.
\end{definition}

(The fact that this functor is strict means that it ignores the 
parametrization of the bigon's source and target edges. Ultimately
one will want to replace the strict 2-group here with a coherent
2-group and the strict 2-functor with some sort of weak 2-functor.
See the discussion section \S\fullref{Discussion and Open Questions}.)

We want to construct a local 2-holonomy from a standard path space
connection 1-form \refdef{standard path space 1-form}. In order
to do so we first construct a `pre-2-holonomy' for any
standard path space connection 1-form and then determine under
which conditions this actually gives a true 2-holonomy. It turns
out that the necessary and sufficient conditions for this is the
vanishing of the fake curvature \refdef{curvature and fake curvature}.

\begin{definition}
  \label{local pre-2-holonomy}
  Given a standard path space connection 1-form 
  \refdef{standard path space 1-form}
  and given any parametrized bigon \refdef{parametrized bigon}
  \[
    \Sigma : [0,1]^2 \to \manifold
  \]
  with 
  source edge
  \[
    \gamma_1 \defas \Sigma\of{\cdot,0}
  \]
  and
  target edge
  \[
    \gamma_2 \defas \Sigma\of{\cdot,1}\,,
  \]
  the triple
  \[
    (g_1, h, g_2 ) \in G\times H \times G
  \]
  with
  \begin{eqnarray}
    \label{assignments of pre-2-holonomy}
    g_i &\defas& W_A\of{\gamma_i}
    \nonumber\\
    h &\defas& \mathcal{W}^{-1}_{\mathcal{A}}\of{\Sigma\of{1-\cdot,\cdot}}
  \end{eqnarray}
  is called the {\bf local pre-2-holonomy} of $\Sigma$ associated with 
  $\mathcal{A}$.
\end{definition}

(The maybe unexpected inverse and parameter inversion here is just due to 
the interplay of our conventions on signs and orientations, as will become
clear shortly.)

In order for a pre-2-holonomy to give rise to a true 2-holonomy
two conditions have to be satisfied:
\begin{enumerate}
   \item 
     The triple $(g_1,h,g_2)$ has to specify a strict 2-group element.
      By prop. \ref{strict 2-group coming from crossed module} 
      this is the case precisely if
      $g_2 = t\of{h}g_1$ \refer{source and target of strict 2-group element}.
   \item
     The pre-2-holonomy has to be invariant under thin homotopy in
     order to be well defined on bigons.
\end{enumerate}

The solution of this is the content of prop. \ref{from pre-2-holonomy to 2-holonomy}
below. In order to get there the following considerations are necessary:

\paragraph{Compatibility with strict 2-groups.}

In order to analyze the first of the above two points consider
the behaviour of the pre-2-holonomy under changes of the
target edge.

  Given a path space $\paths_s^t\of{\manifold}$
  and a $\g$-valued 1-form with line 
  holonomy holonomy $W_A[X]$ on $X \in \paths_s^t$ 
  \refdef{line holonomy and parallel transport} 
  the {\bf change in holonomy} of $W_A$ as one changes $X$ is well known
  to be given by the following:

\begin{proposition}
  \label{variation of holonomy under variation of the path}
  Let $\rho : \tau \mapsto X\of{\tau}$ be the flow generated by the vector field $D$ on
  $\paths_s^t$, then
  \begin{eqnarray}
   \label{variation of holonomy}
    \left.
    \frac{d}{d\tau}
    W_A^{-1}[X\of{0}]W_A[X\of{\tau}]
    \right|_{\tau = 0}
    &=&
    -
    \left(
      \oint\limits_A (F_A)
    \right)\of{D}
    \,.
  \end{eqnarray}  
\end{proposition}
(Note that the right hand side denotes evaluation of the path space 1-form
$\oint_A(F_A)$ on the path space vector field $D$.)

\Proof The proof is standard. The only subtlety is to take care of the
various conventions for signs and orientations which give rise to the minus 
sign in \refer{variation of holonomy}. \endofproof

\begin{proposition}
  \label{pre-2-holonomy asscoiates 2-group elements}
  For the pre-2-holonomy \refdef{local pre-2-holonomy} 
  of parametrized bigons $\Sigma$ 
  associated with the standard connection 1-form
  $\mathcal{A}_{(A,B)}$
  to specify 2-group elements,
  i.e. for the triples $(g_1,h,g_2)$ to satisfy $g_2 = t\of{h}g_1$,
  we must have
  \[
    dt\of{B} + F_A = 0
    \,.
  \]
\end{proposition}

\Proof
According to def. \ref{local pre-2-holonomy}
the condition $g_2 = t\of{h}g_1$ translates into
\begin{eqnarray}
  \label{an equation}
  t\of{
    h
  }
  &=&
  W_A\of{\gamma_2}
  W^{-1}_A\of{\gamma_1}
  \nonumber\\
  &=&
  W_A^{-1}\of{\gamma_2^{-1}}
  W_A\of{\gamma_1^{-1}}
  \,.
  \nonumber
\end{eqnarray}
Now let 
there be a flow $\tau \mapsto \gamma_\tau$ on $\paths_s^t\of{\manifold}$
generated by a vector field $D$ and choose $\gamma_2^{-1} = \gamma_\tau$
and $\gamma_1^{-1} = \gamma_0$. Then according to
prop. \ref{variation of holonomy under variation of the path}
we have
\begin{eqnarray}
  \frac{d}{d\tau}
  W_A^{-1}\of{\gamma_2^{-1}}
  W_A\of{\gamma_1^{-1}}
  &=&
  + \left(\oint\limits_A\of{F_A}\right)_{\gamma_0}\of{D}
  \,,
  \nonumber
\end{eqnarray}
where the plus sign is due to the fact that $D$ here points opposite to the $D$
in prop. \ref{variation of holonomy under variation of the path}.

Applying the same $\tau$-derivative on the left hand side 
of \refer{an equation} yields
\begin{eqnarray}
  - \left(\oint_A\of{dt\of{B}}\right)\of{D} &=& 
    \left(\oint_A \of{F_A}\right)\of{D}
  \,.
  \nonumber
\end{eqnarray}
(Here the minus sign on the left hand side comes from the fact that 
we have identified $t\of{h}$ with the \emph{inverse} path space holonomy
$\mathcal{W}^{-1}_{\mathcal{A}}$. This is necessary because the ordinary
path space holonomy is path-ordered to the right, while we need $t\of{h}$
to be path ordered to the left.)

This can be true for all $D$ only if $-dt\of{B} = F_A$.
\endofproof

This is nothing but the {\bf nonabelian Stokes theorem}.
(Compare for instance  \cite{KarpMansouriRno:1999} and references given there.)

\vskip 2em

Next it needs to be shown that a pre-2-holonomy with 
$dt\of{B} + F_A = 0$ is invariant under thin-homotopy:

\paragraph{Invariance under thin homotopy.}

\begin{definition}
  \label{r-flat}
  A connection 1-form $\mathcal{A}$ on $\paths_s^t\of{\manifold}$ for all $s,t\in \manifold$ 
  is called
  {\bf r-flat} if its holonomy 
  is invariant under thin homotopy.
\end{definition}
(The notion of r-flatness was introduced by \cite{Alvarez:2004}, based on
\cite{AlvarezFerreiraSanchezGuillen:1998,Schreiber:2004e}.)

\begin{proposition}
  \label{local condition for r-flatness}
  The standard path space connection 1-form 
  $\mathcal{A}_{(A,B)}$ \refdef{standard path space 1-form}
  is r-flat \refdef{r-flat} precisely if
  the path space 2-form
  $
    \oint_A\of{ d\alpha\of{T_a}\of{B}, (F_A + dt\of{B})^a }
  $
  vanishes on all path space vector fields $T$ and $\tilde T$ with components
  $T^{(\mu,\sigma)} = \frac{\partial}{\partial \tau} X_\tau^{(\mu,\sigma)}$
  and
  $\tilde T^{(\mu,\sigma)} = a\of{\tau,\sigma}
                             \frac{\partial}{\partial \tau} X_\tau^{(\mu,\sigma)}$
  for $\tau \mapsto X_\tau$ any smooth curve in any path space $\paths_s^t\of{\manifold}$
  and $(\tau,\sigma)\mapsto a\of{\tau,\sigma}$ any smooth function, i.e.
  iff
  \begin{eqnarray}
    \label{the r-flatness condition}
    \oint\limits_A\of{ d\alpha\of{T_a}\of{B}, (F_A + dt\of{B})^a }\of{T,\tilde T} = 0
  \end{eqnarray}
  for all $T$ and $\tilde T$ of the above form.
 \end{proposition} 

\Proof
For the special case $G=H$ and $t = \mathrm{id},\, \alpha = \mathrm{Ad}$ this was proven 
by \cite{Alvarez:2004}.
The full proof is a straightforward generalization of this special case:

Let 
\begin{eqnarray}
  (0,1) &\to& \paths_s^t\of{\manifold}
  \nonumber\\
  \tau &\mapsto& X_\tau
  \nonumber
\end{eqnarray}
be any smooth curve in $\paths_s^t\of{\manifold}$ with tangent vector field
\begin{eqnarray}
  T\of{\tau} &\defas& \frac{\partial}{\partial\tau} X_\tau^{(\mu,\sigma)}\frac{\delta}{\delta X^{(\mu,\sigma)}}
  \,.
  \nonumber
\end{eqnarray}
(Recall the notation defined in def. \ref{notions on path space}.)

Flows from this curve through a set of curves which map to thin homotopy equivalent
surfaces are generated by vector fields of the form
\begin{eqnarray}
  D\of{\tau} &=& \int_0^1 d\sigma\; 
   \left(a\of{\tau,\sigma} \frac{\partial}{\partial \tau} X_\tau^{\mu}\of{\sigma}
       + b\of{\tau,\sigma} \frac{\partial}{\partial \sigma} X^{\mu}_\tau\of{\sigma}
   \right)
   \frac{\delta}{\delta X^\mu\of{\sigma}}
  \,.
  \nonumber
\end{eqnarray}
The $\mathcal{A}$-holonomy on any $\paths_s^t\of{\manifold}$ under such a shift vanishes
for all choices of $a\of{\sigma,\tau}$ and $b\of{\sigma,\tau}$
iff the curvature of $\mathcal{A}$ vanishes when evaluated on these
two vector fields:
\[
  \mathcal{F}_\mathcal{A}\of{T,D} = 0
  \,.
\]
From corollary \ref{corollary: curvature of standard path space 1-form} we know
that $\mathcal{F} = - \oint \of{\extd_A B} -
 \oint (d\alpha\of{T_A}\of{B}, (F_A + dt\of{B})^a)$. 
One finds
\[
  \oint\of{\extd_A B} \of{T,D} = 0
  \,,
\]
since evaluating this involves evaluating the 3-form $\extd_A B$ on three vectors
with a degenerate span. For a similar reason
\[
  \oint (d\alpha\of{T_A}\of{B}, (F_A + dt\of{B})^a)\of{
   (\partial_\sigma X^{\mu})\of{\sigma}\frac{\delta}{\delta X^\mu\of{\sigma}},\cdot }
  = 0
 \,,
\]
since this involves evaluating target space 2-forms on two vectors proportional to
$\partial_\sigma X^\mu$.

All that remains is hence the condition
\[
  \oint (d\alpha\of{T_A}\of{B}, (F_A + dt\of{B})^a)\of{
    T, \tilde T
  }
  = 0
  \,.
\]
\endofproof

From prop \ref{pre-2-holonomy asscoiates 2-group elements} it is known that
the condition for the path space connection to be compatible with the nature
of 2-groups is a special case of r-flatness:
\begin{definition}
  \label{strict r-flatness}
  An r-flat standard path space connection $\mathcal{A}_{(A,B)}$
  \refdef{standard path space 1-form}
  which solves \refer{the r-flatness condition}
  by satisfying
  \begin{eqnarray}
    F_A + dt\of{B} &=& 0 
  \end{eqnarray}
  is called {\bf strictly r-flat}.
\end{definition}

Now we can finally prove the following:

\paragraph{From pre-2-holonomy to true 2-holonomy.}

\begin{proposition}
  \label{from pre-2-holonomy to 2-holonomy}
  The pre-2-holonomy \refdef{local pre-2-holonomy}
  induces a true local 2-holonomy \refdef{local strict 2-holonomy}
  precisely if it comes from a strictly r-flat 
  \refdef{strict r-flatness}
  standard path space
  connection 1-form.
\end{proposition}
\Proof

We have already shown that for $dt\of{B} + F_A = 0$ the
pre-2-holonomy indeed maps into a 2-group 
(prop. \ref{pre-2-holonomy asscoiates 2-group elements})
and that its values are well defined on bigons
(prop. \ref{local condition for r-flatness}). What remains
to be shown is functoriality, i.e. that the pre-2-holonomy
respects the composition of bigons and 2-group elements.

First of all it is immediate that composition of paths is
respected, due to the properties of ordinary holonomy.
Vertical composition of 2-holonomy
(being composition of ordinary holonomy in path space)
is completely analogous. The fact that pre-2-holonomy involves
the \emph{inverse} path space holonomy takes care of the 
nature of the vertical product in the 2-group, which reverses
the order of factors:
In the diagram
\[
\begin{array}{c|ccc}
\twogroup &
\xymatrix{
   \bullet \ar@/^2pc/[rr]^{W_A[\gamma_1]}_{}="0"
           \ar[rr]^<<<<<<{{}_{W_A[\gamma_2]}}_{}="1"
           \ar@{=>}"0";"1"^{{}_{\mathcal{W}^{-1}_\mathcal{A}[\Sigma_1]}}
           \ar@/_2pc/[rr]_{W_A[\gamma_3]}_{}="2"
           \ar@{=>}"1";"2"^{{}_{\mathcal{W}^{-1}_\mathcal{A}[\Sigma_2]}}
&& \bullet
}
&=&
\xymatrix{
   \bullet \ar@/^1pc/[rr]^{W_A[\gamma_1]}_{}="0"
           \ar@/_1pc/[rr]_{W_A[\gamma_3]}_{}="1"
           \ar@{=>}"0";"1"^{{}_{{}_{\mathcal{W}^{-1}_\mathcal{A}[\Sigma_1\circ \Sigma_2]}}}
&& 
}
\\
\;\;\;\;\Big\Uparrow \hol
&
\Big\Uparrow 
&
&
\Big\Uparrow
\\
\paths_2\of{\manifold}
&
\xymatrix{
   x \ar@/^2pc/[rr]^{\gamma_1}_{}="0"
           \ar[rr]^<<<<<<{\gamma_2}_{}="1"
           \ar@{=>}"0";"1"^{[\Sigma_1]}
           \ar@/_2pc/[rr]_{\gamma_3}_{}="2"
           \ar@{=>}"1";"2"^{[\Sigma_2]}
&& y
}
&=&
\xymatrix{
   x \ar@/^1pc/[rr]^{\gamma_1}_{}="0"
           \ar@/_1pc/[rr]_{\gamma_3}_{}="1"
           \ar@{=>}"0";"1"^{{}_{[\Sigma_1\circ \Sigma_2]}}
&& y
}
\end{array}
\]
the top right bigon must be labeled (according to the properties
of strict 2-groups described in prop. \ref{strict 2-group coming from crossed module})
by
\begin{eqnarray}
  (W_A[\gamma_1], \mathcal{W}^{-1}_{\mathcal{A}}[\Sigma_1])
  \circ
  (W_A[\gamma_2], \mathcal{W}^{-1}_{\mathcal{A}}[\Sigma_2])
  &=&
  (W_A[\gamma_1], \mathcal{W}^{-1}_{\mathcal{A}}[\Sigma_2]\mathcal{W}^{-1}_{\mathcal{A}}[\Sigma_1])  
  \nonumber\\
  &=&
  (W_A[\gamma_1], \mathcal{W}^{-1}_{\mathcal{A}}[\Sigma_1\circ\Sigma_2])
  \nonumber\,,
\end{eqnarray}
which indeed is the label associated by the $\hol$-functor
in the right column of the diagram.

So far we have suppressed in these formulas the 
 reversal \refer{assignments of pre-2-holonomy}
in the first coordinate of $\Sigma$, since it plays no role for the above.
This reversal however is essential in order for the $\hol$-functor
to respect horizontal composition.

In order to see this it is sufficient to consider \emph{whiskering}, i.e. horizontal
composition with identity 2-morphisms.

When whiskering from the left we have
\[
\begin{array}{c|ccc}
\twogroup
&
\xymatrix{
   \bullet \ar[rr]^{W_A[\gamma_1]}_{}
   &&\bullet \ar@/^1pc/[rr]^{W_A[\gamma_2]}_{}="0"
           \ar@/_1pc/[rr]_{W_A[\gamma_2^\prime]}_{}="1"
           \ar@{=>}"0";"1"^{{}_{\mathcal{W}_\mathcal{A}^{-1}[\Sigma]}}
&& \bullet
}
&=&
\hspace{28pt}
\xymatrix{
   \bullet \ar@/^1pc/[rr]^{W_A[\gamma_1\circ\gamma_2]}_{}="0"
           \ar@/_1pc/[rr]_{W_A[\gamma_1\circ\gamma_2^\prime]}_{}="1"
           \ar@{=>}"0";"1"^{{}_{\alpha\of{W_A[\gamma_1]}\of{\mathcal{W}_\mathcal{A}^{-1}[\Sigma]}}}
&& 
}
\\
\;\;\;\;\;\Big\Uparrow \hol
&
\Big\Uparrow
&&
\Big\Uparrow 
\\
\paths_2\of{\manifold}
&
\xymatrix{
   x \ar[rr]^{\gamma_1}_{}
   &&y \ar@/^1pc/[rr]^{\gamma_2}_{}="0"
           \ar@/_1pc/[rr]_{\gamma_2^\prime}_{}="1"
           \ar@{=>}"0";"1"^{[\Sigma]}
&& z
}
&=&
\xymatrix{
   x \ar@/^1pc/[rr]^{\gamma_1\circ\gamma_2}_{}="0"
           \ar@/_1pc/[rr]_{\gamma_1\circ \gamma_2^\prime}_{}="1"
           \ar@{=>}"0";"1"^{[\Sigma]}
&& z
}
\end{array}
\]
Evaluating the surface holonomy here involves exponentiating the integrals
\begin{eqnarray}
  \int\limits_{(\gamma_1\circ\gamma_2)^{-1}}
  d\sigma\;
    W_A[(\gamma_1\circ\gamma_2)^{-1}](\sigma,1)\of{B\of{\sigma}}
  &=&
  W_A[\gamma_1^{-1}]
  \of{
    \int\limits_{\gamma_2^{-1}}
    d\sigma\;
      W_A[\gamma_2^{-1}](\sigma,1)\of{B\of{\sigma}}
  }
  \nonumber\\
  &\equalby{notions of parallel tranport}&
  \alpha\of{W_A[\gamma_1]}
  \of{
    \int\limits_{\gamma_2^{-1}}
    d\sigma\;
      W_A[\gamma_2^{-1}](\sigma,1)\of{B\of{\sigma}}
  }
  \,,
  \nonumber
\end{eqnarray}
where the contraction with a path space vector tangent to the curve in path space
is left implicit.
The result in the last line indeed makes the diagram commute. In this
computation the path reversal is essential, which of course is related
to our convention that parallel transport be to the point with parameter 
$\sigma = 1$. A simple plausibility argument for this was given at
the beginning of \S\fullref{section: The Standard Connection 1-Form on Path Space}.

Finally, whiskering to the right is trivial, since we can simply use reparametrization
invariance to obtain
\begin{eqnarray}
  \int\limits_{(\gamma_1\circ\gamma_2)^{-1}}
  d\sigma\;
    W_A[(\gamma_1\circ\gamma_2)^{-1}](\sigma,1)\of{B\of{\sigma}}
  &=&
  \int\limits_{\gamma_1^{-1}}
  d\sigma\;
    W_A[\gamma_1^{-1}](\sigma,1)\of{B\of{\sigma}}
  \,,
  \nonumber
\end{eqnarray}
because for right whiskers the integrand vanishes on $\gamma_2$.

Since general horizontal composition is obtained by first whiskering and then
composing vertically, this also proves that the $\hol$-functor respects 
general horizontal composition.

In summary, this shows that a pre-2-holonomy with 
vanishing fake curvature \refdef{curvature and fake curvature} $dt\of{B} + F_A = 0$
defines a 2-functor $\hol \maps \paths_2 \to \twogroup$ and hence a local
strict 2-holonomy.
\endofproof

Before concluding this section it is noteworthy that for vanishing fake curvature
the path space curvature 2-form reads 
(prop. \ref{corollary: curvature of standard path space 1-form})
\begin{eqnarray}
  \mathcal{F}_\mathcal{A}
  &=&
  -\oint_A\of{H}
  \,,
\end{eqnarray}
where $H = \extd_A B$ is the curvature 3-form \refdef{curvature and fake curvature}.

\paragraph{Summary.}

The space of based paths over a given manifold $\manifold$ is an infinite
dimensional Frechet manifold on which one can do ordinary
differential geometry. A simple semi-heuristic argument shows that an assignment of
2-group elements to `small surface elements' in $\manifold$ should give
rise to a certain local connection 1-form on path space, called the
`standard connection 1-form' which comes
from a $\g$-valued 1-form $A$ and an $\h$-valued 2-form $B$ on $\manifold$.

This can be made precise by defining an appropriate \emph{Chen 1-form}
on path space and using its path space line holonomy to assign elements
in a strict 2-group $\twogroup$
to (possibly degenerate) surface elements with two corners, called \emph{bigons}. 
Under obvious composition these bigons naturally form a coherent 2-groupoid
$\paths_2\of{\manifold}$ and the assignment of 2-group elements to bigons
using the standard path space connection 1-form can be shown to define
a functor $\hol \maps \paths_2\of{\manifold} \to \twogroup$ precisely if
the `fake curvature' vanishes, i.e. if
the differential forms that the path space connection comes from satisfy
$dt\of{B} + F_A = 0$.

\vskip 2em

In the next section this result is used to define a global 2-connection in
a 2-bundle with strict structure 2-group.

\newpage
\subsection{2-Bundles with 2-Connections}
\label{section 2-bundles with 2-connection}

Just like the transition law 
for the transition functions in a 2-bundle 
comes from the concept of an ordinary transition \emph{internalized}
(\S\fullref{internalization}) in the 2-category $\twoDiff$ of 2-spaces, where it is
called a \emph{2-transition} \refdef{2-transition}, the transition rule
for a local 2-connection is obtained by similarly internalizing
the ordinary transition of an ordinary connection. A local 2-connection
together with its 2-transition then constitues a global 2-connection.

This requires that first of all the ordinary concept of a local connection and
its transition in an ordinary bundle be fomulated arrow-theoretically:

An ordinary local connection is a functor $\hol$ from the groupoid of paths
$\paths_1\of{\manifold}$ \refdef{groupoid of paths}
to the structure group G.

Given some covering $U$ of some base manifold $B$, this 
functor in particular
gives rise to a map
  \begin{eqnarray}
    \loops U
    \stackto{W}
    G
    \,.
    \nonumber
  \end{eqnarray}
from free loops in each patch of the cover to the structure group.

In terms of this map the transition law for the holonomy can be expressed
as follows:

\begin{proposition}
  \label{ordinary transition law}
  The transition law 
  for the connection 1-form in an ordinary principal bundle 
with cover $U$
has the arrow-theoretic description:
\begin{eqnarray}
  \label{arrow theoretic 1-connection}
  &&
  \loops U^{[2]}
  \stackrel{j^\loops_0 }{\longrightarrow}
  \loops U^{[1]}
  \stackto{W}
  G
  \nonumber\\
  &=&
  \nonumber\\
  &&
  \loops U^{[2]}
  \stackto{\stackrel{\vee}{\loops U^{[2]}}}
  \loops U^{[2]} \times \loops U^{[2]}
  \stackto{\stackrel{\vee}{\loops U^{[2]}} \times \loops U^{[2]}}
  \loops U^{[2]} \times \loops U^{[2]} \times \loops U^{[2]}
  \longrightarrow
  \nonumber\\
  &&
  \stackto{d_0 \times j^\loops_1 \times d_0}
  U^{[2]} \times \loops U^{[1]} \times U^{[2]}  
  \stackto{g \times W \times g}
  G \times G \times G
  \stackto{m \times s}
  G \times G
  \stackto{m}
  G
  \,.
  \nonumber\\
\end{eqnarray}
\end{proposition}

(Here the notation follows that of \cite{Bartels:2004}: 
$\vee$ denotes the diagonal embedding of a space in its second tensor
power, $m$ is the multiplication and $s$ the inversion operation in the
group. Moreover
$\gamma_{(x,i,j)} \stackrel{j^L_0}{\mapsto} \gamma_{(x,i)}$
and $\gamma_{(x,i,j)} \stackrel{j^L_1}{\mapsto} \gamma_{(x,j)}$,
\cf \refer{cover projection maps}.)

\Proof

The equality of maps is equivalent to
\begin{eqnarray}
  W_i\of{\gamma_x} &=& g_{ij}\of{x} W_j\of{\gamma_x}g_{ij}^{-1}\of{x}
  \,,\hspace{1cm}
  \forall\, \gamma_x \in \loops U_{ij}\,,\forall\, i,j
  \,.
  \nonumber
\end{eqnarray}

By assumption the holonomies $W_i$ come from 1-forms 
$A_i \in  \Omega^1\of{U_i,\g}$,
$W_i = W_{A_i}$ as in \refer{def line holonomy}. In terms of these
the above is equivalent to 
\begin{eqnarray}
  W_{A_i}\of{\gamma_x}
  &=&
  W_{g_{ij}(d+A_j)g_{ij}^{-1}}\of{\gamma_{x}}
  \,,
  \hspace{1cm}
  \forall\, \gamma_x \in \loops U_{ij}\,,\forall\, i,j
  \,.
  \nonumber
\end{eqnarray}
\endofproof

\subsubsection{2-Transition of 2-Holonomy}

Equation \refer{arrow theoretic 1-connection} is internalized in
$\twoDiff$ by letting all spaces be 2-spaces and all maps be 2-maps.
In particular $\loops U^{[2]}$ becomes a 2-loop 2-space:

\begin{definition}
  \label{2-loop 2-space}
  Given a 2-space $S$ the
  {\bf free 2-loop 2-space} $\loops S$  
  over $S$ is the simple \refdef{simple 2-space} 2-space 
  whose point space is $S^2$ and whose arrow space is the disjoint union of all
  $\paths_\gamma^\gamma\of{S^{2}}$ 
  \refdef{based path space}
  for all $\gamma \in S^2$ 
  with the obvious source and target maps.  
\end{definition}

The local 2-connection functor $\hol \maps \paths_2\of{\manifold} \to \twogroup$
\refdef{local strict 2-holonomy} gives rise to a 2-map
  \begin{eqnarray}
    \label{W map}
    \loops U \stackto{W} \twogroup
    \,.
  \end{eqnarray}
with point and arrow part given by \refdef{local pre-2-holonomy}
\begin{eqnarray}
  \label{local 2-connection in terms of 1,2-forms}
  (\loops U)^1 
  & \stackto{W^1} &
  \twogroup^1
  \nonumber\\
  \gamma_{(x,i)} &\mapsto& W_{A_i}\of{\gamma_{(x,i)}}
  \,,
  \nonumber\\
  \nonumber\\
  (\loops U)^2 
  & \stackto{W^2} &
  \twogroup^2
  \nonumber\\
  \Sigma_{\gamma_{(x,i)}} 
   &\mapsto& 
  \left(
    W_{A_i}\of{\gamma_{(x,i)}},
    \mathcal{W}^{-1}_{\mathcal{A}_i}\of{\Sigma_{\gamma^{-1}_{(x,i)}}}
  \right)
  \,,
\end{eqnarray}
in terms of which a global 2-connection is hence defined as follows:
\begin{definition}
  \label{global 2-connection}
  A {\bf global 2-connection} in a $\twogroup$-2-bundle
  with cover 2-space $U$
  is a local 2-holonomy  $\hol$ \refdef{local strict 2-holonomy}
  on each $U_i$
  giving rise to a map \refer{W map}
  together with a natural transformation $a$
\begin{eqnarray}
  \label{natural transformation defining 2-connection}
  &&
  \loops U^{[2]}
  \stackrel{j^\loops_0 }{\longrightarrow}
  \loops U^{[1]}
  \stackto{W}
  G
  \nonumber\\
  & \stackTo{a} &
  \nonumber\\
  &&
  \loops U^{[2]}
  \stackto{ \stackrel{\vee}{\loops U^{[2]}} }
  \loops U^{[2]} \times \loops U^{[2]}
  \stackto{ \stackrel{\vee}{\loops U^{[2]}} \times \loops U^{[2]}}
  \loops U^{[2]} \times \loops U^{[2]} \times \loops U^{[2]}
  \longrightarrow
  \nonumber\\
  &&
  \stackto{d_0 \times j^\loops_1 \times d_0}
  U^{[2]} \times \loops U^{[1]} \times U^{[2]}  
  \stackto{g \times W \times g}
  \twogroup \times \twogroup \times \twogroup
  \stackto{m \times s}
  \twogroup \times \twogroup
  \stackto{m}
  \twogroup
 \end{eqnarray}
\end{definition}

This somewhat baroque assembly of arrows has the following equivalent 
description:

\begin{proposition}
  \label{nattraf of 2-connection written out}
  The natural transformation \refer{natural transformation defining 2-connection}
  defining a global 2-connection in a 2-bundle \refdef{global 2-connection}
  is equivalent to the existence of a map 
\begin{eqnarray}
  \left(\loops U^{[2]}\right)^1
  \stackrel{\mbox{\tiny def. \ref{2-loop 2-space}}}{=} (U^{[2]})^2
  &\stackrel{\tilde a}{\longrightarrow}&
  \twogroup^2
  \nonumber\\
  \gamma_{(x,i,j)}
  &\mapsto&
  \tilde a\of{\gamma_{(x,i,j)}}
  \nonumber\\
  && \defas 
  \left(
    \tilde a^1_{\gamma_{(x,i,j)}},
    \tilde a^2_{\gamma_{(x,i,j)}}  
  \right)
\end{eqnarray}
such that
\begin{eqnarray}
  \label{nattrafo for aij}
  &&
  \left(
    W_{A_i}\of{\gamma_{(x,i,j)}},
    \mathcal{W}^{-1}_{\mathcal{A}_i}\of{\Sigma_{\gamma^{-1}_{(x,i,j)}}}
  \right)
  \circ
  \left(
    \tilde a^1_{\gamma_{(x,i,j)}},
    \tilde a^2_{\gamma_{(x,i,j)}}  
  \right)
  \nonumber\\
  &=&
  \left(
    \tilde a^1_{\gamma_{(x,i,j)}},
    \tilde a^2_{\gamma_{(x,i,j)}}  
  \right)
  \circ
  \left(
  g_{ij} 
  \cdot
  \left(
    W_{A_j}\of{\gamma_{(x,i,j)}},
    \mathcal{W}^{-1}_{\mathcal{A}_j}\of{\Sigma_{\gamma^{-1}_{(x,i,j)}}}
  \right)
  \cdot
  g_{ij}^{-1}
  \right)
  \,,
  \hspace{1cm}
  \forall \, \gamma_{(x,i,j)} \in U^2_{ij}
  \,.
  \nonumber\\
\end{eqnarray}
\end{proposition}
Here all pairs in brackets denote 2-group elements as in 
\refer{local 2-connection in terms of 1,2-forms}. ``$\circ$'' denotes
the vertical product in the 2-group and ``$\cdot$'' the horizontal one.

\Proof This is just the definition of this natural transformation written out.
\endofproof

We now show what the existence of this map $\tilde a$ amounts to in terms
of local data. We do this for the case where the arrow part $g^2_{ij}$ 
of the transition function \refdef{principal 2-bundle} is trivial.
According to \S\fullref{2-Bundles on base 2-Spaces of infinitesimal Loops}
this implies that in terms of nonabelian gerbe cocycles we now restrict 
attention to the case where \refer{curving transformation 2-forms of nonabelian gerbe}
\begin{eqnarray}
  \label{restriction d_ij = 0}
  d_{ij} = 0
  \,.
\end{eqnarray}

\begin{proposition}
  \label{property of 2-connection 2-transition}
  A global 2-connection 
  \refdef{global 2-connection}
  on 
  a $\twogroup$-2-bundle with base 2-space being 
 $B = \bigcup\limits_{i\in I}(U_i^1,\loops U_i^1 )$
and $\twogroup$ a strict automorphism 2-group
gives rise to the transition laws
\refer{gerbe trans law for connection 1-form} and 
\refer{gerbe trans law for curving 2-forms}
for the connection 1-form $A_i$ and curving 2-form $B_i$ of
a nonabelian gerbe for the special case
\begin{eqnarray}
  &&dt\of{B_i} + F_{A_i} = 0 
  \,,\hspace{1cm} \forall i
  \nonumber\\
  && d_{ij} = 0 = \beta_{ij}
  \,,\hspace{1cm} \forall i,j
  \,.
\end{eqnarray}

\end{proposition}

\Proof

The condition to be studied is the equality \refer{nattrafo for aij}
in prop. \ref{nattraf of 2-connection written out}.

In order to avoid awkward notation in the 
rest of this proof most occurences of $\gamma$ will be left implicit.
Hence first of all \refer{nattrafo for aij} becomes
\begin{eqnarray}
  \label{connection nattraf simplified}
  \left(
    W_{A_i},
    \mathcal{W}^{-1}_{\mathcal{A}_i}
  \right)
  \circ
  \left(
    \tilde a^1_{ij},
    \tilde a^2_{ij}  
  \right)
  &=&
  \left(
    \tilde a^1_{ij},
    \tilde a^2_{ij}
  \right)
  \circ
  \left(
  g_{ij} 
  \cdot
  \left(
    W_{A_j},
    \mathcal{W}^{-1}_{\mathcal{A}_j}
  \right)
  \cdot
  (g_{ij})^{-1}
  \right)
  \,.
  \nonumber\\
\end{eqnarray}

The source/target matching condition on $\tilde a$ is
\begin{eqnarray}
  \tilde a^1_{ij}
  &=&
  W_{A_i}
  \nonumber\\
  t\of{\tilde a^2_{\gamma_{(x,i,j)}}}
  W_{A_i}[\gamma_{(x,i,j)}]
  &=&
  g_{ij}^1 W_{A_j}[\gamma_{(x,i,j)}] (g_{ij}^1)^{-1}
  \,.
\end{eqnarray}
In order to be able to apply
prop. \ref{difference in holonomies wrt different connections}
to that rewrite this equivalently as
\begin{eqnarray}
  W_{A_i}[\gamma^{-1}_{(x,i,j)}]
  t\of{(\tilde a^2_{\gamma_{(x,i,j)}})^{-1}}
  &=&
  g_{ij}^1 W_{A_j}[\gamma^{-1}_{(x,i,j)}] (g_{ij}^1)^{-1}
\end{eqnarray}
(using $W[\gamma^{-1}] = W^{-1}[\gamma]$).

Hence $t\of{\tilde a^2}$ makes up for the difference 
between $A_i$ and $g^1_{ij}(\extd+A_j)(g^1_{ij})^{-1}$.
Denote this difference by
\begin{eqnarray}
  a^1_{ij} &\defas&
  g^1_{ij}(\extd+A_j)(g^1_{ij})^{-1}
  -
  A_i
  \,.
\end{eqnarray}

Using prop. \ref{difference in holonomies wrt different connections}
it follows that $t\of{\tilde a^2}$ can be expressed as 
\begin{eqnarray}
  t\of{(\tilde a^2_{\gamma_{(x,i,j)}} )^{-1}}
  &=&
  \lim\limits_{\epsilon = 1/N \to 0}
  \left(
    1 + \epsilon\oint\limits_{A_i}\of{a_{ij}^1}
  \right)
  \left(
    1 + \epsilon\oint\limits_{A_i + \epsilon a^1_{ij}}\of{a_{ij}^1}
  \right)  
  \cdots
  \left(
    1 + \epsilon\oint\limits_{A_i + (1-\epsilon)a^1_{ij}}\of{a_{ij}^1}
  \right)_{|\gamma^{-1}_{(x,i,j)}}
  \,,
  \nonumber\\
\end{eqnarray}
where the right hand side is evaluated at $\gamma^{-1}_{(x,i,j)}$.

Hence there is an $\h$-valued 1-form $a_{ij}$  with 
\begin{eqnarray}
  \label{A transition law from 2-transition}
  dt\of{a_{ij}} &\defas& a^1_{ij} 
  \nonumber\\
  &=&
  g^1_{ij}(d+A_j)(g^1_{ij})^{-1}
  -
  A_i
  \,.
\end{eqnarray}
This is the transition law \refer{gerbe trans law for connection 1-form} 
for the connection 1-form of a nonabelian gerbe.

It follows $\tilde a^2_{ij}$ itself is given by
\begin{eqnarray}
  (\tilde a^2_{\gamma_{(x,i,j)}})^{-1}
  &=&
  \lim\limits_{\epsilon = 1/N \to 0}
  \left(
    1 + \epsilon\oint\limits_{A_i}\of{a_{ij}}
  \right)
  \left(
    1 + \epsilon\oint\limits_{A_i + \epsilon dt\of{a_{ij}}}\of{a_{ij}}
  \right)  
  \cdots
  \left(
    1 + \epsilon\oint\limits_{A_i + (1-\epsilon)dt\of{a_{ij}}}\of{a_{ij}}
  \right)_{|\gamma_{(x,i,j)}^{-1}}
  \nonumber
\end{eqnarray}
(evaluated at $\gamma_{(x,i,j)}^{-1}$).

Inserting this in \refer{nattrafo for aij} and using 
prop. \ref{effect of finite path space gauge transformations}
one finds
\begin{eqnarray}
  \label{effect of 2-connection transition on surface holonomy}
  \mathcal{W}_{(A_i + dt\of{a_{ij}},B_i - k_{ij})} &=&
  \mathcal{W}_{
    (g^1_{ij} A_i g_{ij}^{-1} + g^1_{ij}(d (g^1_{ij})^{-1}),
    \alpha\of{g^1_{ij}}\of{B_j})
  }
  \,.  
\end{eqnarray}
This implies 
\begin{eqnarray}
  \label{B transition law from 2-transition}
  B_i &=& \alpha\of{g^1_{ij}}\of{B_j} + k_{ij}
\end{eqnarray}
with $k_{ij} = \extd_A a_{ij} + a_{ij} \wedge a_{ij}$,
which is the transition law
\refer{gerbe trans law for curving 2-forms}
for the curving 2-form of a nonabelian
gerbe (for the given special case).
\endofproof

Note that the natural transformation \refer{connection nattraf simplified}
involves a gauge transformation \emph{of the first kind} 
\refer{first kind trafo}
coming from the
(horizontal) conjugation with $g_{ij}$ together with a gauge transformation
\emph{of the second kind} 
\refer{second kind trafo}
coming from a (vertical) conjugation with $\tilde a_{ij}$.

The above natural transformation must be supplemented by a coherence
law which ensures its consistency under a chain of compositions translating
from $U_i$ to $U_j$ to $U_k$ back to $U_i$:

\begin{proposition}
  \label{coherence law for a_ij}
  The coherence law of the natural transformation defining a
  global differentiable 2-connection 
  \refdef{global 2-connection}
  gives
  the coherence law for the transformators of the
  connection of a nonabelian gerbe
  \refer{gerbe coherence law for transformators of connection forms}.
\end{proposition}

\Proof

The coherence law ensures consistency of composition of natural transformations.
In this case the relevant composition is that of transitions 
$U_i \to U_j \to U_k \to U_i$. 

In terms of the connection 1-form this means, using 
\refer{effect of 2-connection transition on surface holonomy}

\hspace{-3cm}\parbox{20cm}{
\begin{eqnarray}
  A_i &=&
  g^1_{j}A_j (g^1_{j})^{-1}
  +
  g^1_{j}(\extd (g^1_{j})^{-1})
  -
  dt\of{a_{ij}}
  \nonumber\\
  &=&
  g^1_{j}\left(
    g_{jk}A_k g_{jk}^{-1}
    +
    d_{jk}(\extd g_{jk}^{-1})
    -
    a_{jk}
  \right)
   (g^1_{j})^{-1}
  +
  g^1_{j}(\extd (g^1_{j})^{-1})
  -
  dt\of{a_{ij}}  
  \nonumber\\
  &=&
  g^1_{j}\left(
    g_{jk}
    \left(
       g_{ki}A_i g_{ki}^{-1}
       +
       g_{ki}(\extd g_{ki}^{-1})
       -
       dt\of{a_{ki}}
    \right)
    (g^1_{jk})^{-1}
    +
    d_{jk}(\extd g_{jk}^{-1})
    -
    dt\of{a_{jk}}
  \right)
   (g^1_{j})^{-1}
  +
  g^1_{j}(\extd (g^1_{j})^{-1})
  -
  dt\of{a_{ij}}  
  \nonumber\\
  &=&
  (g^1_{ij}g^1_{jk}g^1_{ki})
  A_i
  (g^1_{ij}g^1_{jk}g^1_{ki})^{-1}
  +
  (g^1_{ij}g^1_{jk}g^1_{ki})
  \extd
  (g^1_{ij}g^1_{jk}g^1_{ki})^{-1}  
  -
  dt\of{a_{ij}}
  -
  g^1_{j}dt\of{a_{jk}}(g^1_{j})^{-1}
  -
  g^1_{ij}g^1_{jk}dt\of{a_{ki}}(g^1_{ij}g^1_{jk})^{-1}
  \nonumber\\
  &\equalby{source/target matching in proof for 2-transition}&
  A_i 
  +
  t\of{f_{ijk}}
  \commutator
  {A_i}
  {t\of{f_{ijk}}^{-1}}
  +
  t\of{f_{ijk}}
  \extd
  \left(t\of{f_{ijk}}^{-1}\right)  
  -
  dt\of{a_{ij}}
  -
  g^1_{j}dt\of{a_{jk}}(g^1_{j})^{-1}
  -
  t\of{f_{ijk}}g^1_{ik} \,dt\of{a_{ki}}\,g^1_{ik} (t\of{f_{ijk}})^{-1}  
  \nonumber\\
  &=&
  A_i
  + 
  dt\of{
    f_{ijk}d\alpha\of{A_i}\of{f_{ijk}^{-1}}
    +
    f_{ijk}\extd f_{ijk}^{-1}
    -
    a_{ij}
    -
    g^1_{ij}\of{a_{jk}}
    -
    f_{ijk}\,d\alpha\of{g^1_{ik}}\of{a_{ki}}\, f_{ijk}^{-1}
  }
  \,.
  \nonumber
\end{eqnarray}
}
It follows that
\begin{eqnarray}
    f_{ijk}d\alpha\of{A_i}\of{f_{ijk}^{-1}}
    +
    f_{ijk}\extd f_{ijk}^{-1}
    -
    a_{ij}
    -
    g^1_{ij}\of{a_{jk}}
    -
    f_{ijk}\,d\alpha\of{g^1_{ik}}\of{a_{ki}}\, f_{ijk}^{-1}
   &=& 
   -\alpha_{ijk}
  \nonumber
\end{eqnarray}
with $\alpha_{ijk} \in \ker\of{dt}$.
This can be simplified a little: For
$j=i$ this equation reduces to
\begin{eqnarray}
    a_{ik}
    +
    d\alpha\of{g^1_{ik}}\of{a_{ki}}
   &=& 
   0
   \,.
\end{eqnarray}
Reinserting this result yields
\begin{eqnarray}
    f_{ijk}d\alpha\of{A_i}\of{f_{ijk}^{-1}}
    +
    f_{ijk}\extd f_{ijk}^{-1}
    -
    a_{ij}
    -
    g^1_{ij}\of{a_{jk}}
    +
    f_{ijk} a_{ki} f_{ijk}^{-1}
   &=& 
   -\alpha_{ijk}
  \,.
\end{eqnarray}
This is indeed the gerbe coherence law 
\refer{gerbe coherence law for transformators of connection forms}.
\endofproof

With the transition law for the connection 1-form and `curving' 2-form
understood, the transition law for the curvature 3-form 
\refdef{curvature and fake curvature} follows:

\subsubsection{2-Transition of Curvature}
\label{2-Transition of Curvature}

Since curvature is the first order term in the holonomy around a small
loop, the 2-transition prop. \ref{nattraf of 2-connection written out}
of 2-holonomy immediately implies a transition law for the
path space connection 1-form $\mathcal{F}_\mathcal{A} = - \oint_A (H)$
\refer{path space curvature in fake flat case} and hence of the
curvature 3-form $H = \extd_A B$ \refdef{curvature and fake curvature}.

First of all one notes the following:

\begin{proposition}
 The curvature 3-form \refdef{curvature and fake curvature}
 $H = \extd_A B$ transforms \emph{covariantly}
  under gauge transformations of the first kind \refer{first kind trafo}.
  Moreover, it is \emph{invariant} under gauge transformations of the second kind
  \refer{second kind trafo} if and only if the fake curvature vanishes.
\end{proposition}

\Proof

The covariant transformation under gauge transformations of the first kind
follows from simple standard reasoning. The invariance under 
infinitesimal transformations of the second kind
with $A \to A + \epsilon dt\of{a}$ and $B \to B - \epsilon \extd_A a$
follows from noting the invariance under `infinitesimal' shifts:
\begin{eqnarray}
  H = \extd_A B &\to& \extd_{A + \epsilon dt\of{a}}(B - \epsilon \extd_A a)
  \nonumber\\
  &=&
  \extd_A B 
   - \epsilon 
  \left(
     \extd_A \extd_A a
     -
     d\alpha\of{dt\of{a}}\of{B}
  \right)
  +
  \order{\epsilon^2}
  \nonumber\\
  &\equalby{notation for action of h on h'}&
  H -\epsilon
  \left(
    d\alpha\of{F_A}\of{a}
    +
    d\alpha\of{dt\of{B}}\of{a}
  \right)
  +
  \order{\epsilon^2}
  \nonumber\\
  &\equalby{definition fake curvature}&
  H - \epsilon d\alpha\of{\tilde F}\of{B} + \order{\epsilon^2}
  \nonumber\\
  &\stackrel{\tilde F = 0}{=}&
  H + \order{\epsilon^2}
  \,.
\end{eqnarray}
\endofproof
(\cf equation (3.43) of \cite{GirelliPfeiffer:2004}).

Note that the invariance of $H$ under transformations of the second
kind does \emph{not} imply invariance of the path space curvature
2-form $\mathcal{F}_\mathcal{A}$. Instead, this transforms as
\begin{eqnarray}
  \mathcal{F}_\mathcal{A}
  =
  -\oint\limits_A (H) 
  &\to&
  -\oint\limits_{A + dt\of{a}}(H)
  \,.
\nonumber
\end{eqnarray}

Using this expression it is clear that, at least locally, it is always possible
to `gauge away' (by transformations of the second kind) that part of $A$ which takes
values in the image $\image\of{dt}$ of $dt$. 
In this sense it is the algebra $\g / \image\of{dt}$ that is of relevance
for the parallel transport that enters the construction of surface holonomy
discussed here. (A related issue is briefly mentioned in the discussion
\S\fullref{Discussion and Open Questions}.)

The transition law for $H_i \defas \extd_{A_i} B_i$ is now a simple corollary:

\begin{corollary}
  The local curvature 3-form $H_i = \extd_{A_i} B_i$ of the local standard path
  space connection of a 2-bundle with 2-connection has the transition law
  \[
    H_i = \alpha\of{g^1_{ij}}\of{H_j}
  \]
  on double intersections $U_{ij}$.
\end{corollary}
This is the transition law \refer{gerbe curvature transition law}
of the curvature 3-form of a nonabelian gerbe for vanishing
fake curvature and the special case $d_{ij} = 0$ that we 
restricted attention to \refer{restriction d_ij = 0}. 

One should note that also the fake curvature 
\refdef{curvature and fake curvature}
transforms covariantly
and can therefore indeed consistently be chosen to vanish:
The transition law for $F_{A_i}$ following from 
\refer{A transition law from 2-transition} is
\[
  F_{A_i} = g_{ij}F_{A_j}g_{ij}^{-1} - dt\of{k_{ij}}
\]
and that of $dt\of{B}$ following from \refer{B transition law from 2-transition}
\[
  dt\of{B_i} = g_{ij}\,dt\of{B_j} g_{ij}^{-1} + dt\of{k_{ij}}
  \,,
\]
so that
\[
  \tilde F_i = g_{ij} \tilde F_j g_{ij}^{-1}
  \,.
\]

\vskip 2em

\paragraph{Summary.}

An ordinary local connection gives rise to a local holonomy
functor that assigns group elements
to paths. These assignments of group elements transform in a well-known
way from one patch $U_i$ to another, $U_j$. 
In the context of 2-bundles and using the path space technology
developed in the previous section, 
using the concept of a 2-honolomy
$\hol \maps \paths_2\of{\manifold} \to \twogroup$ this transition can be 
\emph{internalized} in $\mathfrak{C}$, the 2­-category of 2-spaces, to yield
a 2-transition for a 2-connection. When this is worked out in terms of local
data the transition law for the connection 1-form and curving 2-form of 
nonabelian gerbes are obtained, together with their coherence law.
From this the gerbe transition law for the curvature 3-form directly
follows. 

\vskip 2em

This concludes our analysis of the local description of 2-bundles with
2-connections and the demonstration of the relation to nonabelian
gerbes with connection and curving. The main steps of our discussion 
are summarized in the following section.

\newpage

\section{Summary and Discussion}
\label{Summary and Discussion}


\subsection{Summary of the Constructions and Results}
\label{section: Summary of the constructions and results}

A 2-bundle is the categorification of the notion of an ordinary bundle by 
internalizing it in the 2-category $\twoDiff$ of 2-spaces, where
a 2-space is again a category internalized in $\Diff$ (or $\Diff_\infty$), 
the category of finite (infinite) dimensional smooth spaces.
The notion of a 2-bundle without connection and the nature of
2-transitions in 2-bundles have been dissussed recently in 
\cite{Bartels:2004}. 

We showed first that 
under certain conditions
such a 2-transition in terms
of local data reproduce the cocycle description of nonabelian
gerbes (without connection and curving). Then we augmented 2-bundles
with 2-connections by using holonomy on path space to construct
a 2-functor from the 2-groupoid of bigons to the structure 2-group.
Using the categorified transition law we globalized the resulting local
2-holonomy.

\vskip 2em

To begin with, 2-transitions in 2-bundles are characterized by 
a transition 2-function $U^{[2]} \stackto{g} \twogroup$, a 2-map from 
double overlaps of the cover 2-space to the structure 2-group.
The categorified transition law says that there exists a natural 
isomorphism between $g_{ij}\cdot g_{jk}$ and $g_{ik}$.

This means
(prop. \ref{proposition on local meaning of 2-transition})
that there exists a map
\begin{eqnarray}
  (U^{[3]})^1 & \stackto{f} & \twogroup
  \nonumber\\
  (x,i,j,k) &\mapsto& f_{ijk}\of{x} \in \twogroup
  \nonumber\\
  && = (f^1_{ijk}\of{x},f^2_{ijk}\of{x})
  \nonumber
\end{eqnarray}
such that
\begin{eqnarray}
   \label{natural transformation encoded 2-transition}
   (g_{ik}^1,1) \circ f_{ijk}\of{x}
   &=&
   f_{ijk}\of{x}
   \circ
   \left(
   \,
   (
     g^1_{ij}\of{x},
     1 
    )
   \,
   \cdot 
   \,
   (
    g^1_{jk}\of{x},
    1
   )
   \,
  \right)
  \,.
\end{eqnarray}
Here a superscript 1 denotes the point part of a space or map and a superstript
2 the arrow part.
We always denote by `$\circ$' the 
\emph{vertical} product in the 2-group and by `$\cdot$' the
\emph{horizontal} one. 

We restrict attention throughout to the case 
where all morphisms in the base 2-space have coinciding source and target.
In this case,
at the point level the above says that source and target of this equation
must match, which means that
\[
  t\of{f^2_{ijk}} g^1_{ik}
  =
  g^1_{ij}g^1_{jk}
  \,.
\]
This is the first equation 
\refer{gerbe transition law for the transition functions}
of a gerbe cocycle description. 

The natural transformation encoded by $f$ can be thought of as 
modifying the ordinary horizontal product. To emphasize this,
equation \refer{natural transformation encoded 2-transition}
can be equivalently rewritten as
\begin{eqnarray}
   (
     g^1_{ij}\of{x},
     1
    )
   \,
   \cdot 
   \,
   (
    g^1_{jk}\of{x},
    1
   )
   &=&
   (g_{ik}^1,f^2_{ijk}\of{x}g^2_{ik}\of{\gamma_x} (f^2_{ijk}\of{x})^{-1})
   \,.
   \nonumber
\end{eqnarray}
This `weakened' version of the ordinary product between transition functions
should be consistent, in the sense that
\begin{eqnarray}
  \label{associativity of transition functions}
  (g_{ij} \cdot g_{jk}) \cdot g_{kl}
  &=&
  g_{ij} \cdot ( g_{jk} \cdot g_{kl})
\end{eqnarray}
on double overlaps,
where the parentheses indicate in which order we apply the above formula. This
is the \emph{coherence law} for the natural transformation encoded by $f$.

Evaluating this equation yields 
\refer{coherence law for transition transformation}
the condition
\begin{eqnarray}
   \lambda_{ijkl} 
   &=& 
   (f^2_{ikl})^{-1}(f^2_{ijk})^{-1}\alpha\of{g^1_{ij}}\of{f^2_{jkl}}f^2_{ijl}
  \nonumber
  \,,
\end{eqnarray}
for some $\ker\of{t} \subset H$-valued twist 0-form $\lambda_{ijkl}$.
This is the gerbe transition law \refer{gerbe coherence law for transformators 
of transition functions}.

So far this assumes that the arrow part $g^2_{ij}$ of the transition
function $g$ is trivial. It can be shown 
(\S\fullref{2-Bundles on base 2-Spaces of infinitesimal Loops}) that
when $g^2_{ij}$ is allowed to take nontrivial values on `infinitesimal'
loops it encodes the `curving transformation 2-forms' $d_{ij}$
of a nonabelian gerbe
\refer{curving transformation 2-forms of nonabelian gerbe} as well as
their transition law \refer{gerbe trans law for curving transformation 2-forms}.
When discussing the local description of 2-connections on 2-bundles 
we assume the $g^2_{ij}$ to be trivial in the following.

\skiph{2em}

Now we turn to 2-connections:

Once the notion of 2-holonomy is available,
the transition laws for the 2-connection can be dealt with in complete
analogy to the above analysis of the transition law for the 
transition function.

2-holonomy itself is obtained by categorifying ordinary holonomy,
which, in a trivial bundle, is a functor $\paths_1\of{M} \to G$.
Accordingly, local 2-holonomy should be a 2-functor
\[
  \hol \maps \paths_2\of{M} \to \twogroup
\] 
that assigns 2-group elements to bigons in $M$. 
Since this functor is supposed to be differentiable there is
a $\g$-valued 1-form $A$ and an $\h$-valued 2-form $B$ such that
a small bigon with source edge $\gamma$ and surface $\Sigma$ is
mapped to the 2-group element
\[
  \approx (\exp\of{\int_\gamma A}, \exp\of{\int_\Sigma B})
  \,.
\]
This was discussed in \cite{GirelliPfeiffer:2004}.

We want to find a connection on path space which computes such a 
2-holonomy:
Using the rule for horizontal products in a 2-group
\[
  (g,h) \cdot (g^\prime, h^\prime)
  =
  (gg^\prime, h \, \alpha\of{g}\of{h^\prime})
\]
and iterating it for a long chain of small bigons whose size tends to zero,
one finds 
(\S\fullref{section: The Standard Connection 1-Form on Path Space})
that $H$-labels of the 2-group elements obtained from the
2-holonomy 2-functor are equivalently computed by the ordinary holonomy of 
a 1-form $\standardpathspaceconnection$ on path space, given by
(see def. \ref{standard path space 1-form})
the integral over a given path of the pullback of $B$ to that path,
parallel transported with respect to $A$ to the source vertex of the path.

Precise formulas capturing this simple idea are given in 
\S\fullref{path space differential calculus}. 
It turns out that the above path space connection defines a 2-functor
$\paths_2\of{\manifold} \stackto{\hol} \twogroup$ precisely if the 
`fake curvature' vanishes:
\[
  F_A + dt\of{B} = 0
  \,.
\]
(This condition is essentially implicit in the very nature of strict 
2-groups, as has first been noted in \cite{GirelliPfeiffer:2004}.)

The importance of
this path space formulation is that it allows to express 2-holonomy
differentiably in terms of differential forms and to compute the effect on these
differential forms when path space holonomy is conjugated by
group elements, as in the
transition law for the 2-holonomy:

Denote by $\gamma_{(x,i,j)}$ any element of $U_{ij}^2 = LU_{ij}^1$, a closed loop
based at $x$ in the double overlap $U_{ij}^1$. Let $\Sigma_{\gamma_{(x,i,j)}}$
be a bigon with source and target equal to $\gamma_{(x,i,j)}$ (i.e. a closed
loop of loops in $U^2_{ij}$) and denote by
\[
  \left(
    W_{A_i}\of{\gamma_{(x,i,j)}},
    \mathcal{W}^{-1}_{\mathcal{A}_i}\of{\Sigma_{\gamma^{-1}_{(x,i,j)}}}
  \right)
\]
the 2-group element associated to $\Sigma$ by the 2-holonomy,
where $W_{A_i}$ is the ordinary holonomy of $A_i$ along $\gamma$
and 
$\mathcal{W}_{\mathcal{A}_i}$ is the above mentioned path space
holonomy of $\Sigma$.
(The appearance of the inverse is just due to a couple of conventions 
of signs and orientations that go into this.)

The categorified version of the transition law for this 2-holonomy is
\refer{nattrafo for aij}
\begin{eqnarray}
  &&
  \left(
    W_{A_i}\of{\gamma_{(x,i,j)}},
    \mathcal{W}^{-1}_{\mathcal{A}_i}\of{\Sigma^i_{\gamma^{-1}_{(x,i,j)}}}
  \right)
  \circ
  \left(
    \tilde a^1_{\gamma_{(x,i,j)}},
    \tilde a^2_{\gamma_{(x,i,j)}}  
  \right)
  \nonumber\\
  &=&
  \left(
    \tilde a^1_{\gamma_{(x,i,j)}},
    \tilde a^2_{\gamma_{(x,i,j)}}  
  \right)
  \circ
  \left(
  g_{ij} 
  \cdot
  \left(
    W_{A_j}\of{\gamma_{(x,i,j)}},
    \mathcal{W}^{-1}_{\mathcal{A}_j}\of{\Sigma^j_{\gamma^{-1}_{(x,i,j)}}}
  \right)
  \cdot
  g_{ij}^{-1}
  \right)
  \,,
  \nonumber
\end{eqnarray}
where
\begin{eqnarray}
  \loops U^{[2]} & \stackto{\tilde a} & \twogroup
  \nonumber\\
  \gamma_{(x,i,j)} & \mapsto & \tilde a_{ij}\of{x}
  \nonumber\\
  && = (\tilde a^1_{ij}\of{x}, \tilde a_{ij}^2\of{x}) 
  \nonumber
\end{eqnarray}
encodes the natural transformation.

One should note that beneath the host of symbol decorations this
equation has the simple structure of an ordinary 
`horizontal' gauge transformation (`of the first kind' \refer{first kind trafo})
by $g_{ij}$ together with an additional 
`vertical' gauge transformation (`of the second kind' \refer{second kind trafo}) 
by $\tilde a$. 

This is again an equation between 2-maps that splits into its
point-part and its arrow-part, from which we can extract the transition laws
for the 1-form $A_i$ and the 2-form $B_i$.

At the point level the target matching condition of this equation
says that
\[
  W_{A_i}t\of{(\tilde a^2_{ij})^{-1}}
  =
  g^1_{ij}W_{A_j}(g^1_{ij})^{-1}
  \,,
\]
where we suppress the index $\gamma_{(x,i,j)}$ for convenience. At the arrow level
the equation says that
\[
  \tilde a^2_{ij} \mathcal{W}_{\mathcal{A}_i} (\tilde a^2_{ij})^{-1}
  =
  \alpha\of{g_{ij}}\of{\mathcal{W}_{\mathcal{A}_j}}
  \,.
\]
This are essentially our transition laws. It remains to express these
equalities between group elements as equalities between the generating
algebra elements. This is straightforward but requires a couple of
results on path space differential calculus which are derived in
\S\fullref{path space differential calculus}. 
In prop. \ref{property of 2-connection 2-transition} 
it is shown how using these results the above two
equations can be seen to be equivalent to the equations
\begin{eqnarray}
  &&
  A_i + dt\of{a_{ij}} = g^1_{ij}(\extd + A_j)(g_{ij}^1)
  \nonumber\\
  &&
  B_i = d\alpha\of{g^1_{ij}}\of{B_j} + k_{ij}
  \,,
  \nonumber
\end{eqnarray}
where
\[
  a_{ij} \in \Omega^1\of{U_{ij},\h}
\]
is a 1-form taking values in the Lie algebra $\h$ of $\H$, and
\[
  k_{ij} \defas \extd_{A_i} a_{ij} + a_{ij}\wedge a_{ij}
\]
is its field strength (relative to $A_i$).

The first of these is precisely the gerbe transition law
\refer{gerbe trans law for connection 1-form}
for the connection 1-form $A_i$. The second is the
gerbe transition law
\refer{gerbe trans law for curving 2-forms} for the
`curving' 2-form $B_i$ for the case that the gerbe data
\refer{gerbe twist p-forms} and 
\refer{curving transformation 2-forms of nonabelian gerbe}
is such that
\[
  d_{ij} = -\beta_{ij}
  \,.
\]

Next, there is again a coherence law for the natural transformation
$\tilde a$. Here we need to require that transforming a holonomy from
patch $i$ to $j$ to $k$ and then back to $i$ reproduces the
original result - up to a twist that arises when
when $\gamma_i \neq \gamma_i \mapsto \gamma_j \mapsto \gamma_k \mapsto \gamma_i$.

A straightforward calculation (see prop. \ref{coherence law for a_ij})
shows that at the point-space level the coherence law implies that
\[
  dt\of{
    f_{ijk}d\alpha\of{A_i}\of{f_{ijk}^{-1}}
    +
    f_{ijk}\extd f_{ijk}^{-1}
    -
    a_{ij}
    -
    g^1_{ij}\of{a_{jk}}
    +
    f_{ijk} a_{ki} f_{ijk}^{-1}
  }
  = 
  0
  \,.
\]
This again implies that there is a twist 2-form
\[
  \alpha_{ijk} \in (\ker\of{dt}) \times \Omega^2\of{U_{ijk}}
\]
taking values in the kernel of $dt$ such that
\[
    f_{ijk}d\alpha\of{A_i}\of{f_{ijk}^{-1}}
    +
    f_{ijk}\extd f_{ijk}^{-1}
    -
    a_{ij}
    -
    g^1_{ij}\of{a_{jk}}
    +
    f_{ijk} a_{ki} f_{ijk}^{-1}
  =
  - 
  \alpha_{ijk}
  \,.
\]
This is the gerbe transition law 
\refer{gerbe coherence law for transformators of transition functions}.

Finally, the path space curvature $\mathcal{F}_\mathcal{A}$
gives rise to a curvature 3-form $H_i = \extd_{A_i} B_i$
\refdef{curvature and fake curvature} which 
(as shown in \S\fullref{2-Transition of Curvature})
has the covariant trasition law
\[
  H_i = \alpha\of{g^1_{ij}}\of{H_j}
  \,.
\]
This gives us the last two gerbe laws
\refer{gerbe curvature 3-form} and 
\refer{gerbe curvature transition law} subject to the constraints
considered here.

\vspace{2em}

In conclusion, a 2-bundle with 2-connection under certain conditions
gives rise to a nonabelian gerbe with connection and curving equipped
with a notion of nonabelian surface holonomy.

\subsection{Discussion and Open Questions}
\label{Discussion and Open Questions}

Our central result is that a locally trivializable 
2-bundle with simple base 2-space and strict structure 2-group
induces the cocylce description of a nonabelian gerbe with connection
and curving for vanishing `fake curvature' 
and augments the latter with a notion of surface holonomy.

Together with this unifying result
we have found a curious dichotomy between generalizations on the 2-bundle
side and on the gerbe side. In order to derive the nonabelian gerbe we
had to restrict the possible freedom available in 2-bundles by
\begin{itemize}
  \item
    restricting the base 2-space to be `simple',
  \item restricting the arrow part of the transition functions to be non-trivial
        only on `infinitesimal' loops.
\end{itemize}
In the context of 2-bundles these restrictions appear artificial and should
be relaxed. But, as we have discussed, doing so seems to lead to local 
transition laws that have no counterpart in the literature on nonabelian gerbes.

On the other hand, from the point of view of the study of nonabelian gerbes it is the
requirement on the `fake curvature' to vanish that looks artificial and 
unsatisfactory. 

It is clear, however, that this condition is tightly related to the existence of
a notion of 2-holonomy in $\twogroup$-2-bundles for strict $\twogroup$. One important
open question is therefore: 

\begin{itemize}
\item
\emph{How does the discussion in this paper generalize when the standard fiber $F$ 
of the 2-bundle is allowed to be something which is not a strict 2-group?}
\end{itemize}

The case of most immediate interest is that where the standard fiber is 
not a strict, but a
\emph{coherent} 2-group \cite{BaezLauda:2003}. 
2-bundles with coherent 2-groups are the most general
result of the categorification procedure considered here \cite{Bartels:2004}. 
They are also 
strongly suggested by the fact that the 2-groupoid of bigons that we
use in \S\fullref{section: Local 2-Holonomy from local Path Space Holonomy} 
for construcing 2-holonomy
is a coherent 2-groupoid. Consequently 2-holonomy should take values
in a coherent 2-group and our discussion should be correspondingly generalized.
While allowing coherent structure 2-groups in 2-bundles again drives one
out of the realm of the known local description of nonabelian gerbes, 
it seems possible to generalize this appropriately.

But even with a more general understanding of 2-holonomy currently lacking,
the known notion of 2-holonomy (i.e. for strict structure 2-groups) 
deserves to be better understood. Recall that the path space curvature of
the local connection 1-form on path space that we found was
\[
  \mathcal{F} = \oint_A (H) = \oint_A \of{\extd_A B}
\]
which takes values in the \emph{abelian} ideal $\ker\of{dt} \in \h$.

When the curvature of an ordinary bundle with connection takes values in a proper
ideal of the total algebra the structure group can be reduced
to that generated by this ideal. One is therefore led to ask:

\begin{itemize}
\item
  \emph{
  Is the nonabelian surface holonomy in 2-bundles with strict structure
  2-group `reducible' in some appropriate sense to ordinary \emph{abelian}
  surface holonomy?}
\end{itemize}

The answer to this, either way, seems to be nontrivial due to the fact that
even though the path space curvature takes values in an abelian subalgebra it
involves nonabelian parallel transport.

Apart from these and similar questions revolving around a better understanding
of the general formalism, very interesting questions are related to
applications of this formalism to physics. In particular, given that for the
first time we now have a notion of global nonabelian surface holonomy, a central question 
in the context of the topics mentioned in 
\S\fullref{section: Nonabelian 2-Holonomies and Physics} is:
\begin{itemize}
\item
  \emph{
  Does this notion of gobal nonabelian surface holonomy correctly describe the 
  coupling of membrane boundaries to stacks of 5-branes? 
  }

  \emph{
  In other words, is this
  the correct notion of surface holonomy for the `nonabelian gerbe theories' expected
  to describe the six-dimensional decompactification of ordinary four-dimensional
  gauge theory? 
  }

  \emph{
  What is the physical interpretation of `fake curvature' here?
  }
\end{itemize}

\acknowledgments{
We are grateful to
Orlando Alvarez,
Paolo Aschieri,
Toby Bartels,
Jens Fjelstad,
Branislav Jur{\v c}o,
Amitabha Lahiri,
Thomas Larsson
and
Hendryk Pfeiffer
for comments and helpful discussion. As we hope to have made
clear in the text, the r-flatness condition for the case 
$G=H,t=\mathrm{id}, 
\alpha = \mathrm{Ad}$ is due to
Orlando Alvarez. Special thanks go to Eric Forgy for preparing
the figure on curves in loop space.
The second author was supported by SFB/TR 12.

\newpage

\newpage

\bibliography{2conn}

\providecommand{\href}[2]{#2}\begingroup\raggedright\begin{thebibliography}{10}

\bibitem{BaezDolan:1998}
J.~Baez and J.~Dolan, {\it Categorification},  in {\em Higher Category Theory},
  Contemp. Math. 230.
\newblock Am. Math. Soc., 1998.
\newblock \Math{QA}{9802029}.

\bibitem{BaezLauda:2003}
J.~Baez and A.~Lauda, {\it Higher-dimensional algebra {V}: 2-groups},  {\em
  Theory and Applications of Categories} {\bf 12} (2004) 423.
  \Math{QA}{0307200}.

\bibitem{BaezCrans:2003}
J.~Baez and A.~Crans, {\it Higher-dimensional algebra {VI}: {L}ie 2-algebras},
  {\em Theory and Applications of Categories} {\bf 12} (2004) 492.
  \Math{QA}{0307263}.

\bibitem{Baez:2002}
J.~Baez, {\it Higher {Y}ang-{M}ills theory}, . \hepth{0206130}.

\bibitem{Bartels:2004}
T.~Bartels, {\it Categorified gauge theory: two-bundles}, . \Math{CT}{0410328}.

\bibitem{Pfeiffer:2003}
H.~Pfeiffer, {\it Higher gauge theory and a non-{A}belian generalization of
  2-form electromagnetism},  {\em \ap{308}{2003}{447}}. \hepth{0304074}.

\bibitem{GirelliPfeiffer:2004}
F.~Girelli and H.~Pfeiffer, {\it Higher gauge theory - differential versus
  integral formulation}, . \hepth{0309173}.

\bibitem{AlvarezFerreiraSanchezGuillen:1998}
O.~Alvarez, L.~Ferreira, and J.~S{\'a}nchez~Guill{\'en}, {\it A new approach to
  integrable theories in any dimension},  {\em \npb{529}{1998}{689}} (1998).
  \hepth{9710147}.

\bibitem{Schreiber:2004e}
U.~Schreiber, {\it Nonabelian 2-forms and loop space connections from 2d {SCFT}
  deformations}, . \hepth{0407122}.

\bibitem{Alvarez:2004}
O.~Alvarez, 2004.
\newblock (private communication).

\bibitem{BreenMessing:2001}
L.~Breen and W.~Messing, {\it Differential geometry of gerbes}, .
  \Math{AG}{0106083}.

\bibitem{AschieriCantiniJurco:2003}
P.~Aschieri, L.~Cantini, and B.~Jur{\v c}o, {\it Nonabelian bundle gerbes,
  their differential geometry and gauge theory}, . \hepth{0312154}.

\bibitem{AschieriJurco:2004}
P.~Aschieri and B.~Jur{\v c}o, {\it Gerbes, {M5}-brane anomalies and {$E_8$}
  gauge theory},  {\em \jhep{10}{2004}{068}}. \hepth{0409200}.

\bibitem{Murray:1994}
M.~Murray, {\it Bundle gerbes}, . \Math{DG}{9407015}.

\bibitem{MackaayPicken:2000}
M.~Mackaay and R.~Picken, {\it Holonomy and parallel transport for abelian
  gerbes}, . \Math{DG}{0007053}.

\bibitem{Schreiber:2004}
U.~Schreiber, {\it On deformations of 2d {SCFT}s},  {\em \jhep{06}{2004}{058}}.
  \hepth{0401175}.

\bibitem{Witten:2002}
E.~Witten, {\it Talk at `{T}opology, {G}eometry and {Q}uantum {F}ield
  {T}heory'},  2002.
\newblock slides available at
  \href{http://www.maths.ox.ac.uk/notices/events/special/tgqfts/photos/witten/%
}{http://www.maths.ox.ac.uk/notices/events/special/tgqfts/photos/witten/}.

\bibitem{AharonyHananyKol:1997}
O.~Aharony, A.~Hanany, and B.~Kol, {\it Webs of $(p,q)$ 5-branes, five
  dimensional field theories and grid diagrams},  {\em \jhep{01}{1998}{002}}.
  \hepth{9710116}.

\bibitem{KolRahmfeld:1998}
B.~Kol and J.~Rahmfeld, {\it {BPS} spectrum of 5 dimensional field theories,
  {$(p,q)$} webs and curve counting},  {\em \jhep{08}{1998}{006}}.
  \hepth{9801067}.

\bibitem{BerensteinLeigh:1998}
D.~Berenstein and R.~Leigh, {\it String junctions and bound states of
  intersecting branes},  {\em \prd{60}{1990}{026005}}. \hepth{9812142}.

\bibitem{EckmannHilton}
B.~Eckmann and P.~Hilton, {\it Group-like structures in categories},  {\em
  Math. Ann.} {\bf 145} (1962) 227--255.

\bibitem{Forrester-Barker:2002}
M.~Forrester-Barker, {\it Group objects and internal categories}, .
  \Math{CT}{0212065},
  \href{http://www.bangor.ac.uk/~map601/catgrp.abs.html}{http://www.bangor.ac.%
uk/~map601/catgrp.abs.html}.

\bibitem{MacLane}
S.~{Mac Lane}, {\em Categories for the Working Mathematician}.
\newblock Springer, 1998.

\bibitem{Brylinski}
J.-L. Brylinski, {\em Loop Spaces, Characteristic Classes and Geometric
  Quantization}.
\newblock Birkhauser, 1993.

\bibitem{CareyJohnsonMurray}
S.~J. A.~Carey and M.~Murray, {\it Holonomy on {D}-branes}, . \hepth{0204199}.

\bibitem{CMR}
M.~Caicedo, I.~Mart{\'\i}n, and A.~Restuccia, {\it Gerbes and duality},  {\em
  \ap{300}{2002}{32}}. \hepth{0205002}.

\bibitem{Hitchin}
N.~Hitchin, {\it Lectures on special lagrangian submanifolds}, .
  \Math{DG}{9907034}.

\bibitem{Keurentjes}
A.~Keurentjes, {\it Flat connections from flat gerbes},  {\em
  \forp{50}{2002}{916}}. \hepth{0201072}.

\bibitem{Ehresmann:1966}
C.~Ehresmann, {\it Introduction to the theory of structured categories},  1996.
\newblock Technical Report Univ. of Kansas at Lawrence.

\bibitem{Borceux:1994}
F.~Borceux, {\em Handbook of Categorical Algebra 1: Basic Category Theory}.
\newblock Cambridge U. Press, 1994.

\bibitem{Moerdijk:2002}
I.~Moerdijk, {\it Introduction to the language of stacks and gerbes}, .
  \Math{AT}{0212266}.

\bibitem{Chatterjee:1998}
D.~Chatterjee, {\em On {G}erbs}.
\newblock PhD thesis, University of Cambridge, 1998.
\newblock
  \href{http://www.ma.utexas.edu/~hausel/hitchin/hitchinstudents/chatterjee.pd%
f}{http://www.ma.utexas.edu/~hausel/hitchin/hitchinstudents/chatterjee.pdf}.

\bibitem{CareyJohnsonMurrayStevensonWang:2004}
A.~Caray, S.~Johnson, M.~Murray, D.~Stevenson, and B.~Wang, {\it Bundle gerbes
  for {C}hern-{S}imons and {W}ess-{Z}umino-{W}itten theories}, .
  \Math{DG}{0410013}.

\bibitem{GetzlerJonesPetrack:1991}
E.~Getzler, D.~Jones, and S.~Petrack, {\it Differential forms on loop spaces
  and the cyclic bar complex},  {\em Topology} {\bf 30} (1991), no.~3 339.

\bibitem{CaetanoPicken:1993}
A.~Caetano and R.~Picken, {\it An axiomatic definition of holonomy},  {\em Int.
  J. Math.} {\bf 5} (1993), no.~6 835--848.

\bibitem{Hofman:2002}
C.~Hofman, {\it Nonabelian 2-forms}, . \hepth{0207017}.

\bibitem{Chepelev:2002}
I.~Chepelev, {\it Non-abelian {W}ilson surfaces},  {\em \jhep{02}{2002}{013}}.

\bibitem{Polchinski:1998}
J.~Polchinski, {\em String Theory}.
\newblock Cambridge University Press, 1998.

\bibitem{KarpMansouriRno:1999}
R.~Karp, F.~Mansouri, and J.~Rno, {\it Product integral formalism and
  non-abelian {S}tokes theorem},  {\em \jmp{40}{1999}{6033}}. \hepth{9910173}.

\end{thebibliography}\endgroup

\end{document}